\documentclass[a4paper,twocolumn,unpublished,11pt]{quantumarticle}
\pdfoutput=1
\usepackage[utf8]{inputenc}
\usepackage[numbers,sort&compress]{natbib}
\usepackage{graphicx,amsmath,amsfonts,enumerate,amsthm,amssymb,mathtools,enumitem,thmtools,hyperref,mathdots,enumitem,centernot,bm,soul,bbm}
\usepackage[capitalise, noabbrev]{cleveref}
\usepackage{float}
\usepackage{physics}
\usepackage{xcolor,colortbl}
\usepackage{mathrsfs}  
\graphicspath{{Figures/}}
\hypersetup{colorlinks=true,linkcolor=blue,citecolor=blue,urlcolor=blue}
\definecolor{mygreen}{rgb}{0.01, 0.31, 0.59}
\definecolor{myblue}{rgb}{0.01, 0.31, 0.59}
\definecolor{bluecyan}{rgb}{0.27, 0.66, 0.88}
\hypersetup{
  colorlinks=true,   
  linkcolor=myblue,  
  citecolor=mygreen, 
  urlcolor =myblue    
}

\def\B{ {\cal B} }

\def\E{ {\cal E} }
\def\S{ {\cal S} }
\def\O{ {\cal O} }

\def\M{ {\cal M} }
\def\F{ {\cal F} } 
\def\H{ {\cal H} }

\def\>{\rangle}
\def\<{\langle}

\renewcommand{\v}[1]{\ensuremath{\boldsymbol #1}}

\definecolor{ppblue}{RGB}{46,117,182}
\definecolor{ppred}{RGB}{197, 90, 17}

\theoremstyle{plain}
\newtheorem{thm}{Theorem}

\newtheorem{obs}[thm]{Observation}
\theoremstyle{definition}

\begin{document}

\title{Resource theory of Absolute Negativity}

\author{Roberto Salazar}
\email{roberto.salazar@uj.edu.pl}
\affiliation{Faculty of Physics, Astronomy and Applied Computer Science, Jagiellonian University, 30-348 Kraków, Poland}

\author{Jakub Czartowski}
\email{jakub.czartowski@doctoral.uj.edu.pl}
\affiliation{Faculty of Physics, Astronomy and Applied Computer Science, Jagiellonian University, 30-348 Kraków, Poland}

\author{A. de Oliveira Junior}
\email{alexssandre.oliveirajunior@uj.edu.pl}
\affiliation{Faculty of Physics, Astronomy and Applied Computer Science, Jagiellonian University, 30-348 Kraków, Poland}


\begin{abstract} 
A crucial goal of quantum information is to find new ways to exploit the properties of quantum devices as resources.    One of the prominent properties of quantum devices of particular interest is their negativity in quasi-probability representations, intensively studied in foundational and practical investigations. In this article, we introduce the concept of Absolute Negativity to characterise the negativity of sets of quantum devices in a basis-independent way. Moreover, we provide a resource theory for our relational notion of Absolute Negativity, which applies to sets of quantum state-measurement pairs.
Additionally, we determine a complete hierarchy of upper bounds for resource measures, which allows for estimating the resources of a set of devices. We demonstrate operational interpretations of our resource theory for communication and output-estimation advantages over state-measurement pairs with a classical probability representation. Furthermore, we illustrate the newly introduced concepts with an exhaustive analysis of a simple case of four qubit state-measurement pairs. Finally, we outline possible generalisations, applications and open questions.
\end{abstract}

\maketitle
\vspace{-0.4 cm}
\section{Introduction}
\label{sec:introduction}
\vspace{-0.2 cm}
Resource theories provide a robust theoretical framework to characterise the value of quantum devices, widely used in the field of quantum information \cite{Gour2019, Horodecki2003,Horodecki2009,Streltsov2017,TakagiRegula2019,Oszmaniec2019,Paul2019,Regula2020}. Traditionally, a resource theory focuses on devices represented by either states, measurements or channels separately as the object of study, which it orders according to their usefulness for a practical task. On the other hand, fields that involve a diversity of multiple interconnected devices, such as  Quantum Computers and Quantum Internet, are developing rapidly. Such structures allow a wide range of tasks to be solved and usually are classified according to their functional capabilities \cite{Wehner2018}. Naturally, such networks exploit the properties of sets of devices, the most elementary being information encode-decode pairs. The challenge of harnessing the properties of such composite devices demands an extension of the standard scope of resource theories.

Indeed, recent research has formulated theories for composite devices such as pairs of measurements \cite{Paul2019,Roope2019,Buscemi2020}, state-measurement pairs \cite{Ducuara2020}, state and context of observables  \cite{Martins2020} as well as finite sets of states (or measurements) \cite{Designole2020}. The corresponding resource theories study composite devices --denoted multi-objects \cite{Ducuara2020}-- to determine how to take advantage of relational properties, such as incompatibility \cite{Paul2019,Roope2021} or set coherence \cite{Designole2020} as resources for operational tasks.

However, a relational property that resource theories of composite devices have so far disregarded is negativity in a given quasi-probability representation \cite{Ferrie2009,Ferrie2011}. Precisely, it ties into quantum optics \cite{Glauber1965}, measures of nonclassicality \cite{Kenfack2004,Chuan2020}, quantum computation \cite{Eisert2012}, and quantum probability estimation \cite{Pashayan2015}.  Without studying quasi-probabilities, one might think that the intersection of all these subjects would be the empty set. On the contrary, the above representations are fruitful for the various fields in which they are applied, the Wigner function being the best known and studied \cite{Wigner1932,Kenfack2004,Ferrie2009,Ferrie2011}.
 
When dealing with a single information encoding device represented by a quantum state, there always exists a specific basis in which it is representable by a classical probability distribution. On the other hand, when considering a set of encoding devices, it may be impossible to simultaneously make all their quasi-probability functions non-negative by choosing a specific basis. Similarly, it could happen for a set of encoding-decoding devices with quantum state-measurement representations. The above motivates us
to introduce a notion of negativity for composite devices, independent of any specific basis: \emph{Absolute Negativity}. Concretely, we present two
quantifiers that minimise the average negativity of a set of encoding-decoding devices over all unitarily equivalent  representations.

Moreover, we show that each of the introduced quantifiers determines a different operational advantage for devices with non-zero Absolute Negativity over those without Absolute Negativity. In the case of the quantifier based on robustness, the advantage translates into augmented performance in discrimination tasks, while the one based on sum-negativity provides a measure of the higher sampling cost for estimating outputs. Additionally, we present a complete hierarchy of measures that provide upper bounds approximating the previous quantifiers. Furthermore, we illustrate the efficiency of our methods by studying the Absolute Negativity in selected families of sets of  qubit system devices .

Since Absolute Negativity is an independent property from distributed entanglement, it is potentially valuable in prepare-and-measure quantum networks. The advantages in quantum communication and computing demonstrated in this work contribute to the development of applications of this quantum resource in such networks.

\vspace{-0.4 cm}


\section{Negativity of single devices} \label{sec:single_ob_neg}

Quantum formalism associates a Hilbert space $\mathcal{H}$ to every physical system, and describes the state of the system by a density operator $\rho\in\mathcal{B}(\mathcal{H})$~\cite{peres1995quantum}. 
These are self-adjoint and positive linear operators on $\mathcal{H}$ that are also normalised to a unit trace. In this work, we will limit ourselves to the study of finite-dimensional systems. Additionally, to describe measurements on a system, the formalism employs positive-operator-valued measures (POVMs) with the possible outcomes given by non-negative operators called effects $E_i$ comprising a POVM. The effects sum to identity, belong to the dual space $\mathcal{B}^*(\mathcal{H})$ and the probability of observing a given outcome $i$ when measuring a system in a state $\rho$ is given by $p_i = \Tr(\rho E_i)$~\cite{Keyl2002}. Moreover, quantum systems evolve under the action of completely positive trace-preserving (CPTP) maps~$\E$, named quantum channels.

A complementary description is given by quasi-probability representations on phase space defined as normalised distributions allowed to take on negative values \cite{hillery1984,Lee1995}. Here we introduce the above description through the unified formalism of frames \cite{Ferrie2009,Ferrie2011}.

A frame is a generalisation of the notion of a basis for $\mathbb{H}(\mathcal{H})$, which is the set of Hermitian operators on $\H$ equipped with the usual Hilbert-Schmidt norm $\|\! \cdot\!\|$. Thus, for a measure space $\Omega$ with cardinality $d^2 \leq |\Omega| < \infty$, a frame is formally defined as a set of operators $\mathbb{V} := \{V(\alpha) : \alpha \in \Omega\} \subset \mathbb{H}(\mathcal{H})$ satisfying
\begin{equation}
    a \| A\|^2 \leq \sum_{\alpha \in \Omega} \tr{V(\alpha) A}^2 \leq b \| A\|^2 \, ,
\end{equation}
for all $A \in \mathbb{H}(\H)$ and certain $a, b > 0$. The mapping $A \mapsto \tr{V(\alpha) A}$ is called a frame representation of $\mathbb{H}(\H)$. Moreover, a frame $\mathbb{G} = \{G(\alpha) \}$ satisfying a linear reconstruction formula
\begin{equation}
\label{eq:frameoperator}
    A = \sum_{\alpha \in \Omega} \tr{V(\alpha) A} G(\alpha) \, ,
\end{equation}
for all $A \in \mathbb{H}(\H)$ is a dual frame to $\mathbb{V}$.

Following Ref.~\cite{Pashayan2015}, a quasi-probability distribution on $\Omega$ associated with a quantum state $\rho$ and a measurement effect $E$ is defined as 
\begin{align}
\label{eq-general_wigner}
    W_{\alpha} (\rho) &= \tr{V(\alpha) \rho} , \\
    W(E | \alpha)  &= \tr{E G(\alpha)} ,
    \end{align}
where $W_{\alpha}(\rho)$ and $W(E | \alpha)$ are real-valued and satisfy the normalisation conditions. In this context, the Born rule defining the probability $\mathrm{Pr}(E|\rho)$ of observing an outcome associated with the effect $E$ for a system in a state $\rho$ takes the form of

\vspace{-0.5 cm}

 \begin{equation}
    \mathrm{Pr}(E|\rho) = \sum_{\alpha \in \Omega} W(E|\alpha) W_{\alpha}(\rho).
 \end{equation}
 
 \vspace{-0.2 cm}

A concrete construction of a discrete quasi-probability for the case of a prime dimension $d$ with measure space $\Omega= \mathbb{Z}_d \times \mathbb{Z}_d$ follows by adopting the frame operators of the form

\vspace{-0.7 cm}

\begin{equation}
V(\alpha)=\frac{1}{d^2}\sum_{j,m=0}^{d-1}
\omega^{pj-qm+\frac{jm}{2}}  X^j  Z^m,  \label{eq:wootters-operator}
\end{equation}
where $\alpha = (q,p)$ is a point in the phase space, $\omega = e^{2\pi i/d}$, and $X$ and $Z$ are generalised Pauli operators defined as the clock operator \mbox{$X\ket{i} = \ket{i\oplus1\,\text{mod}\,d}$} and the phase operator \mbox{$Z\ket{i} = \omega^{i}\ket{i}$} with \mbox{$\omega = \exp(2\pi i/d)$}. Correspondingly, the dual is given by $G(\alpha) = d V(\alpha)$. The previous example of quasi-probability representation is the extensively studied \emph{Wigner function}~\cite{Leonhardt1996,Gibbons2004, Klimov2006,Gross2007}.

\begin{figure}[H]
    \centering
    \includegraphics[width = \columnwidth]{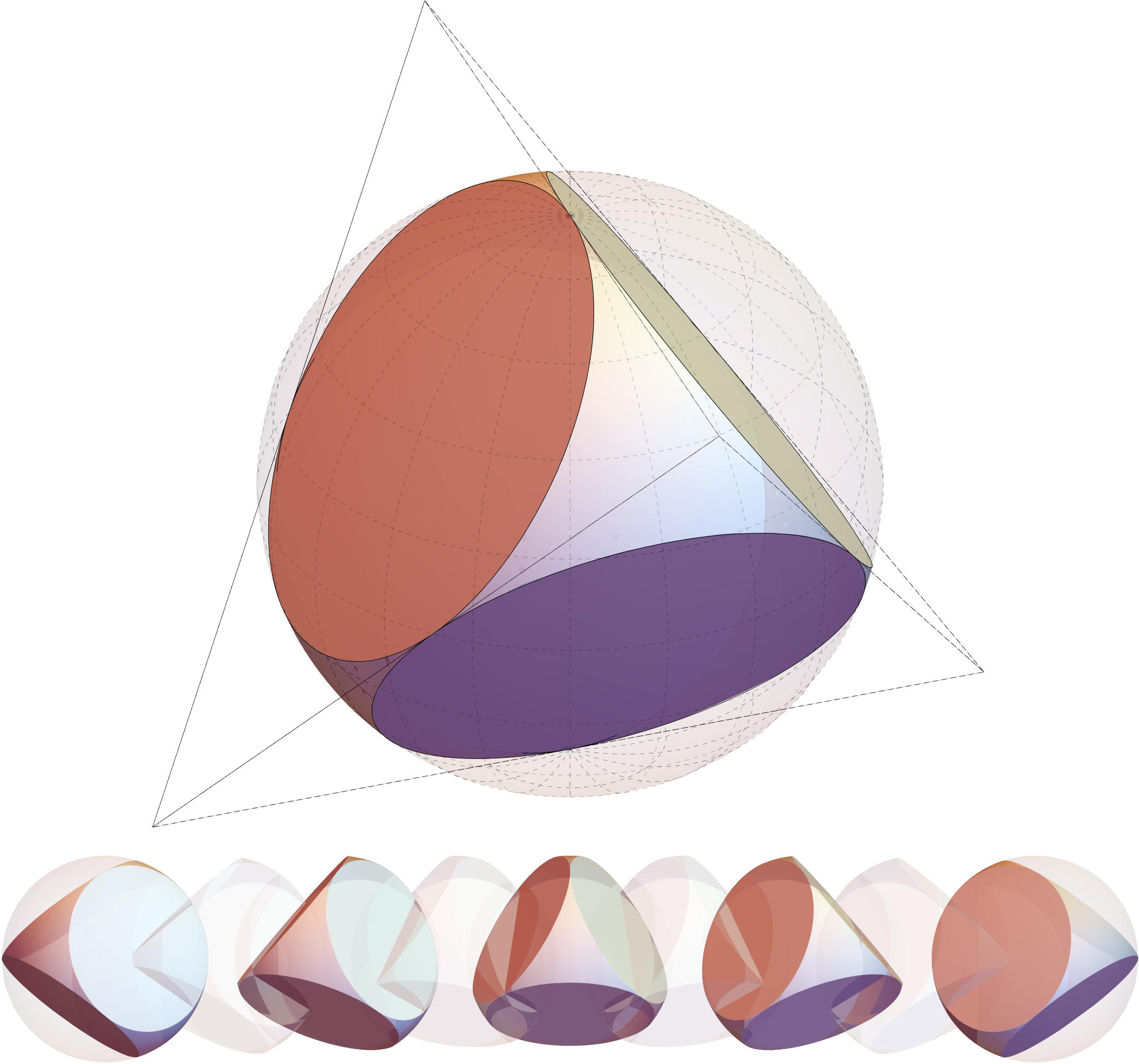}
    \caption{
    Set of free states with a positive Wigner function represented in the Bloch sphere.
    Geometrically, this region is given by the intersection of a tetrahedron and a ball tangent to its edges.
    For convenience in the lower part of the figure we provide the view of the free set from varying angles, equivalent to different choice of basis, rotated by a given angle with respect to the $z$ axis.}
    \label{fig_Wignershape}
\end{figure}

Given that quasi-probability distributions may contain intrinsic negativity, it is natural to ask for the set of states with a positive Wigner function. Focusing on the qubit case and considering a general state with Bloch vector $\v{r} = (x,y,z)$, one first applies Eq.~\eqref{eq-general_wigner} with the frame specified by Eq.~\eqref{eq:wootters-operator}, and imposes that the components of the Wigner function are non-negative for each point of the phase space, i.e., $W_{\alpha}(\rho) \geq 0$. As a result, we obtain:

\vspace{-0.4 cm}
 
\begin{align}
\begin{split}
\label{eq:dotproduct_rvectors}
 \v{r}\cdot \hat{\v{\beta}}_i  &\leq \frac{1}{\sqrt{3}} \quad \text{for}\! \quad 
 i = 1,...,4 ,
\end{split}
\end{align}
where $||\v{r}||^2 \leq 1 $ and $\hat{\v{\beta}}_i$ corresponding to the set of unit vectors given by $\hat{\v{\beta}}_1\! = \!\frac{1}{\sqrt{3}}(-1,-1,-1)$, $\hat{\v{\beta}}_2 \!=\! \frac{1}{\sqrt{3}}(-1,1,1)$, $\hat{\v{\beta}}_3 \!=\!\frac{1}{\sqrt{3}} (1,1,-1)$ and $\hat{\v{\beta}}_4 \!=\! \frac{1}{\sqrt{3}}(1,-1,1)$. The set of qubit states described by a non-negative Wigner function forms a convex subset depicted in the Bloch ball in Fig.~\ref{fig_Wignershape}. 
Physically, for the set of states satisfying (\ref{eq:dotproduct_rvectors}), the Wigner function realises a classical simulation through the phase space $\alpha = (q,p)$ and distribution (\ref{eq-general_wigner}). In this initial setting, the states with Bloch vector $\v{r}=\hat{\v{\beta}}_i$ are the harder to simulate, since their Wigner function (\ref{eq-general_wigner}) has the largest negative terms. However, the situation changes once we grant access to all unitarily equivalent classical distributions. In the latter case, we can choose any unitary $U\in \mathrm{SU(2)}$ and simulate with  non-negative Wigner functions after rotating frame operators $\{ V(\alpha),\, G(\alpha) \}$  by $U$ into  $\{ U^{\dagger}V(\alpha)U,\,  U^{\dagger}G(\alpha)U \}$. In this extended setting, the states with $\v{r}=\hat{\v{\beta}}_i$ are easy to simulate, even simultaneously, illustrating the drastic difference between the simple simulation based on the standard Wigner function and an adaptive strategy including unitary transformations.
\section{Resource theories}
\label{sec:RT}
\vspace{-0.2 cm}
Resource theory is a framework applied to explore the practical advantages of objects in a given set $\mathcal{S}$.
Formally, it is defined as a triple $\{\mathcal{F}, \mathcal{O}, \mathcal{M}\}$,  where the \emph{free set} $\F \subset \mathcal{S}$ is the subset of objects lacking the property of interest. Likewise, the set of \emph{free operations} $\mathcal{O}$ is a set of maps $\phi : \mathcal{S} \rightarrow \mathcal{S}$ preserving the set of free objects and the \emph{monotone} $\M: \S \mapsto [0, +\infty[$ is a resource quantifier satisfying $\M(\v{f}) = 0$ for all $\v{f} \in \F$ and is monotonically non-increasing under free operations: $\M(\phi(\v{s})) \leq \M(\v{s})$ for all $\phi \in \O$ and $\v{s} \in \S$. In what follows, we will distinguish resource theories of states and measurements by the superscripts $(s)$ and $(e)$, for the corresponding sets and measures.

A typical resource quantifiers are the generalised robustness of states \cite{Takagi2019},
\begin{equation}
\label{robustness}
\mathcal{R}_{\mathcal{F}^{(s)}}(\rho) \equiv  \min_{\sigma \in \B(\H)} \left \{ \lambda \geq 0 \: \Big{|} \: \frac{\rho + \lambda\sigma}{1+\lambda} \in \mathcal{F}^{(s)}\right  \},
\end{equation}
and the generalised robustness of measurement, similarly defined as
\begin{equation}    
\mathcal{R}_{\mathcal{F}^{(e)}}(\mathbb{M}) \equiv  \min_{\mathbb{N} \in \mathrm{POVM}} \left \{ \lambda\geq 0 \: \Big{|} \: \frac{\mathbb{M} + \lambda\mathbb{N}}{1+\lambda} \in \mathcal{F}^{(e)}\right \} , \label{robustnesseffect}
\end{equation}
with $\mathbb{N}\in \mathrm{POVM}$ being a measurement acting on the same system as $\mathbb{M}=\left\{ M_{a}\right\} _{a}$. A natural set of free operations on measurements are classical simulations, also known as  classical post-processing (CPP) operations:
\begin{equation}
\Xi_{\left\{ p\left(x\mid a\right)\right\} }\left(\mathbb{M}\right)=\left\{  \sum_{a}p\left(x\mid a\right)M_{a}\right\} _{x} .
\end{equation}
Remarkably the generalised robustness of measurement is known to be non-increasing under CPP 
when $\mathcal{F}^{(e)}$ is closed under CPP operations \cite{Paul2019}.

The choice of monotone depends exclusively on the task under consideration.  Here, we will use the former monotone, but also the state and effect sum-negativity:
 \begin{eqnarray}
 \mathcal{N}^{(s)}(\rho) & = & \sum_{\alpha \in \Omega}  |W_{\alpha}(\rho)|-1\,,\label{eq:StaNeg}\\
 \mathcal{N}^{(e)}(E) & = & \sum_{\alpha \in \Omega}| W(E|\alpha)| \,. \label{eq:EfecNeg}
 \end{eqnarray}
The effect sum-negativity can be extended to measurements $\mathbb{M}=\left\{ M_{a}\right\} _{a}$ as: 
 \begin{equation}
\mathcal{N}^{(\mathbf{e})}\left(\mathbb{M}\right)=\max_{a}\left[\frac{\mathcal{N}^{(e)}\left(M_{a}\right)}{\mathrm{Tr}\left(M_{a}\right)}\right]-1 .\label{eq:MeasNeg}
\end{equation}
Furthermore, we will denote by $\mathcal{N}_{[U]}^{(s)}(\rho)$, $\mathcal{N}_{[U]}^{(e)}(E)$ and $\mathcal{N}_{[U]}^{\mathbf{(e)}}\left(\mathbb{M}\right)$ the previous sum-negativities, but with the initial frame operators rotated into $\{ U^{\dagger}V(\alpha)U,\,  U^{\dagger}G(\alpha)U \}$  by  an  unitary $U\in\mathrm{SU(d)}$.  This notation will be helpful to present our results in a basis-independent way.


\section{Absolute Negativity}
\label{sec:AN}
\vspace{-0.2 cm}
An encoding-decoding pair of devices form the essential equipment for communication tasks, and quantum communications are no exception. They play a crucial role in various quantum technologies, including prepare-and-measure quantum networks \cite{Wehner2018}, cryptography \cite{Ekert1991}, shallow quantum circuits \cite{bravy2018} and dimension witnesses \cite{Brunner2013}. Due to the aforementioned transversal relevance in quantum information, encoding-decoding pairs will be our primary object of study.

In quantum formalism, an encoding-decoding pair is represented mathematically by a state-measurement pair $\left(\rho,\mathbb{M}\right)$. Each pair already forms a composite device and, therefore, can be designated as a multi-object; however, we intend to investigate the joint value of larger communication equipment, precisely a finite and ordered set  $\{\left(\rho_{1},\mathbb{M}_{1}\right),\left(\rho_{2},\mathbb{M}_{2}\right),...,\left(\rho_{n},\mathbb{M}_{n}\right)\}$ of encoding-decoding pairs. Thus, our theory considers multi-objects of the form:

\vspace{-0.6 cm}
\begin{eqnarray}
\left(\vec{\rho},\vec{\mathbb{M}}\right)  =  \left\{ \left(\rho_{j},\mathbb{M}_{j}\right)\right\} _{j=1}^{n} , \label{statemesmultiobj}
\end{eqnarray}
where all states $\rho_j$ are defined on a Hilbert space of dimension $d$ and measurements  $\mathbb{M}_{j}=\left\{ M_{a| j}\right\} _{a}$ act on the same Hilbert space of $\rho_j$ when in the pair $(\rho_{j},\mathbb{M}_{j})$. Additionally, we will demonstrate the resourcefulness of the state multi-object $\vec{\rho} \! = \! \left\{ \rho_{j}\right\} _{j=1}^{n}$  and measurement multi-object $\vec{\mathbb{M}} \! = \! \left\{ \mathbb{M}_{j}\right\} _{j=1}^{n}$ which we can select from (\ref{statemesmultiobj}) keeping the same order as in $(\vec{\rho},\vec{\mathbb{M}})$. 

Our resource theory aims to characterise the minimal amount of negativity intrinsic to a set $(\vec{\rho},\vec{\mathbb{M}})$ of available encoding-decoding pairs and show how to exploit it for concrete tasks.
To achieve the above goal, we first define free sets of objects containing zero Absolute Negativity. Intuitively, the above free set consists of multi-objects, such that there exists a basis in which the corresponding quasi-probabilities are proper probability distribution for each encoding or decoding device composing the multi-object. Equivalently we can express the free set in the standard quasi-probability representation by allowing an arbitrary unitary transformation $U$ of the set:

\vspace{-0.1 cm}
\begin{equation}
\! \! \!\!\!\!\!\!\!\!\!\!\!\!\!\!\!\!\!\!\!\!\!\!\!\!\mathcal{W}^{(s)}_n[U]: \{\vec{\rho} \: | \:  \forall j, \alpha, W_{\alpha} (U \rho_j U^{\dagger}) \geq 0 \} ,
\end{equation}
\vspace{-0.5  cm}
\begin{equation}
\!\!\mathcal{W}^{(e)}_n[U]\!:\!\! \{\vec{\mathbb{M}} \: | \: \forall j,\:\! a,\:\!\! \alpha, W( UM_{a| j} U^{\dagger}|\alpha)\! \geq 0 \} ,
\end{equation}
   
\vspace{-0.8  cm}
\begin{equation}
\!\!\mathcal{W}_n[U]\!:\!\! \{(\vec{\rho},\vec{\mathbb{M}}) \: | \: \vec{\rho} \in \mathcal{W}^{(s)}_n[U], \vec{\mathbb{M}} \in  \mathcal{W}^{(e)}_n[U] \} .
\end{equation}
with $n$ standing for the number of states, measurements or state-measurement pairs in the corresponding  multi-object.
Moreover, whenever there exists a unitary $U$, such that the multi-object has a proper probability representation, we will simply write $ \vec{\rho} \in \mathcal{W}^{(s)}_n$,  $\vec{\mathbb{M}} \in  \mathcal{W}^{(e)}_n $ and $(\vec{\rho} ,\vec{\mathbb{M}}) \in  \mathcal{W}_n $ respectively. If no such unitary can be found, then the multi-object features nonzero Absolute Negativity (AN).

To find an appropriate measure, we focus first on single devices with a probability distribution representation in the given frame. For instance, 
\vspace{-0.1 cm}
\begin{equation}
    \mathcal{W}^{(s)}_1[I]: \{\sigma |\forall \alpha, W_{\alpha} (\sigma) \geq 0\} .
\end{equation}
are the states with a probability distribution on an initially chosen arbitrary basis
Similarly, $ \mathcal{W}^{(e)}_1[I]$ determine classical measurement devices in the previous basis. Since both sets $ \mathcal{W}^{(s)}_1[I]$ and $\mathcal{W}^{(e)}_1[I]$ are convex, we have several measures -- such as the generalised robustness -- that are well defined, as it was mentioned in Section \ref{sec:RT}. The above fact will be useful in constructing appropriate multi-object monotones $\M_{\mathcal{W}_n[U]}$ in every basis. In order to remove the dependency on a particular basis, we minimise the monotonic measures $\M_{\mathcal{W}_n[U]}$ with respect to any possible basis choice:
\vspace{-0.3 cm}
\begin{equation}
    \M_n(\vec{\rho},\vec{\mathbb{M}}) = \underset{U}{\text{inf}} \: \M_{\mathcal{W}_n[U]} (\: \vec{\rho}\:,\! \:\vec{\mathbb{M}}\:) ,\label{inv:mon3}
\end{equation}

\vspace{-0.3 cm}

where the minimisation is performed over all unitaries $U$ acting on $\mathcal{H}$. Clearly, the above quantities are basis-independent and correspond to the intrinsic (or minimal) amount of negativity present in the multi-object $(\vec{\rho},\vec{\mathbb{M}})$. In this sense, our notion of negativity adheres to the tradition of defining physical properties as invariant under groups of unitary transformations, such as absolute separability \cite{Karol2001}.

Concerning the free operations, we can apply simultaneously any unitary $U$ over every object: 

\vspace{-0.6 cm}

\begin{equation}
   U^{\otimes n}[(\vec{\rho},\vec{\mathbb{M}})]=\left\{ \left(U \rho_j U^{\dagger},U\mathbb{M}_{j}U^{\dagger}\right)\right\} _{j=1}^{n}, 
\end{equation}

\vspace{-0.2 cm}

 and leave the monotones (\ref{inv:mon3}) invariant. Additionally, we include  depolarising channels $\Theta_{\epsilon}$  on states or measurements as  free operations: 
\vspace{-0.2 cm}
\begin{eqnarray}
\Theta_{\epsilon}\left(\vec{\rho}\right) & = & \left\{ \left(1-\epsilon\right)\rho_{j}+\epsilon\frac{1}{d}I\right\} _{j},\\
\Theta_{\epsilon}(\vec{\mathbb{M}}) & = & \left\{ \left(1-\epsilon\right)\mathbb{M}_{j}+\epsilon\mathbb{I}\right\} _{j},
\end{eqnarray}
with $\frac{1}{d}I$ the maximally mixed state, $\mathbb{I}=\left\{ I_a\right\} _{a}$ the
identity measurement (with $I_a=I/d$ for all $a$) and $\epsilon\in\left[0,1\right]$ a noise applied
simultaneously to every state and measurement in the multi-object. These operations are sufficient to include consumption of the resource, model different noises and/or simulation scenarios
in our theory.

Additionally, for measurements, we include a simultaneous CPP operation over every $\mathbb{M}_{j}$:

\vspace{-0.6 cm}

\begin{equation}
\Xi_{\left\{ p\left(x\mid a\right)\right\} }\left(\vec{\mathbb{M}}\right)=\left\{ \sum_{a}p\left(x\mid a\right)M_{a\mid j}\right\} _{x,j}.
\end{equation}

\vspace{-0.2 cm}

The monotonicity under CPP is crucial for the operational interpretation of the robustness based quantifier, but irrelevant for the application of quantifiers based on sum-negativity. However, we include a proof that measurement sum-negativity is also non-increasing for CPP operations.
\vspace{-0.4 cm}


\section{Results}
\label{sec:Results}
\vspace{-0.2 cm}

Below we present our general results, valid for any quasi-probability representation and multi-objects $(\vec{\rho},\vec{\mathbb{M}})$ with $n$ pairs of devices acting on systems of finite dimension  $d$. The first result consists of the technique we developed for approximating basis-independent quantifiers, the effectiveness of which we demonstrate in the examples in Section \ref{sec:regular_cones}. Second, we demonstrate the operational value of a resource $(\vec{\rho},\vec{\mathbb{M}})$ in communication protocols and Third, we quantify the cost overload of $(\vec{\rho},\vec{\mathbb{M}})$ in circuit samplings for the output estimation task.

\subsection{Hierarchy of monotones} 
\label{ssec:hierarchy}
\vspace{-0.2 cm}

From the definition of Absolute Negativity of  multi-object (\ref{inv:mon3}), it follows that considering only unitaries $U$ in a discrete set $\mathcal{U}_{1}$ provides an upper bound to the exact measure:

 \vspace{-0.4 cm}
 
    \begin{equation}
        \mathcal{M}_{n}^{(1)}=\inf_{U\in\mathcal{U}_{1}}\M_{\mathcal{W}_n[U]} \geq\mathcal{M}_{n},
    \end{equation}
    
     \vspace{-0.2 cm}

\noindent where for simplicity, we dropped the argument $(\vec{\rho},\vec{\mathbb{M}})$ from the monotones $\mathcal{M} $ since it plays no role in the description of the technique, a license in notation, we use only by this subsection. In the same way, if we have another discrete superset $\mathcal{U}_{2}\supseteq\mathcal{U}_{1}$
then, 

   \vspace{-0.8 cm}
  
    \begin{equation}
        \mathcal{M}_{n}^{(1)}\geq\mathcal{M}_{n}^{(2)}=\inf_{U\in\mathcal{U}_{2}}\M_{\mathcal{W}_n[U]} \geq\mathcal{M}_{n}.
    \end{equation}
    
     \vspace{-0.4 cm}
    
Consequently, for any sequence of supersets $\mathcal{U}_{m}\supseteq\mathcal{U}_{m-1}...\supseteq\mathcal{U}_{1}$,
we have a hierarchy of upper bounds for $\mathcal{M}_{n}$:

 \vspace{-0.4 cm}

    \begin{equation}
        \mathcal{M}_{n}^{(1)}\geq\mathcal{M}_{n}^{(2)}\geq...\mathcal{M}_{n}^{(m)}\geq\mathcal{M}_{n}.
    \end{equation}
    
     \vspace{-0.2 cm}
    
To provide a complete hierarchy of upper bounds, our strategy is to construct a sequence $\mathcal{U}_{m}\supseteq\mathcal{U}_{m-1}...\supseteq\mathcal{U}_{1}$
such that the hierarchy approaches $\mathcal{M}_{n}$
in the infinity:

 \vspace{-0.4 cm}

    \begin{equation}
        \lim_{m\rightarrow\infty}\mathcal{M}_{n}^{(m)}=\mathcal{M}_{n}.
    \end{equation}

 \vspace{-0.2 cm}

A well-known approximation of the sphere is by means of the so-called
\emph{Geodesic polyhedrons} which are convex polyhedrons made of triangles
\cite{Popko2012}. A geodesic polyhedron can approximate the sphere by
starting from an inscribed regular solid, choosing an appropriate
division of the faces of the solid into regular triangles and then
projecting the vertices of each triangle to the sphere. To construct
our hierarchy, we start with a specific choice of regular solid as
step $0$ and in each step $m$ generate a strictly larger geodesic
polyhedron than in step $m-1$. Finally, the set of unitaries in each
set $\mathcal{U}_{m}$ are all different unitaries transforming vertices of the geodesic polyhedron at step $m$ into other vertices. 

We illustrate the construction of this hierarchy for the case of qubit states. First, note that the 3 mutually unbiased bases defined by vector pairs $\qty{\ket{0},\ket{1}}$, $\qty{\ket{+} = (\ket{0}+\ket{1})/\sqrt{2}, \ket{-} = (\ket{0}-\ket{1})/\sqrt{2}}$, and $\qty{\ket{\odot}\! =\! (\ket{0}\!+\!i\ket{1})/\sqrt{2}, \ket{\otimes}\! = \!(\ket{0}\!-\!i\ket{1})/\sqrt{2}}$. 
define the vertices of a regular octahedron $\diamondsuit$ inscribed in the Bloch sphere.

\begin{figure}[H]
    \centering
    \includegraphics[width=\columnwidth]{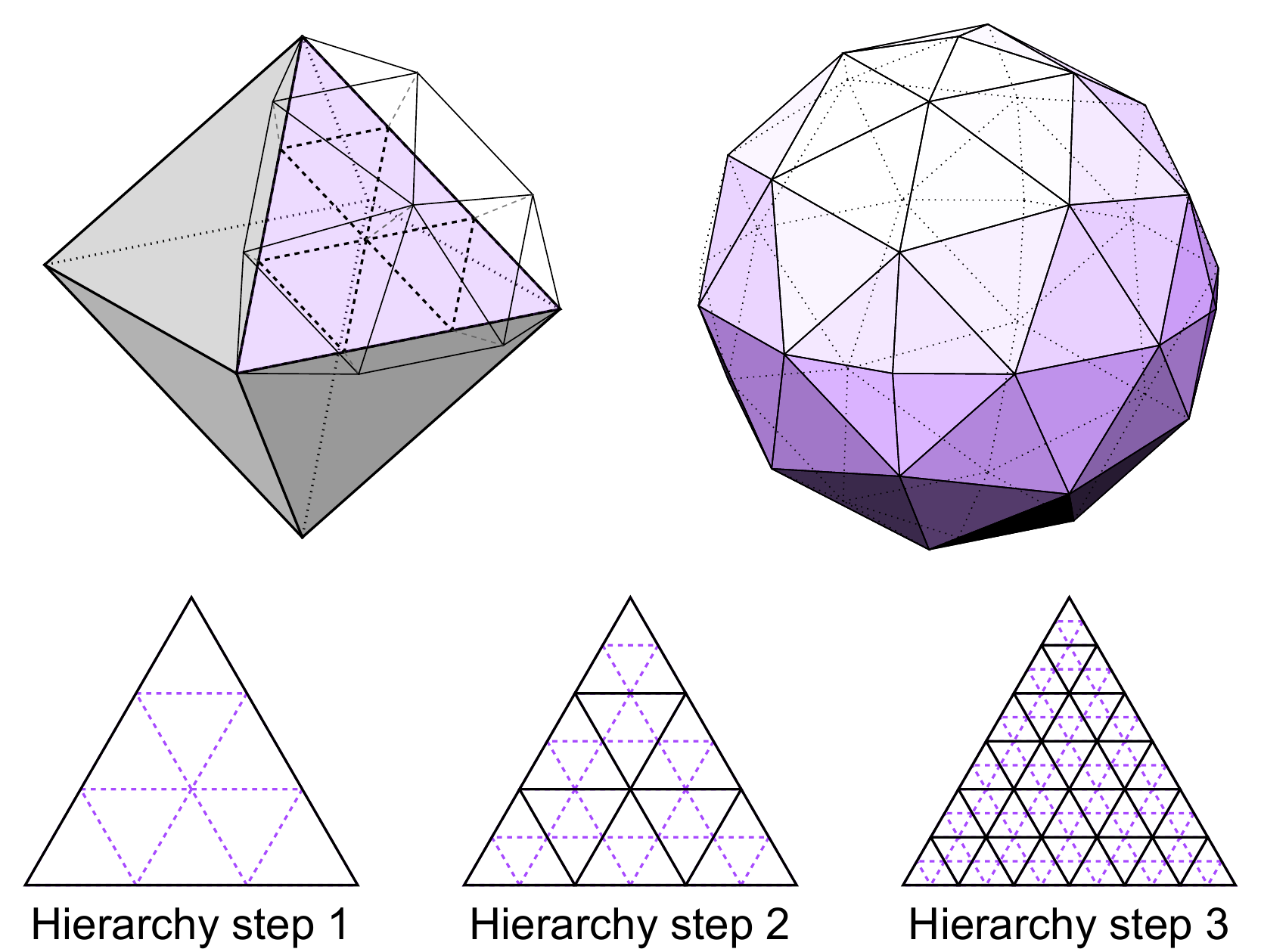}
    \caption{(a) In the first step of the hierarchy, a face is divided into nine equal triangles to cover its barycenter. (b) All subsequent steps involve further division into four times more parts. The resulting geodesic polyhedra give a good approximation of the sphere already for early steps.}
    \label{fig_hier_visualization}
\end{figure}

To construct our hierarchy we start with $\diamondsuit$
at step $0$ and then continue with the following iterative method to generate vertices
from the faces of $\diamondsuit$:

 \vspace{-0.1 cm}
    \begin{enumerate}
        \item In the first step $m=1$ we select a face of $\diamondsuit$ spanned by vertices $\hat{\mathbf{x}}_{1},\hat{\mathbf{x}}_{2},\hat{\mathbf{x}}_{3}$. Later we determine all points   $\mathbf{r}=(q_1/3)\mathbf{x}_{1}+(q_2/3)\mathbf{x}_{2}+(q_3/3)\mathbf{x}_{3}$ with barycentric coordinates  $q_1,q_2,q_3\in\mathbb{N}_{0}$, such that $q_{1}+q_{2}+q_{3}=3$. Because of the orthonormality of $\hat{\mathbf{x}}_{1},\hat{\mathbf{x}}_{2},\hat{\mathbf{x}}_{3}$ a  projection $\mathbf{r}\rightarrow\tilde{\mathbf{r}}$ of the vertices onto the surface of the Bloch sphere gives:
        \begin{equation}
\tilde{\mathbf{r}}=\frac{q_{1}}{\sqrt{\sum q_{k}^{2}}}\hat{\mathbf{x}}_{1}+\frac{q_{2}}{\sqrt{\sum q_{k}^{2}}}\hat{\mathbf{x}}_{2}+\frac{q_{3}}{\sqrt{\sum q_{k}^{2}}}\hat{\mathbf{x}}_{3}.\label{eq:vertexsphere}
\end{equation}
  The application of this operation to all faces of $\diamondsuit$ provides the vectors $\tilde{\mathbf{r}}$ that form the vertices of the first geodesic polyhedra, as depicted in Fig.~\ref{fig_hier_visualization}.

        \item Now, on step $m\geq2$ we add a vertex in the middle of each edge
        connecting a two barycentric vectors $\mathbf{r}$ from step $m-1$. The resulting set of vertices in the face of $\diamondsuit$ spanned by $\hat{\mathbf{x}}_{1},\hat{\mathbf{x}}_{2},\hat{\mathbf{x}}_{3}$ has barycentric coordinates:
        \vspace{-0.3 cm}
            \begin{equation}
                \mathbf{r}=\frac{q_{1}}{2^{m-1}3}\hat{\mathbf{x}}_{1}+\frac{q_{2}}{2^{m-1}3}\hat{\mathbf{x}}_{2}+\frac{q_{3}}{2^{m-1}3}\hat{\mathbf{x}}_{3},\nonumber
            \end{equation}
            
            \vspace{-0.2 cm}
            
        with the numbers $q_{1},q_{2},q_{3}\in\mathbb{N}_{0}$ taking all
        values satisfying $q_{1}+q_{2}+q_{3}=2^{m-1}3$. The corresponding
        set of projected vertices $\tilde{\mathbf{r}}$ is also described
        by equation (\ref{eq:vertexsphere}), but with the discrete values
        $q_{k}$ allowed in step $m$.
        
        \item We define the set $O_{m}$ as the unitaries inducing a rotation around the $\hat{\mathbf{z}}$ axis by angle $\phi = n\cdot \frac{\pi}{3\cdot2^{m}}$ with $n\in\mathbb{N}$ composed with   the group of unitaries that rotate the vector $\hat{\mathbf{z}} $ -- which is a vertex of $\diamondsuit$ by construction -- to every other vertex of the geodesic polyhedra at step $m$.
        
        \item Let us call $S_{\mathcal{W}}$ the group of unitaries that
        leave the free set invariant, then the unitaries $\mathcal{U}_{m}$ of the hierarchy at step $m$
        are $\mathcal{U}_{m}\cong O_{m}/S_{\mathcal{W}}$. In our case, the
        symmetries $S_{\mathcal{W}}$ are those of a simplex,
        which for qubit states is a regular tetrahedron.
    \end{enumerate}
    
We describe the generalisation of this construction of unitaries to arbitrary dimensions in the appendix \ref{app:ndimHier}, where we also show that for higher dimensions, keeping a similar approximation requires the level $m$ of the hierarchy to increase like  $  m(d) \propto \frac{2}{\log 2} \log (\frac{d}{12})$ for the Wigner representation. Additionally, in the same appendix we include a discussion distinguishing our hierarchy from analogous techniques, such as those in \cite{Oszmaniec20017}. 
\vspace{-0.4 cm}
\subsection{Advantage in discrimination tasks}
\label{ssec:discrim_adv}

\vspace{-0.2 cm}

We now present the operational interpretation of Absolute Negativity
in terms of a quantum discrimination task. Let us choose
the monotone $\mathcal{M}_{\mathcal{W}_{n}[U]}$ for a multi-object $(\vec{\rho},\vec{\mathbb{M}})$ in terms of the two generalized robustness quantifiers for states and measurements:
\begin{widetext}
\vspace{-0.4 cm}

\begin{align}
\mathcal{R}_{\mathcal{W}_{n}\left[U\right]}\left(\vec{\rho},\vec{\mathbb{M}}\right)&=\mathbb{E}_{j}\left[ \left(1+\mathcal{R}_{\mathcal{W}_{1}^{(s)}\left[U\right]}\left(\rho_{j}\right)\right)
\left(1+\mathcal{R}_{\mathcal{W}_{1}^{(e)}\left[U\right]}\left(\mathbb{M}_{j}\right)\right)\right]-1 ,\label{eq:SEmon1}
\end{align}
\end{widetext}
with $\mathcal{R}_{\mathcal{W}^{(s)}_1[U]}$ a generalized robustness defined
as in Eq.(\ref{robustness}) with free set $\mathcal{F}^{(s)}\rightarrow\mathcal{W}_{1}^{(s)}\left[U\right]$, similarly $\mathcal{R}_{\mathcal{W}_{1}^{(e)}\left[U\right]}$ is a generalized robustness of measurements with the free set given by
$\mathcal{F}^{(e)}\rightarrow\mathcal{W}_{1}^{(e)}\left[U\right]$ and $ \mathbb{E}_{j}$ indicates average over the $n$ elements of $(\vec{\rho},\vec{\mathbb{M}})$. Note that from the respective definitions follows the convexity of sets $\mathcal{W}_{1}^{(s)}\left[U\right]$ and $\mathcal{W}_{1}^{(e)}\left[U\right]$. Therefore, the corresponding robustness measures are efficiently computable and also have proper operational interpretations \cite{TakagiRegula2019}.
Taking the infimum over the unitary $U$ as in (\ref{inv:mon3}), we obtain the measure $\mathcal{R}_{n}$. Moreover, we show that if $\mathcal{R}_{n}(\vec{\rho},\vec{\mathbb{M}})>0$ there exists a specific discrimination game for which $(\vec{\rho},\vec{\mathbb{M}})$
provides an advantage over any multi-object in $\mathcal{W}_{n}$. Precisely, the value of $\mathcal{R}_{n}(\vec{\rho},\vec{\mathbb{M}})$
quantifies the relative advantage provided by $(\vec{\rho},\vec{\mathbb{M}})$
over any multi-object $(\vec{\tau},\vec{\mathbb{L}})$ with zero Absolute Negativity.

Such a task consists of subchannel discrimination, i.e., distinguishing
between different branches of a time evolution. The branches are modeled
as sets of completely positive maps $\Lambda=\left\{ \Gamma_{a}\right\} _{a}$
with the property that $\sum_{a}\Gamma_{a}$ is trace-preserving.
Given $\Lambda$ and a final measurement $\mathbb{N}=\left\{ N_{a}\right\} _{a}$,
the goal is to identify which subchannel $\Gamma_{a}$ has been applied.
If the resource is the initial state $\varrho$, the success probability
is given by:
\vspace{-0.4 cm}
\begin{equation}
p_{succ}\left(\varrho,\Lambda,\mathbb{N}\right)=\sum_{a}\mathrm{tr}\left(\Gamma_{a}\left(\varrho\right)N_{a}\right).
\end{equation}

\vspace{-0.2 cm}

Indeed, in Appendix \ref{app:op_proofs} we show the following:

\vspace{-0.1 cm}

\begin{thm}
\label{theorem1}
For any multi-object $(\vec{\rho},\vec{\mathbb{M}})$ and quantifier $\mathcal{R}_{\mathcal{W}_{n}\left[U\right]}$ defined as in Eq. (\ref{eq:SEmon1}) we have:

\vspace{-0.5 cm}

\begin{equation}
\inf_{U}\mathbb{E}_{j}\!\!\left[ \sup_{\Lambda_{j}}
\frac
{p_{succ}\left(\rho_{j},\!\Lambda_{j},\!\mathbb{M}_{j}\right)}
{\underset{(\tau,\mathbb{L})\in\mathcal{W}_{1}\left[U\right]}{\sup}\!\!p_{succ}\left(\tau,\!\Lambda_{j},\!\mathbb{L}\right)}\right] \!\!\!=\!1+\mathcal{R}_{n}(\vec{\rho},\vec{\mathbb{M}}),\label{eq:SEadv11}
\end{equation} 

\vspace{-0.2 cm}
\noindent
with $\mathcal{R}_{n}(\vec{\rho},\vec{\mathbb{M}})$ given by, 
\begin{equation}
    \mathcal{R}_{n}(\vec{\rho},\vec{\mathbb{M}}) = \inf_{U} \: \mathcal{R}_{\mathcal{W}_{n}\left[U\right]} (\: \vec{\rho}\:, \:\vec{\mathbb{M}}\:). \label{inv:monR}
\end{equation}

\end{thm}

\vspace{-0.2 cm}

Note that here the discrimination procedure over $\left\{ \Lambda_{j}\right\} $
depends on $U$. In other words, for any reference basis given by
$U$, there exists a set of subchannel discrimination tasks $\left\{ \Lambda_{j}\right\} $,
in which the multi-object $(\vec{\rho},\vec{\mathbb{M}})$ outperforms any multi-object
with a proper probability distribution representation in this reference frame, with
relative advantage given by $\mathcal{R}_{\mathcal{W}_{n}\left[U\right]}(\vec{\rho},\vec{\mathbb{M}})$.
Then, minimising this advantage over unitarily equivalent reference frames provides the
mean-robustness of the multi-object $\mathcal{R}_{n}(\vec{\rho},\vec{\mathbb{M}})$.

Additionally, in Appendix \ref{app:op_proofs} we explore particular cases of multi-objects where either the state or the measurement is free for all unitary $U$. In these particular cases, we also show how the monotones in Eq. (\ref{eq:SEmon1}) lead to independent AN measures for measurements $\vec{\mathbb{M}}$ or states $\vec{\rho}$ proportional to advantages in sub-channel discrimination tasks. The latter results allow us to imagine as though  $(\vec{\rho},\vec{\mathbb{M}})$ could  decompose into two multi-objects $\vec{\rho}$ and $\vec{\mathbb{M}}$. 

Theorem \ref{theorem1} characterises  Absolute Negativity
as an advantage for discrimination tasks, which are an essential subroutine in communication protocols. Consequently, multi-objects with non-zero AN provide better performance in some communication
protocols than those with zero AN. 

In particular, a channel $\Lambda$ with subchannels 
$\text{\ensuremath{\left\{ \Gamma_{x}\right\} }}_{x\in X}$ naturally induces a communication scheme that exploits an encoding-decoding device represented by a pair $(\rho,\mathbb{M})$. Precisely, if we identify the set X with a random variable, the application of $\Lambda$ to the state $\rho$ encodes $X$ in the ensemble $\left\{ \mu_{x},p(x)\right\} $ with $p(x)=\mathrm{Tr}\left[\Gamma_{x}\left(\rho\right)\right]$
and $\mu_{x}=\Gamma_{x}\left(\rho\right)/p(x)$. To access the information in $X$, we decode it using the measurement $\mathbb{M}=\left\{ M_{g}\right\} _{g\in G}$ with its outcomes generating a guess random variable $G$. For the sake of clarity, we add subindexes $X\rightarrow X_{\Lambda,\rho}$ and $G\rightarrow G_{\mathbb{M}}$  to the random variables of the protocol, which keep track of the device's dependence.  
A well-studied figure of merit for this communication protocol is the infinity mutual information \cite{wilde2017}:

\vspace{-0.6 cm}

\begin{equation}
I_{+\infty}\!\left(\!X_{\Lambda_{j},\rho_{j}}\!:\!G_{\mathbb{M}_{j}}\!\right)=H_{+\infty}\left(X_{\Lambda,\rho}\right)-H_{+\infty}\left(X_{\Lambda,\rho}\!\mid\! G_{\mathbb{M}}\right) ,
\end{equation}

\vspace{-0.2 cm}

\noindent
where,\begin{eqnarray*}
H_{+\infty}\left(X_{\Lambda,\rho}\right) & = & -\log\left\{ \max_{x}p(x)\right\} , \\
H_{+\infty}\left(X_{\Lambda,\rho}\mid G_{\mathbb{M}}\right) & = & -\log\left\{ \sum_{g}\max_{x}p(x)p(g\mid x)\right\} ,
\end{eqnarray*}

\vspace{-0.2 cm}

\noindent
with the 
probability $p(g\!\!\mid\!\! x)\!=\!\mathrm{Tr}\left[M_{g}\mu_{x}\right]$. The
$H_{+\infty}\left(X_{\Lambda,\rho}\right) $,  $H_{+\infty}\left(X_{\Lambda,\rho}\mid G_{\mathbb{M}}\right) $ are denoted by the min-entropy and min-conditional entropy, respectively. Using these quantities, we show in Appendix \ref{app:op_proofs} the following lower and upper bounds:

\begin{widetext}

\begin{equation}
\log\left[1\!+\!\mathcal{R}_{n}(\vec{\rho},\vec{\mathbb{M}})\right]\!\leq\inf_{U}\mathbb{E}_{j}\!\left[\sup_{\Lambda_{j}}\left\{ \!I_{+\infty}\!\left(\!X_{\Lambda_{j},\rho_{j}}\!:\!G_{\mathbb{M}_{j}}\!\right)\!-\!\!\!\sup_{(\tau,\mathbb{L})\in\mathcal{W}_{1}\left[U\right]}\!I_{+\infty}\!\left(\!X_{\Lambda_{j},\tau}\!:\!G_{\mathbb{L}}\!\right)\!\right\} \right]\!\leq\mathcal{R}_{n}(\vec{\rho},\vec{\mathbb{M}}) .
\label{commadv}
\end{equation}
\end{widetext}

Bounds~(\ref{commadv}) demonstrate how the robustness quantifier $\mathcal{R}_{n}(\vec{\rho},\vec{\mathbb{M}})$ determines the advantage of a pair of resources $(\vec{\rho},\vec{\mathbb{M}})$ against all free state-measurement pairs in the class of communication protocols induced by subchannels $\Lambda=\text{\ensuremath{\left\{ \Gamma_{x}\right\} }}_{x\in X}$ discrimination in terms of the amount of mean accessible information.

The above communication advantage could be crucial for early quantum networks~\cite{Wehner2018}, since AN does not necessarily involve distributed entanglement, and therefore it is exploitable in setups with access only to prepare-and-measure devices. 

\vspace{-0.3 cm}

\subsection{Estimation cost}
\label{ssec:est_cost}

\vspace{-0.2 cm}

The previous section presented how multi-objects $(\vec{\rho},\vec{\mathbb{M}})$ and multi-objects $\vec{\rho}$, $\vec{\mathbb{M}}$ with AN greater than
zero could be advantageous for communication protocols. Here, we show
a complementary value for multi-objects $(\vec{\rho},\vec{\mathbb{M}})$ in terms of independent contributions of $\vec{\rho}$, $\vec{\mathbb{M}}$ to the cost of output estimation tasks. In this way, we
obtain another operational meaning for our AN theory by showing
that a resource multi-object is more difficult to reproduce with classical estimation methods than a resourceless multi-object. 

A concrete example of estimation cost consists
of the number of samples needed to estimate the probability of a fixed
outcome $o$ (with  associated effect $E_o$) in a preparation-and-measure circuit. Suppose that we had access
to a quantum computer that implemented a circuit in this class. In
this case, we could use it to estimate the probability of a fixed
outcome by computing the observed frequency $f(o)$ of the outcome $o$
over $s(E_o)$ samples. By Hoeffding inequality \cite{Hoeff1963} $f(o)$ will be
within $\varepsilon$ of the quantum probability $p\left(o\right)$
with probability $1-\delta$ provided that the number of samples $s_{\varepsilon,\delta}(E_o)$
satisfies:

\vspace{-0.5 cm}

\begin{equation}
s_{\varepsilon,\delta}(E_o)\geq(1/2\varepsilon)\log\left(2/\delta\right).
\end{equation}

\vspace{-0.2 cm}

Reference~\cite{Pashayan2015} provides two efficient upper bound estimators for $s_{\varepsilon,\delta}(E_o)$,
based on the quasi-probability representation of the elements in the
circuit. The first estimator $s_{\varepsilon,\delta}^{(\rightarrow)}$ considers a forward circuit evolution, starting from a state and ending on a measurement, while the second estimator $s_{\varepsilon,\delta}^{(\leftarrow)}$ gives the cost in a time-reversal simulation, which interchanges the effect with the state up to a normalisation constant (see Appendix \ref{app:op_proofs} for explicit definitions). Since the AN optimizes over unitarily equivalent quasi-probability representations, we write $s_{\varepsilon,\delta}^{(\rightarrow)}(\rho, E_o)\left[U\right]$  and $s_{\varepsilon,\delta}^{(\leftarrow)}(\rho, E_o)\left[U\right]$
to emphasise the dependence of the sampling on the representation
associated with $U$ and include the dependence on the initial state $\rho$.

To quantify the cost of prepare-and-measure
circuits generated from a multi-object $(\vec{\rho},\vec{\mathbb{M}})$
a natural choice is the outcome $o$ that maximises the sampling
relative cost overload $s_{\varepsilon,\delta}(\mathrm{Resource})\left[U\right]/s_{\varepsilon,\delta}(\mathrm{Free})\left[U\right]$ in both forward or time-reversal estimation
and average over the available circuits $(\rho_{j},\mathbb{M}_{j})$.
We provide measures related to the above quantifier of the output estimation by selecting frame operators $\{ V(\alpha),\, G(\alpha) \}$ and applying the measures~(\ref{eq:StaNeg}), (\ref{eq:MeasNeg}) in frame operators rotated by $U$.  
Without loss of generality, we define the following  mean sum-negativity measures for all frames unitarily equivalent to $\{ V(\alpha),\, G(\alpha) \}$:
\vspace{-0.3 cm}
\begin{equation}\label{neg1}
\mathcal{N}_{n}^{(s)}(\vec{\rho})=\inf_{U}\mathbb{E}_{j}\left[\mathcal{N}^{(s)}_{\left[U\right]}\left(\rho_{j}\right)\right],
\end{equation}

\vspace{-0.6 cm}

\begin{equation}\label{neg2}
\mathcal{N}_{n}^{(\mathbf{e})}(\vec{\mathbb{M}})=\inf_{U}\mathbb{E}_{j}\left[\mathcal{N}_{[U]}^{(\mathbf{e})}\left(\mathbb{M}_j\right)\right].
\end{equation}

\vspace{-0.2 cm}

The estimators in reference~\cite{Pashayan2015} allow us to demonstrate the following theorem (see Appendix \ref{app:op_proofs}):
\begin{thm}
\label{theorem2}
For any multi-object $(\vec{\rho},\vec{\mathbb{M}})$, quantifiers  $\mathcal{N}_{n}^{(s)}\left(\vec{\rho}\right)$, $\mathcal{N}_{n}^{(\mathbf{e})}\left(\vec{\mathbb{M}}\right) $  defined as in Eq. (\ref{neg1}) and Eq. (\ref{neg2}) respectively, we have:

\vspace{-0.5 cm}
\begin{equation}
1\!+\!\mathcal{N}_{n}^{(s)}\left(\vec{\rho}\right)\!=\!\inf_{U}\mathbb{E}_{j}\!\left[\sqrt{\frac{\max_{o}s_{\varepsilon,\delta}^{(\rightarrow)}\left(\rho_{j},M_{o\mid j}\right)\left[U\right]}{\max_{o}s_{\varepsilon,\delta}^{(\rightarrow)}\left(\tau_{j},M_{o\mid j}\right)\left[U\right]}}\right]\!\!,\label{eq:estbound1}
\end{equation}
\vspace{-0.3 cm}
\begin{equation}
1+\mathcal{N}_{n}^{(\mathbf{e})}\left(\vec{\mathbb{M}}\right)\!\geq\inf_{U}\mathbb{E}_{j}\!\left[\sqrt{\max_{o}\frac{s_{\varepsilon,\delta}^{(\leftarrow)}\left(\rho_{j},M_{o\mid j}\right)\left[U\right]}{s_{\varepsilon,\delta}^{(\leftarrow)}\left(\rho_{j},L_{o\mid j}\right)\left[U\right]}}\right] ,\label{eq:estbound2}
\end{equation}
with each $\tau_{j}\in \mathcal{W}_{1}^{(s)}\left[U\right] $  and $ L_{o|j} \in \mathcal{W}_{1}^{(e)}\left[U\right] $ being the most costly free states and free effects in the corresponding sampling cost for a given $M_{o|j} $ or $\rho_{j}$, respectively.
\end{thm}
\vspace{-0.3 cm}
 The identity (\ref{eq:estbound1}) demonstrates $\mathcal{N}_{n}^{(s)}\left(\vec{\rho}\right)$ as a good quantifier of the relative cost increase in simulating circuits $(\rho_{j},\mathbb{M}_{j})$ over those in which states are consistently replaced by free states $\tau_{j}$. 
 Similarly, the bound~(\ref{eq:estbound2}) shows how $\mathcal{N}_{n}^{(\mathbf{e})}\left(\vec{\mathbb{M}}\right)$ could be applied to quantify the relative cost overload of the circuits in $(\vec{\rho},\vec{\mathbb{M}})$ over those in $(\vec{\rho},\vec{\mathbb{L}})$  with $\vec{\mathbb{L}}$ free for all unitarily equivalent frames to $\{ V(\alpha),\, G(\alpha) \}$. In a sense, the relations~(\ref{eq:estbound1})-(\ref{eq:estbound2}) correspond to a bottom-up approach to the multi-object advantages, in contrast to the top-down approach used in the communication scenarios. Furthermore, in Appendix \ref{app:op_proofs} we show how to build a monotone combining the contributions to the estimation cost of both $\vec{\rho}$ and $\vec{\mathbb{M}}$:
\begin{widetext}

\begin{equation}
1+\mathcal{N}_{n}(\vec{\rho},\vec{\mathbb{M}})\!\geq\inf_{U}\mathbb{E}_{j}\!\left[\sqrt{\!\max_{o^{\prime}}\frac{s_{\varepsilon,\delta}^{(\leftarrow)}\left(\rho_{j},M_{o^{\prime}\mid j}\right)\left[U\right]}{s_{\varepsilon,\delta}^{(\leftarrow)}\left(\rho_{j},L_{o^{\prime}\mid j}\right)\left[U\right]}\frac{\max_{o}s_{\varepsilon,\delta}^{(\rightarrow)}\left(\rho_{j},M_{o\mid j}\right)\left[U\right]}{\max_{o}s_{\varepsilon,\delta}^{(\rightarrow)}\left(\tau_{j},M_{o\mid j}\right)\left[U\right]}}\right],\label{eq:estbound3}
\end{equation}
\noindent
where  $\mathcal{N}_{n}(\vec{\rho},\vec{\mathbb{M}}) $ is defined as: 
\vspace{-0.1 cm}
\begin{equation}
\mathcal{N}_{n}(\vec{\rho},\vec{\mathbb{M}}) \!= \! \inf_{U}\mathbb{E}_{j}\!\left[\!\left(1+\!\mathcal{N}^{(s)}_{\left[U\right]}\left(\rho_{j}\right)\right)\!\left(1+\!\mathcal{N}_{[U]}^{(\mathbf{e})}\left(\mathbb{M}_j\right)\right)\!\right]\! - 1 . \label{defNegmult}
\end{equation}
\end{widetext}
However, we remark that the identity~(\ref{eq:estbound1}) and upper bound~(\ref{eq:estbound2}) are tighter than the one provided by $\mathcal{N}_{n}(\vec{\rho},\vec{\mathbb{M}})$, and consequently are more significant for the output estimation task. 
\vspace{-0.4 cm}


\section{Resources from regular polyhedral cones in the Bloch sphere}
\label{sec:regular_cones}
\begin{figure}[htb!]
    \centering
    \includegraphics{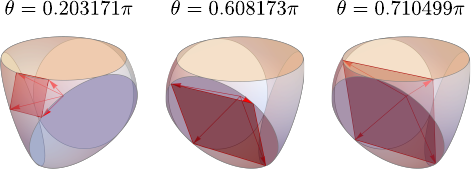}
    \caption{Depiction of state quadruplets for critical values of the opening angle $\theta$ in the free set.}
    \label{fig_crit_arrs}
\end{figure}

To illustrate the features of the Absolute Negativity, we study selected qubit multi-objects in the Wigner representation introduced in Section \ref{sec:single_ob_neg}, which from now on we denote as Absolute Wigner Negativity (AWN). Since we wish to highlight the invariance of our measures under unitary transformations, we simplify further and consider multi-objects whose resource comes exclusively from the encoding devices. To do this, we select multi-objects in the form $(\vec{\rho},\vec{\mathbb{L}})$ with $\vec{\mathbb{L}} \in  \mathcal{W}^{(e)}_n[U] $ for all $U$. A concrete example is a case in which each encoding-decoding pair applies the same measurement $\mathbb{L}_1= \mathbb{L}_2=...= \mathbb{L}_n= \mathbb{K}=\left\{ K_{i}\right\} _{i=1}^{4}$ , with effects $K_{i}=\left(1/4\right)I+\left(1/4\sqrt{3}\right)\hat{\beta}_{i}\cdot\vec{\sigma}$ (with every $\hat{\beta}_{i}$ same as in Eq.(\ref{eq:dotproduct_rvectors})) on the standard basis. Since each effect is inside the sphere inscribed in $\mathcal{W}^{(e)}_1[I]$, they also belong to each $\mathcal{W}^{(e)}_1[U]$. 

Following the above, the contribution to the AWN in $(\vec{\rho},\vec{\mathbb{L}})$ comes exclusively from the set of states $\vec{\rho}$ which in Appendix \ref{sapp:geom_exp_rob} we demonstrate to have a robustness measure: 

\vspace{-0.7 cm}

\begin{equation}
    \mathcal{R}_{\mathcal{W}_n[U]}(\vec{\rho}) \!=\! \frac{\sqrt{3}}{1\!+\!\sqrt{3}} \mathbb{E}_{j}\!\left[ \max_k\! \left\{\mathcal{I}^{+}\! \left[  \v{r}_j\!\cdot\! \hat{\v{\beta}}_k[U]\!   - \!\frac{1}{\sqrt{3}} \right]\right\} \right], \label{GeomRob}
\end{equation}
with $\mathcal{I}^{+}[x]=\max\{x,0\}$ the non-negative part of the identity function and $\hat{\v{\beta}}_k[U] $ the vectors from (\ref{eq:dotproduct_rvectors}) but in the standard basis rotated by $U$. Moreover, we also show in \ref{sapp:geom_exp_rob} that such robustness is proportional to the sum-negativity in the case of qubit states, i.e $\mathbb{E}_{j}\left[\mathcal{N}_{\left[U\right]}^{(s)}\left(\rho_{j}\right)\right]=\left(1+\sqrt{3}\right)\mathcal{R}_{\mathcal{W}_{n}\left[U\right]}\left(\vec{\rho}\right)$. Because of the  proportionality of measures, we continue our discussion of examples in terms of the robustness quantifier alone. 
The sets of qubit states $\vec{\rho}$ we investigate form a regular polygon in their Bloch representations. We say that such sets of states form regular polyhedral cones.
\begin{obs}
    A single-qubit state and sets of 2 and 3 pure-qubit states forming regular polyhedral cones have zero Absolute Wigner Negativity. 
\end{obs}
The set of states with a positive Wigner function, with respect to a fixed basis, is convex and includes pure qubits on the surface. Consequently, for every single qubit-state $\ket{\psi}$, one can always find a unitary $U$ such that $U\ket{\psi}$ will match one of the free pure qubit states.

\begin{figure}[htb!]
    \centering
    \includegraphics{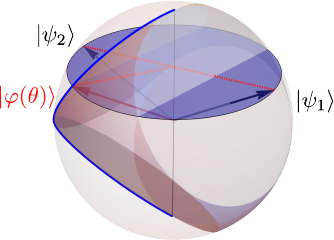}
    \caption{
    Continuous pole-to-pole path belonging to the surface of free set and the Bloch sphere together with an arbitrary pair of states and the rotation determined by $\ket{\varphi(\theta)}$.}
    \label{fig_continuous_path}
\end{figure}

For a pair of states, without loss of generality, one can assume that they are arranged in a regular cone, $\ket{\psi_1} = \cos \frac{\theta}{4} \ket{0} + \sin \frac{\theta}{4} \ket{1}$ and $\ket{\psi_2} = \sigma_z \ket{\psi_1}$,
where $\theta$ is the opening angle of the common cone. First, we note that the free set of Wigner negativity is symmetric with respect to the $z$ axis; therefore, $\ket{\psi_2}$ is free whenever $\ket{\psi_1}$ is. This allows us to restrict ourselves to $\ket{\psi_1}$. Next, we note that for every plane of constant $\theta$ there exists at least one free pure state $\ket{\varphi(\theta)}$. Finally, one can always find a rotation $U_z$ around the $z$ axis that $U_z \ket{\psi_1} = \ket{\varphi(\theta)}$. 

We present an analogous proof for three pure states arranged in a regular triangular cone under appropriate rotation in Appendix \ref{app:3_vec_proof}, together with the complete resource analysis for triplets of arbitrary pure states.\qed

Therefore, the simplest non-trivial case is that of four vectors forming a regular polyhedral cone, 

\vspace{-0.2 cm}

\begin{equation}
    \ket{\psi_j} = \begin{pmatrix} \cos \frac{\theta}{4} \\   i^{j-1}\sin \frac{\theta}{4}  \end{pmatrix},
\end{equation}

\noindent
with $j\in\{1, 2, 3, 4\}$ and $\theta$ being the opening angle of the cone. The first step to characterise the resource content of the states is to find for what opening angles $\theta$ the quadruplet of states is within the free set. It turns out that they are free only for $\theta\in\qty[0,0.2032\hdots\pi]\cup\qty[\arccos(-1/3),0.7105\hdots\pi]$. 
Details of the derivation of the extreme points are presented in Appendix \ref{app:ext_pts}, while in Fig. \ref{fig_crit_arrs} we present the extremal configurations and their orientation with respect to the free set.

\begin{figure}[H]
    \centering
    \includegraphics{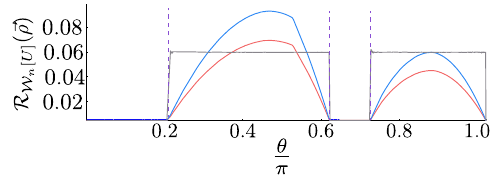}
    \includegraphics[width=\columnwidth]{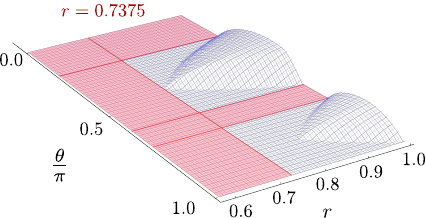}
   \caption{\footnotesize{ 
    Mean robustness (red) and mean sum-negativity (blue) plot for four pure states arranged in a regular polyhedral cone versus the opening angle $\theta$ (upper plot). The plot exhibits two disjoint regions of zero-resource sets, $[0,0.2032\pi]$ and $[\arccos(-1/3),0.7105\pi]$ highlighted by dashed purple vertical lines. Furthermore, sum-negativity and robustness  exhibit constant ratio (gray line), which is equal to $1 + \sqrt{3}$. Varying the length $r$ of the Bloch vectors (lower plot), one finds a free region for $r < 0.7375$ which is much larger than $r_\circ = \frac{1}{\sqrt{3}}$, the radius of the maximal ball inscribed in the free set.
    }}
    \label{fig_mean_dist_plot_2d}
\end{figure}

In the regimes $\theta\in]0.2032\pi,\arccos(-1/3)[$ and $ \theta\in]0.7105\pi,\pi[$ we estimate the resource of the set $\{\ket{\psi_j}\}_{j=1}^{4}$, resorting to a numerical annealing procedure, which converges to the optimal unitary rotation of the set. The resulting dependence of the mean robustness $\mathcal{R}_{\mathcal{W}_n[U]}(\vec{\rho})$ on the opening angle $\theta$ is depicted in Fig. ~\ref{fig_mean_dist_plot_2d}. It is important to note that the curves exactly predict zero in the free regimes, with non-zero values of the robustness only in the complementary regimes. 

Since the method is not limited to pure states, we have performed numerical annealing for states corresponding to Bloch vectors of length $r\in[0,1]$. As a result, we found that all sets of states with $r < 0.7375$ are free. This region is much larger than expected $r_\circ = \frac{1}{\sqrt{3}}$, the radius of the maximal inscribed ball in the free set.

\begin{figure}[H]
    \centering
    \includegraphics{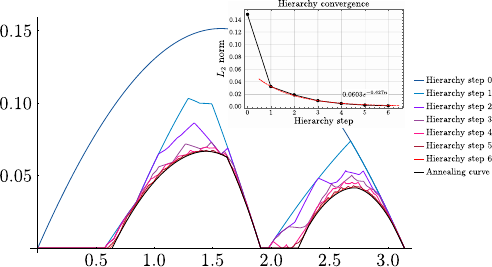}
    \caption{Depiction of the hierarchy for the mean robustness (colors from red to blue) compared to the actual curve obtained through the numerical annealing procedure (black). 
    The $L_2$ norm  between the actual curve and the  curve from the annealing shows convergence, following the exponential $0.0603 e^{-0.627 n}$ from step 1.
    }
    \label{fig_hier_plot}
\end{figure}

Now, we show the capacity of the upper bound hierarchy described in Section \ref{ssec:hierarchy}
to estimate the resource and further confirm the curves obtained from numerical annealing. We list the number of unitary operations in $O_m$ and $\mathcal{U}_m$ for each  step  $m\leq 6$   in Table \ref{tab:hierarchy_sizes}.
\vspace{-.4 cm}
Based on these discrete subsets of the unitary group, we calculated the hierarchical approximations for the AWN of four states arranged in a regular polyhedral cone. Qualitatively, \cref{fig_hier_plot} shows that every step of the hierarchy is above the value obtained from the numerical annealing approximation. In fact, computing the $L_2$ distance between the curves witnesses an exponential convergence of the hierarchy.

To summarize, when considering regular arrangements of qubit states, a minimum of four states is necessary to find configurations possessing nonzero Absolute Negativity. We analytically derive the extremal values of the opening angle of the regular polyhedral cone lying on the boundary between free and non-free regimes of quadruplets. Using the numerical annealing procedure, we determined actual values of mean robustness of Absolute Wigner Negativity for non-free regimes.

\begin{table}[H]
    \centering
    \small{
    \begin{tabular}{|c|r|r|r|}
    \hline
        $m$ & $\overline{O_m}$ & $\overline{\mathcal{U}_m}$ & $\frac{\overline{O_m}}{\overline{\mathcal{U}_m}}$  \\ \hline
        0 &         24 &       2 & 12.00 \\
        1 &        672 &      26 & 25.85 \\
        2 &      4 368 &     198 & 22.06\\
        3 &     30 960 &   1 410 & 21.96\\
        4 &    234 720 &  10 304 & 22.78\\
        5 &  1 823 232 &  78 204 & 23.31\\
        6 & 14 370 048 & 607 976 & 23.53\\\hline
    \end{tabular}
    }
    \caption{Comparing the number of unitary operations in the group before ($\overline{O_n}$) and after ($\overline{\mathcal{U}_n}$) reduction with respect to the symmetry group of the free set $\mathcal{S}_{\mathcal{W}}$, in this case equivalent to the symmetry group of the regular tetrahedron $\mathcal{S}_\Delta$. In the limit $m\rightarrow\infty$ one finds that the ratio tends to $\overline{\mathcal{S}_{\mathcal{W}}} = 4! = 24$.}
    \label{tab:hierarchy_sizes}
\end{table}

\vspace{-.5cm}
The resulting values, together with the hierarchy of unitary transformations introduced in Section \ref{ssec:hierarchy} demonstrate the exponential convergence of the hierarchical approximation, proving it to be a viable strategy to determine upper bounds on Absolute Negativity of arbitrary arrangement of states.

\vspace{-1 cm}


\section{Discussion}
\label{sec:discussion}
\vspace{-0.4 cm}

This article introduces a novel quantum resource: the Absolute Negativity of a set of quantum devices given an arbitrary quasi-probability representation. Moreover, our results reveal the value of Absolute Negativity for practical purposes. First, the demonstrated advantages in discrimination are essential for subroutines in communication protocols \cite{Bae2015}, and second, the shown cost of outcome estimation is relevant for the certification of quantum supremacy claims \cite{Bravyi2016,Bravyi2019,pashayan2021}. Furthermore, considering the minimum resource intrinsic to a set makes our results reliable in worst-case scenarios such as specific classes of noise \cite{Shap2003} or adversary attacks \cite{Govia2020}. Additionally, at the foundational level, Absolute Negativity is a signature of non-classicality stronger than standard negativity, but weaker than contextuality \cite{Spekkens2008}. On the other hand, AN quantifiers could lead to upper bounds to compute similar measures of contextuality for the same devices; this is the case of generalised robustness \cite{LiLu2020}.

A natural extension of our work is the investigation of Absolute Negativity in sets of quantum channels. An open question for such research would be to determine the practical advantages of non-zero AN channels over channels described by a stochastic transition matrix for a given frame representation (see \cite{Pashayan2015,Ferrie2011}). However, we believe that the extension of our study to quantum channels deserves independent investigation, because of the phenomenological richness of quantum channels. For example, an investigation of channel resources can vary dramatically depending on whether its composition is sequential or parallel \cite{liuwinter2019}. Some fields that could benefit from such a study include magic state computation and shallow quantum circuits whose quantum advantages require contextuality and therefore negativity \cite{Praka2020,bravy2018}.

In the general context of resource theories, our work contribution is twofold. First, it demonstrates the applicability of powerful general results in a concrete setting \cite{TakagiRegula2019,Ducuara2020} and compels the community to apply other critical general results \cite{PaulLind2019,DucaraPaul2020,Regula2021,Ducuara2021}. Secondly, it develops new methods to quantify and analyse multi-object resources. In particular, our hierarchy of upper bounds can be applied to a large class of basis-independent quantifiers, such as in Refs.~\cite{Designole2020,Yu2021}, to derive analytical or numerical conclusions. Indeed, our study of sets of four qubit states shows that the hierarchy converges exponentially to the exact value and quickly identifies critical configurations at the boundary of free multi-objects set. Therefore, we firmly believe that our research represents an essential step in extending the applicability of resource theory to highly complex devices.

\begin{acknowledgments}

\vspace{-0.2 cm}

The authors thank K. Życzkowski and K. Korzekwa for useful discussions and comments. RS and AOJ acknowledge the financial support of the Foundation for Polish Science through the TEAM-NET project (contract no. POIR.04.04.00-00-17C1/18-00). Furthermore, JC acknowledges the financial support from NCN DEC-2019/35/O/ST2/01049.
\end{acknowledgments}


\appendix

\section{Wigner phase space representation}
\label{app:Wigner}
This appendix is devoted to reviewing the formulation of the Wigner function defined on a discrete phase-space. Although several analogues of the Wigner function representation for finite-dimensional quantum systems have been proposed over the past few years, we will focus on the representation presented by Wooters~\cite{Wootters1987}. 

Consider a Hilbert space of prime dimension $d$, with a phase-space denoted by $\Phi_d$ consisting of a $d \times d$ array of points $\alpha = (q,p) \in \mathbb{Z}_d \times \mathbb{Z}_d$. A line $\lambda$ is defined as a set of $d$ points satisfying the linear equation $a q + bp = c$, where $a,b$ and $c$ are elements of the finite field, and thus the addition and multiplication is modulo $d$. Note that $a$ and $b$ cannot be zero simultaneously. Two lines are parallel if their linear equations differ only in the value of $c$. The phase-space contains $d+1$ sets of $d$ parallel lines called striations.

If the Hilbert space has composite dimension $d= d_1 d_2 ..., d_k$, where $d_k$ is prime, one can associate a discrete phase-space $\Phi_{d_i}$ for each, such that the entire $d$ dimensional system is the Cartesian product of two-dimensional phase spaces of the subsystems. The phase space is thus $(d_1 \times d_1) \times ... \times (d_k \times d_k)$ array. The same definitions can be extended for the composite dimension, i.e., now a point is the $k$-tuple $\alpha = (\alpha_1, ..., \alpha_k)$ of points, a line is the $k$-tuple $\lambda = (\lambda_1, ..., \lambda_k)$ of lines in the phase-space, and in the same way, parallel lines are the striations.

The next step is to translate the Hilbert space into the discrete phase-space formalism. This is done by choosing the particular basis for the space of Hermitian operators,
\begin{equation}
    A_{\alpha} = \frac{1}{d} \sum_{j,m=0}^{d-1} \omega^{p j - q m + (jm)/2}X^j Z^m , 
\end{equation}
where $\omega = e^{2\pi i/d}$, and $X$ and $Z$ are the generalised Pauli operators. Formally, $Z$ is an operator with the spectrum $\operatorname{spec}(Z) = \{ \omega^k : k \in \mathbb{Z}_d \}$. Its eigenvectors form a basis for $\mathcal H$ and are denoted by $\{ \ket{\phi_k} \}$. Consider also the operators defined by the action $X \ket{\phi_k} = \ket{\phi_{k+1}}$, where all arithmetic is modulo $d$, and $Y$ implicitly through $[X,Z] = 2iY$. Note that when $d=2$, these operators take the form of the usual Pauli operators. 

The operators $\{ A_{\alpha} : \alpha \in \Phi_d \}$ acting on a $d$ dimensional Hilbert space are called phase point operators and they are defined in such a way that they satisfy three properties:
\begin{enumerate}
    \item For each point $\alpha$, the operator $A_{\alpha}$ is Hermitian. 
    \item For any two points $\alpha$ and $\beta$ the operators are orthogonal under the standard Hilbert-Schmidt scalar product 
        \begin{equation}
        \label{eq_AaAb}
        \tr{A_{\alpha} A_{\beta}} = d \delta_{\alpha \beta} . 
        \end{equation}    
    \item For each line $\lambda$ in a given striation, the operators
        \begin{equation}
            P_{\lambda} = \frac{1}{d} \sum_{\alpha \in \lambda} A_{\alpha} ,
        \end{equation}    
    form a projective valued measurement (PVM), i.e., a set of $d$ orthogonal projectors that sum up to identity. 
\end{enumerate}
These properties are discrete analogues of the continuous set of operators in the original Wigner function. For a composite system $d$, the phase point operator in $\Phi_d$ associated with the point $\alpha=(\alpha_1, ..., \alpha_k)$ is given by $A_{\alpha} = A_{\alpha_1} \otimes ... \otimes A_{\alpha_k} $. 

The $d^2$ phase point operators are linearly independent and form a basis for the space of Hermitian operators. Thus, any density operator $\rho$ can be expressed as
\begin{equation}
    \rho = \sum_{\alpha} W_{\alpha}(\rho) A_\alpha ,
\end{equation}
where the real coefficients can be explicitly calculated using Eq.~\eqref{eq_AaAb},
\begin{equation}
    W_{\alpha}(\rho) = \frac{1}{d} \tr{\rho A_\alpha} .
\end{equation}
This quasi-probability function is the Wootters' discrete Wigner function. Moreover, one can show that it satisfies the same properties as its continuous version,
\begin{enumerate}
    \item For all $\rho$, $W_{\alpha}(\rho)$ is real .
    \item For all $\rho_1$ and $\rho_2$,
    \begin{equation}
        \tr{\rho_1 \rho_2} = d \sum_\alpha W_{\alpha}(\rho_1) W_{\alpha}(\rho_2) .
    \end{equation}
    \item For all $\rho$, summing $W_{\alpha}(\rho)$ along the line $\lambda$ produces the probability that a measurement of PVM associated with the striation containing $\lambda$ has the outcome associated with $\lambda$.
\end{enumerate}

After defining the phase-space and stating its properties, we may turn to showing Eq.~\eqref{eq-general_wigner} in more details. The first step is to consider a qubit system prepared in a general state with the Bloch vector $\v{r} = (x,y,z)$, one first applies Eq.~\eqref{eq-general_wigner} with the frame given by
\begin{equation}
V(\alpha)=\frac{1}{d^2}\sum_{j,m=0}^{d-1}
\omega^{pj-qm+\frac{jm}{2}}  X^j  Z^m,  
\end{equation}
and imposing that the components of the Wigner function are non-negative for each point of the phase space, i.e., $W_{\alpha}(\rho) \geq 0$, we obtain the following conditions
\begin{equation}
\label{eq:conditions}
\begin{matrix}
(1 + x + y + z) \geq 0 \, ,\\ 
(1 + x - y - z) \geq 0\, ,\\ 
(1 - x - y + z) \geq 0\, ,\\ 
(1 - x + y - z) \geq 0 \,, \\
x^2+y^2 + z^2 \leq 1 \, ,
\end{matrix} 
\end{equation}
which determines the set of states with a positive Wigner function (see Fig.\ref{fig_Wignershape}). These inequalities can be recast in terms of unity vectors as in Eq.~\eqref{eq:dotproduct_rvectors}.

\begin{widetext}
\section{Proofs of operational interpretations}
\label{app:op_proofs}
\subsection{Discrimination game for state-measurement pair}
\label{sapp:disc_game_sm_pair}
The Result 2 of reference \cite{Ducuara2020} states that for any convex QRT of
states with free set $\mathcal{F}^{\left(s\right)}$  and a convex
QRT of measurements with free set $\mathcal{F}^{\left(e\right)}$
closed under classical post-processing (CPP), any state-measurement
pair $(\rho,\mathbb{M})$ satisfies:

\begin{equation}
\sup_{\Lambda}\frac{p_{succ}\left(\rho,\Lambda,\mathbb{M}\right)}{\underset{\tau\in\mathcal{F}^{\left(s\right)},\mathbb{L}\in\mathcal{F}^{\left(e\right)}}{\sup}p_{succ}\left(\tau,\Lambda,\mathbb{L}\right)}\!=\!\left[1\!+\!\mathcal{R}_{\mathcal{F}^{\left(s\right)}}\!\left(\rho\right)\right]\!\!\left[1\!+\!\mathcal{R}_{\mathcal{F}^{\left(e\right)}}\!\left(\mathbb{M}\right)\right]. \label{eq:productrelation}
\end{equation}
where $\Lambda$ is a set of subchannels (i.e. completely positive
maps that together preserve trace) and $p_{succ}$ is the optimal
success probability in discriminating the subchannels in $\Lambda$.
In our case, for each fixed unitary $U$ we can define a convex QRT
of states with $\mathcal{F}^{\left(s\right)}=\mathcal{W}_{1}^{(s)}\left[U\right]$
and a convex QRT of measurements with $\mathcal{F}^{\left(e\right)}=\mathcal{W}_{1}^{(e)}\left[U\right]$
closed under CPP. As a consequence, for every pair $(\rho_{j},\mathbb{M}_{j})$
in the multi-object $(\vec{\rho},\vec{\mathbb{M}})$ we can apply
(\ref{eq:productrelation}):
\begin{align}
\sup_{\Lambda_{j}}\frac{p_{succ}\left(\rho_{j},\Lambda_{j},\mathbb{M}_{j}\right)}{\underset{(\tau,\mathbb{L})\in\mathcal{W}_{1}\left[U\right]}{\sup}\:p_{succ}\left(\tau,\Lambda_{j},\mathbb{L}\right)} =\left[1+\mathcal{R}_{\mathcal{W}_{1}^{(s)}\left[U\right]}\left(\rho_{j}\right)\right]\!\!\left[1+\mathcal{R}_{\mathcal{W}_{1}^{(e)}\left[U\right]}\left(\mathbb{M}_{j}\right)\right] ,\label{eq:Uproduct}
\end{align}
where we used the definition of $\mathcal{W}_{1}\left[U\right]$
to simplify the notation and to remark that the optimal $\Lambda_{j}$ depends
on $U$. Taking the average over all pairs $(\rho_{j},\mathbb{M}_{j})$
in $(\vec{\rho},\vec{\mathbb{M}})$ we obtain:
\begin{align}
\mathbb{E}_{j}\left[\sup_{\Lambda_{j}}\frac{p_{succ}\left(\rho_{j},\Lambda_{j},\mathbb{M}_{j}\right)}{\underset{(\tau,\mathbb{L})\in\mathcal{W}_{1}\left[U\right]}{\sup}\:p_{succ}\left(\tau,\Lambda_{j},\mathbb{L}\right)}\right]  =\mathbb{E}_{j}\left[\left(1+\mathcal{R}_{\mathcal{W}_{1}^{(s)}\left[U\right]}\left(\rho_{j}\right)\right)\left(1+\mathcal{R}_{\mathcal{W}_{1}^{(e)}\left[U\right]}\left(\mathbb{M}_{j}\right)\right)\right]. \label{eq:Uproduct2}
\end{align}
Then, minimizing over unitaries $U$ and using the definitions of
$\mathcal{R}_{\mathcal{W}_{n}\left[U\right]}(\vec{\rho},\vec{\mathbb{M}})$
and $\mathcal{R}_{n}(\vec{\rho},\vec{\mathbb{M}})$
on the right-hand side of (\ref{eq:Uproduct2}):
\begin{equation}
\inf_{U}\mathbb{E}_{j}\left[ \sup_{\Lambda_{j}}\frac{p_{succ}\left(\rho_{j},\Lambda_{j},\mathbb{M}_{j}\right)}{\underset{(\tau,\mathbb{L})\in\mathcal{W}_{1}\left[U\right]}{\sup}\:p_{succ}\left(\tau,\Lambda_{j},\mathbb{L}\right)}\right] \!\!\!=1+\mathcal{R}_{n}(\vec{\rho},\vec{\mathbb{M}}) , \label{eq:SEadv112}
\end{equation}
which is the desired result Eq. (\ref{eq:SEadv11}) from the main text. 

Now, to prove bound (\ref{commadv}), we start rewriting the min conditional entropy: 

\begin{eqnarray}
H_{+\infty}\left(X_{\Lambda,\rho}\mid G_{\mathbb{M}}\right) & = & -\log\left\{ \sum_{g}\max_{x}p(x)\mathrm{Tr}\left[M_{g}\mu_{x}\right]\right\} \nonumber \\
 & \overset{(a)}{=} & -\log\left\{ \sum_{g}\max_{\left\{ q(x\mid g)\right\} }\sum_{x}q(x\mid g)p(x)\mathrm{Tr}\left[M_{g}\mu_{x}\right]\right\} \nonumber \\
 & \overset{(b)}{=} & -\log\left\{ \max_{\left\{ q(x\mid g)\right\} }\sum_{x}\mathrm{Tr}\left[\left(\sum_{g}q(x\mid g)M_{g}\right)\mu_{x}\right]p(x)\right\} \nonumber \\
 & \overset{(c)}{=} & -\log\left\{ \max_{\Xi_{\left\{ q\left(x\mid g\right)\right\} }\left(\mathbb{M}\right)}\sum_{x}\mathrm{Tr}\left[\Xi_{\left\{ q\left(x\mid g\right)\right\} }\left(\mathbb{M}\right)_{x}\mu_{x}\right]p(x)\right\} \nonumber \\
 & = & -\log\left\{ \max_{\Xi_{\left\{ q\left(x\mid g\right)\right\} }}\sum_{x}\mathrm{Tr}\left[\Xi_{\left\{ q\left(x\mid g\right)\right\} }\left(\mathbb{M}\right)_{x}\Gamma_{x}\left(\rho\right)_{x}\right]\right\} \nonumber \\
 & \overset{(d)}{=} & -\log\left\{ p_{succ}\left(\rho,\Lambda,\mathbb{M}\right)\right\}, \label{eq:if1}
\end{eqnarray}
where in $(a)$ we use the identity:
\begin{equation}
\max_{x}h_{g}(x)=\max_{\left\{ q(x\mid g)\right\} }\sum_{x}q(x\mid g)h_{g}(x) ,
\end{equation}
with $\left\{ q(x\mid g)\right\}$ a probability distribution and $h_{g}(x)$ some function of $x$. In $(b)$, we use the fact that the sum of maximization of independent non-negative functions is equal to the maximization of the sum of those functions. Then, in $(c)$ we use the definition of CPP operations and $(d)$ uses the definition of the optimal subchannel discrimination probability when accessing $\rho$, $\mathbb{M} $ and any CPP $\Xi_{\left\{ q\left(x\mid g\right)\right\} }$. Later, applying the definition of infinity mutual information and (\ref{eq:if1}) we have: 

\begin{eqnarray}
I_{+\infty}\!\left(\!X_{\Lambda_{j},\rho_{j}}\!:\!G_{\mathbb{M}_{j}}\!\right)\!-\!\!\!\sup_{(\tau,\mathbb{L})\in\mathcal{W}_{1}\left[\bar{U}\right]}\!I_{+\infty}\!\left(\!X_{\Lambda_{j},\tau}\!:\!G_{\mathbb{L}}\!\right) & = & -H_{+\infty}\!\left(\!X_{\Lambda_{j},\rho_{j}}\!:\!G_{\mathbb{M}_{j}}\!\right)\!-\!\!\!\sup_{(\tau,\mathbb{L})\in\mathcal{W}_{1}\left[\bar{U}\right]}\!-H_{+\infty}\!\left(\!X_{\Lambda_{j},\tau}\!:\!G_{\mathbb{L}}\!\right)\nonumber \\
 & = & \log\left\{ p_{succ}\left(\rho_{j},\Lambda_{j},\mathbb{M}_{j}\right)\right\} -\sup_{(\tau,\mathbb{L})\in\mathcal{W}_{1}\left[\bar{U}\right]}\log\left\{ p_{succ}\left(\tau,\Lambda_{j},\mathbb{L}\right)\right\} \nonumber \\
 & = & \log\left\{ \frac{p_{succ}\left(\rho_{j},\Lambda_{j},\mathbb{M}_{j}\right)}{\sup_{(\tau,\mathbb{L})\in\mathcal{W}_{1}\left[\bar{U}\right]}p_{succ}\left(\tau,\Lambda_{j},\mathbb{L}\right)}\right\} . \label{eq:inf2}
\end{eqnarray}

Taking the supremum on both sides of (\ref{eq:inf2}) and using (\ref{eq:Uproduct}) we obtain the following. 
\begin{equation}
\log\left\{ \left[1\!+\!\mathcal{R}_{\mathcal{W}_{1}^{(s)}\left[\bar{U}\right]}\left(\!\rho_{j}\!\right)\right]\!\!\left[1\!+\!\mathcal{R}_{\mathcal{W}_{1}^{(e)}\left[\bar{U}\right]}\left(\!\mathbb{M}_{j}\!\right)\!\right]\right\} =\sup_{\Lambda_{j}}\left\{ \!I_{+\infty}\!\left(\!X_{\Lambda_{j},\rho_{j}}\!:\!G_{\mathbb{M}_{j}}\!\right)\!-\!\!\!\sup_{(\tau,\mathbb{L})\in\mathcal{W}_{1}\left[\bar{U}\right]}\!I_{+\infty}\!\left(\!X_{\Lambda_{j},\tau}\!:\!G_{\mathbb{L}}\!\right)\!\right\} . \label{eq:inf3}
\end{equation}

Now, we start by showing the upper bound in (\ref{commadv}): 
\begin{eqnarray*}
\mathcal{R}_{n}(\vec{\rho},\vec{\mathbb{M}}) & = & \inf_{U}\mathbb{E}_{j}\!\left[\left[1\!+\!\mathcal{R}_{\mathcal{W}_{1}^{(s)}\left[U\right]}\left(\!\rho_{j}\!\right)\right]\!\!\left[1\!+\!\mathcal{R}_{\mathcal{W}_{1}^{(e)}\left[U\right]}\left(\!\mathbb{M}_{j}\!\right)\!\right]\right]\!-\!1\\
 & = & \mathbb{E}_{j}\!\left[\left[1\!+\!\mathcal{R}_{\mathcal{W}_{1}^{(s)}\left[\bar{U}\right]}\left(\!\rho_{j}\!\right)\right]\!\!\left[1\!+\!\mathcal{R}_{\mathcal{W}_{1}^{(e)}\left[\bar{U}\right]}\left(\!\mathbb{M}_{j}\!\right)\!\right]\!-\!1\right]\\
 & \overset{(a)}{\geq} & \mathbb{E}_{j}\!\left[\log\left\{ \left[1\!+\!\mathcal{R}_{\mathcal{W}_{1}^{(s)}\left[\bar{U}\right]}\left(\!\rho_{j}\!\right)\right]\!\!\left[1\!+\!\mathcal{R}_{\mathcal{W}_{1}^{(e)}\left[\bar{U}\right]}\left(\!\mathbb{M}_{j}\!\right)\!\right]\right\} \right]\\
 & \overset{(b)}{=} & \mathbb{E}_{j}\!\left[\sup_{\Lambda_{j}}\left\{ \!I_{+\infty}\!\left(\!X_{\Lambda_{j},\rho_{j}}\!:\!G_{\mathbb{M}_{j}}\!\right)\!-\!\!\!\sup_{(\tau,\mathbb{L})\in\mathcal{W}_{1}\left[\bar{U}\right]}\!I_{+\infty}\!\left(\!X_{\Lambda_{j},\tau}\!:\!G_{\mathbb{L}}\!\right)\!\right\} \right]\\
 & \overset{(c)}{\geq} & \inf_{U}\mathbb{E}_{j}\!\left[\sup_{\Lambda_{j}}\left\{ \!I_{+\infty}\!\left(\!X_{\Lambda_{j},\rho_{j}}\!:\!G_{\mathbb{M}_{j}}\!\right)\!-\!\!\!\sup_{(\tau,\mathbb{L})\in\mathcal{W}_{1}\left[U\right]}\!I_{+\infty}\!\left(\!X_{\Lambda_{j},\tau}\!:\!G_{\mathbb{L}}\!\right)\!\right\} \right],
\end{eqnarray*}
where a unitary $\bar{U}$ minimises the infimum of the first line due to the convexity and compactness of the free sets, in $(a)$ we use the inequality $\log(x)\leq x-1$ when $x\geq 1$, in $(b)$ we apply identity (\ref{eq:inf3}) and $(c)$ follows from the definition of infimum. 

Following with the lower bound, we have:

\begin{eqnarray*}
\inf_{U}\mathbb{E}_{j}\!\left[\sup_{\Lambda_{j}}\left\{ \!I_{+\infty}\!\left(\!X_{\Lambda_{j},\rho_{j}}\!:\!G_{\mathbb{M}_{j}}\!\right)\!-\!\!\!\sup_{(\tau,\mathbb{L})\in\mathcal{W}_{1}\left[U\right]}\!I_{+\infty}\!\left(\!X_{\Lambda_{j},\tau}\!:\!G_{\mathbb{L}}\!\right)\!\right\} \right] & \overset{(a)}{=} & \mathbb{E}_{j}\!\left[\log\left\{ \left[1\!+\!\mathcal{R}_{\mathcal{W}_{1}^{(s)}\left[\tilde{U}\right]}\left(\!\rho_{j}\!\right)\right]\!\!\left[1\!+\!\mathcal{R}_{\mathcal{W}_{1}^{(e)}\left[\tilde{U}\right]}\left(\!\mathbb{M}_{j}\!\right)\!\right]\right\} \right]\\
 & \overset{(b)}{\geq} & \log\left\{ \mathbb{E}_{j}\!\left[\left[1\!+\!\mathcal{R}_{\mathcal{W}_{1}^{(s)}\left[\tilde{U}\right]}\left(\!\rho_{j}\!\right)\right]\!\!\left[1\!+\!\mathcal{R}_{\mathcal{W}_{1}^{(e)}\left[\tilde{U}\right]}\left(\!\mathbb{M}_{j}\!\right)\!\right]\right]\right\} \\
 & \overset{(c)}{\geq} & \log\left\{ \inf_{U}\mathbb{E}_{j}\!\left[\left[1\!+\!\mathcal{R}_{\mathcal{W}_{1}^{(s)}\left[U\right]}\left(\!\rho_{j}\!\right)\right]\!\!\left[1\!+\!\mathcal{R}_{\mathcal{W}_{1}^{(e)}\left[U\right]}\left(\!\mathbb{M}_{j}\!\right)\!\right]\right]\right\} \\
 & = & \log\left[1\!+\!\mathcal{R}_{n}(\vec{\rho},\vec{\mathbb{M}})\right],
\end{eqnarray*}
where in $(a)$ we apply identity (\ref{eq:inf3}) with the unitary $\tilde{U}$ achieving the infimum of the left hand side, in $(b)$ we use Jensen's inequality for the logarithm, and in $(c)$ we consider the definition of infimum. 

This demonstrates the upper and lower bounds in~(\ref{commadv}) of the main text. Here, we also remark that in the article~\cite{Ducuara2020} a similar expression to (\ref{eq:inf3}) is derived; however, in their case, the communication task is implemented with a channel ensemble since this is a strict subset of the subchannel encoding they have an inequality rather than an identity. If we want to apply the encoding in channel ensembles, our upper bound still holds, but not the lower bound.  

\subsection{Relative estimation cost and negativity}
\label{sapp:rel_est_cost}
The forward estimator presented in Eq.(11) of reference \cite{Pashayan2015} can efficiently estimate the probability $\mathrm{Pr}(M_{a\mid j}|\rho_j)$ at the following sampling cost: 

\begin{equation}
s_{\varepsilon,\delta}^{(\rightarrow)}\!\left(\rho_{j},M_{a\mid j}\right)\!\left[U\right]\!=\!c_{\varepsilon,\delta}\!\!\left[\sum_{\alpha}\!\left|W_{\alpha}\left(U\rho_{j}U^{\dagger}\right)\right|\!\max_{\alpha^{\prime}}\!\left|W\!\left(UM_{a\mid j}U^{\dagger}\!\mid\!\alpha^{\prime}\right)\right|\right]^{2},\label{eq:for1}
\end{equation}
with constant $c_{\varepsilon,\delta}=\frac{2}{\epsilon^{2}}\ln\left(\frac{2}{\delta}\right)$. Similarly, if for the same output, we estimate the probability $\mathrm{Pr}(M_{a\mid j}|\tau_j)$ with any $\tau_j\in \mathcal{W}_{1}^{(s)}\left[U\right]$ the sampling cost is: 
\begin{equation}
s_{\varepsilon,\delta}^{(\rightarrow)}\!\left(\tau_{j},M_{a\mid j}\right)\!\left[U\right]\!=\!c_{\varepsilon,\delta}\!\left[\max_{\alpha^{\prime}}\left|W\!\left(UM_{a\mid j}U^{\dagger}\!\mid\!\alpha^{\prime}\right)\right|\right]^{2}.\label{eq:for2}
\end{equation}
The quotient between $s_{\varepsilon,\delta}^{(\rightarrow)}\!\left(\rho_{j},M_{a\mid j}\right)\!\left[U\right]$  and $s_{\varepsilon,\delta}^{(\rightarrow)}\!\left(\tau_{j},M_{a\mid j}\right)\!\left[U\right]$ gives a measure of the relative cost overload of estimating $\mathrm{Pr}(M_{a\mid j}|\rho_j)$ over $\mathrm{Pr}(M_{a\mid j}|\tau_j)$. Using (\ref{eq:for1}) and (\ref{eq:for2}), we have: 

\begin{eqnarray}
\frac{s_{\varepsilon,\delta}^{(\rightarrow)}\left(\rho_{j},M_{a\mid j}\right)\left[U\right]}{s_{\varepsilon,\delta}^{(\rightarrow)}\left(\tau_{j},M_{a\mid j}\right)\left[U\right]} & = & \left[\sum_{\alpha}\left|W_{\alpha}\left(U\rho_{j}U^{\dagger}\right)\right|\right]^{2}\nonumber \\
 & = & \left[1+\mathcal{N}_{[I]}^{(s)}\left(U\rho_{j}U^{\dagger}\right)\right]^{2}\nonumber \\
 & \overset{(a)}{=} & \left[1+\mathcal{N}_{[U]}^{(s)}\left(\rho_{j}\right)\right]^{2},\label{eq:negstate1}
\end{eqnarray}
where in $(a)$ we used the equivalence of computing the negativity of a state after a unitary transformation $U$ and the negativity of the state in the frame rotated by the same unitary $U$. We remark,  since the choice of effect does not influence the relative cost overload in the forward sampling, the natural choice for $ M_{a\mid j}$ among all operators in $\mathrm{M}_j $ is the output requiring the largest sampling cost, representing the worst case scenario when measuring $\mathrm{M}_j $. With the above choice and using the fact that both sides of (\ref{eq:negstate1}) are positive we have: 
\begin{equation}
1+\mathcal{N}_{[U]}^{(s)}\left(\rho_{j}\right)=\sqrt{\frac{\max_{o}s_{\varepsilon,\delta}^{(\rightarrow)}\left(\rho_{j},M_{o\mid j}\right)\left[U\right]}{\max_{o}s_{\varepsilon,\delta}^{(\rightarrow)}\left(\tau_{j},M_{o\mid j}\right)\left[U\right]}}.\label{eq:relstatecost}
\end{equation}
Now, averaging over $j$ and minimising over $U$ on both sides of the equality (\ref{eq:relstatecost}) we recover (\ref{eq:estbound1}) from the main text. 

Another estimator for $\mathrm{Pr}(M_{a\mid j}|\rho_j)$ presented in Equation~(5) of the Supplementary Material of \cite{Pashayan2015} uses the time-reversal picture of the circuit computing $\mathrm{Pr}(M_{a\mid j}|\rho_j)$. Such a reversal estimator has the following sampling cost \cite{Pashayan2015}: 

\begin{equation}
s_{\varepsilon,\delta}^{(\leftarrow)}\!\!\left(\rho_{j},M_{a\mid j}\right)\!\left[U\right]\!=\!c_{\varepsilon,\delta}\!\!\left[\!\sum_{\alpha}\!\left|W\!\left(UM_{a\mid j}U^{\dagger}\!\!\mid\!\alpha\!\right)\right|\!\max_{\alpha^{\prime}}\!\left|W_{\alpha^{\prime}}\!\!\left(U\rho_{j}U^{\dagger}\right)\right|\right]^{2}. \label{eq:rev1}
\end{equation}

In this case, we compare the sampling cost of $\mathrm{Pr}(M_{a\mid j}|\rho_j)$ with that of  $\mathrm{Pr}(L_{a\mid j}|\rho_j)$ where  $L_{a|j} \in \mathcal{W}_{1}^{(e)}\left[U\right]$ is the effect of a free measurement chosen to have the largest sampling contribution accessible by free effects. Hence, the sampling cost of the reversal estimator for $\mathrm{Pr}(L_{a\mid j}|\rho_j)$ is: 
\begin{equation}
s_{\varepsilon,\delta}^{(\leftarrow)}\!\left(\rho_{j},L_{a\mid j}\right)\!\left[U\right]\!=\!c_{\varepsilon,\delta}\!\left[\mathrm{Tr}\left[L_{a\mid j}\right]\!\max_{\alpha^{\prime}}\!\left|W_{\alpha^{\prime}}\left(U\rho_{j}U^{\dagger}\right)\right|\right]^{2}.\label{eq:rev2}
\end{equation}

Here note that the hardest estimation for a free effect happens when $\mathrm{Tr}\left[L_{a\mid j}\right]=d$.

Analogously to the forward estimator, we quantify the relative sampling cost overload of the reversal estimator by the quotient of $s_{\varepsilon,\delta}^{(\leftarrow)}\!\!\left(\rho_{j},M_{a\mid j}\right)\!\left[U\right] $ and $s_{\varepsilon,\delta}^{(\leftarrow)}\!\!\left(\rho_{j},L_{a\mid j}\right)\!\left[U\right] $: 

\begin{eqnarray}
\frac{s_{\varepsilon,\delta}^{(\leftarrow)}\left(\rho_{j},M_{a\mid j}\right)\left[U\right]}{s_{\varepsilon,\delta}^{(\leftarrow)}\left(\rho_{j},L_{a\mid j}\right)\left[U\right]} & = & \left[\frac{\sum_{\alpha}\left|W\left(UM_{a\mid j}U^{\dagger}\mid\alpha\right)\right|}{\mathrm{Tr}\left[L_{a\mid j}\right]}\right]^{2}\nonumber \\
 & \overset{(b)}{\leq} & \left[\frac{\sum_{\alpha}\left|W\left(UM_{a\mid j}U^{\dagger}\mid\alpha\right)\right|}{\mathrm{Tr}\left[M_{a\mid j}\right]}\right]^{2}\nonumber \\
 & \overset{(c)}{=} & \left[\frac{\mathcal{N}_{[I]}^{(e)}\left(UM_{a\mid j}U^{\dagger}\right)}{\mathrm{Tr}\left[M_{a\mid j}\right]}\right]^{2}\nonumber \\
 & \overset{(d)}{=} & \left[\frac{\mathcal{N}_{[U]}^{(e)}\left(M_{a\mid j}\right)}{\mathrm{Tr}\left[M_{a\mid j}\right]}\right]^{2}\nonumber \\
 & \leq & \left[\max_{o}\left[\frac{\mathcal{N}^{(e)}\left(M_{o\mid j}\right)}{\mathrm{Tr}\left(M_{o\mid j}\right)}\right]\right]^{2}\nonumber \\
 & = & \left[1+\mathcal{N}_{[U]}^{(\mathbf{e})}\left(\mathbb{M}_{j}\right)\right]^{2}. \label{eq:negmea1}
\end{eqnarray}
where in (b) we use $1/\mathrm{Tr}\left[L_{a\mid j}\right]=1/d\leq 1/\mathrm{Tr}\left[M_{a\mid j}\right] $, in (c) the definition of effect sum-negativity, in (d)  the equivalence between rotated effect and rotated frame  for computing the sum-negativity. 

As before, we take square roots on both sides of (\ref{eq:negmea1}) and consider the effect in $\mathrm{M}_j $ that generates the largest relative cost overload: 

\begin{equation}
1+\mathcal{N}_{[U]}^{(\mathbf{e})}\left(\mathbb{M}_{j}\right)\geq\sqrt{\max_{o}\frac{s_{\varepsilon,\delta}^{(\leftarrow)}\left(\rho_{j},M_{o\mid j}\right)\left[U\right]}{s_{\varepsilon,\delta}^{(\leftarrow)}\left(\rho_{j},L_{o\mid j}\right)\left[U\right]}}.\label{eq:relmeascost}
\end{equation}

Again, averaging over $j$ and minimising over $U$ at both sides of the inequality~(\ref{eq:relmeascost}) we recover the second bound~(\ref{eq:estbound2}) from the main text. To obtain the third bound~(\ref{eq:estbound3}) we proceed to multiply both sides of the inequality~(\ref{eq:relmeascost}) by the terms of equality~(\ref{eq:relstatecost}):

\begin{equation}
\left[1\!+\!\mathcal{N}_{[U]}^{(s)}\left(\rho_{j}\right)\right]\!\!\left[1\!+\!\mathcal{N}_{[U]}^{(\mathbf{e})}\left(\mathbb{M}_{j}\right)\right]\!\geq\!\sqrt{\!\frac{s_{\varepsilon,\delta}^{(\leftarrow)}\!\left(\rho_{j},M_{o^{\prime}\mid j}\right)\!\left[U\right]s_{\varepsilon,\delta}^{(\rightarrow)}\!\left(\rho_{j},M_{o^{\prime\prime}\mid j}\right)\!\left[U\right]}{s_{\varepsilon,\delta}^{(\leftarrow)}\!\left(\rho_{j},L_{o^{\prime}\mid j}\right)\!\left[U\right]s_{\varepsilon,\delta}^{(\rightarrow)}\!\left(\tau_{j},M_{o^{\prime\prime}\mid j}\right)\!\left[U\right]}},\label{eq:relmulticost}
\end{equation}
where $o^{\prime} $ and $o^{\prime\prime} $ indicate the effects with maximal contribution to the estimators as indicated on the right-hand sides of (\ref{eq:relstatecost}) and (\ref{eq:relmeascost}). Finally, we perform an average over $j$ and an infimum over $U$ on both sides of (\ref{eq:relmulticost}), then applying the definition (\ref{defNegmult}) on the left side, we recover (\ref{eq:estbound3}) from the main text.  

\subsection{Monotonicity of free operations}\label{sapp:free_op_monoton}

Here, we demonstrate the monotonicity of our quantifiers under the free operations introduced in Section~\ref{sec:AN}. First we consider the robustness-based quantifier when depolarising noise $\Theta_\epsilon$ is applied: 

\begin{eqnarray}
\mathcal{R}_{\mathcal{W}_{1}^{(s)}\left[U\right]}\left(\Theta_{\epsilon}\left[\rho_{j}\right]\right) & = & \mathcal{R}_{\mathcal{W}_{1}^{(s)}\left[U\right]}\left(\left(1-\epsilon\right)\rho_{j}+\epsilon I/d\right)\nonumber \\
 & \overset{(a_{1})}{\leq} & \left(1-\epsilon\right)\mathcal{R}_{\mathcal{W}_{1}^{(s)}\left[U\right]}\left(\rho_{j}\right)+\epsilon\mathcal{R}_{\mathcal{W}_{1}^{(s)}\left[U\right]}\left(I/d\right)\nonumber \\
 & \overset{(a_{2})}{\leq} & \left(1-\epsilon\right)\mathcal{R}_{\mathcal{W}_{1}^{(s)}\left[U\right]}\left(\rho_{j}\right)\nonumber \\
 & \leq & \mathcal{R}_{\mathcal{W}_{1}^{(s)}\left[U\right]}\left(\rho_{j}\right),\label{eq:mon1}
\end{eqnarray}
where in $(a_1)$ we use the convexity of the generalised robustness and in $(a_2)$ the fact that the maximally mixed state is free. The proof of the monotonicity follows in complete analogy when depolarising noise is applied to the measurements: 
\begin{equation}
\mathcal{R}_{\mathcal{W}_{1}^{(e)}\left[U\right]}\left(\Theta_{\epsilon^{\prime}}\left[\mathbb{M}_{j}\right]\right)\leq\mathcal{R}_{\mathcal{W}_{1}^{(e)}\left[U\right]}\left(\mathbb{M}_{j}\right).\label{eq:mondep1}
\end{equation}

Taking the product of both sides of the inequalities (\ref{eq:mon1}) and (\ref{eq:mondep1}) we obtain:

\begin{equation}
\left(1+\mathcal{R}_{\mathcal{W}_{1}^{(s)}\left[U\right]}\left(\Theta_{\epsilon}\left[\rho_{j}\right]\right)\right)\!\!\left(1+\mathcal{R}_{\mathcal{W}_{1}^{(e)}\left[U\right]}\left(\Theta_{\epsilon^{\prime}}\left[\mathbb{M}_{j}\right]\right)\right)\leq\left(1+\mathcal{R}_{\mathcal{W}_{1}^{(s)}\left[U\right]}\left(\rho_{j}\right)\right)\!\!\left(1+\mathcal{R}_{\mathcal{W}_{1}^{(e)}\left[U\right]}\left(\mathbb{M}_{j}\right)\right).\label{eq:mondep2}
\end{equation}
Taking the average, then the infimum over the unitaries and subtracting 1 on both sides of (\ref{eq:mondep2}) lead to the desired result: 
\begin{equation}
\mathcal{R}_{n}\left(\Theta_{\epsilon}\left[\vec{\rho}\right],\Theta_{\epsilon^{\prime}}\left[\vec{\mathbb{M}}\right]\right)\leq\mathcal{R}_{n}\left(\vec{\rho},\vec{\mathbb{M}}\right) . 
\end{equation}
Now, we demonstrate the monotonicity under classical postprocessing (CPP)  in the case that only measurements are potentially resourceful since states only contribute with multiplicative factors invariant under CPP:
\begin{eqnarray*}
\mathcal{R}_{n}\left(\vec{\mathbb{M}}\right) & = & \inf_{U}\mathbb{E}_{j}\left[\mathcal{R}_{\mathcal{W}_{1}^{(e)}\left[U\right]}\left(\mathbb{M}_{j}\right)\right]\\
 & = & \mathbb{E}_{j}\left[\mathcal{R}_{\mathcal{W}_{1}^{(e)}\left[\bar{U}\right]}\left(\mathbb{M}_{j}\right)\right]\\
 & \overset{(b_{1})}{\geq} & \mathbb{E}_{j}\left[\mathcal{R}_{\mathcal{W}_{1}^{(e)}\left[\bar{U}\right]}\left(\Xi_{\left\{ p(x\mid a)\right\} }\left[\mathbb{M}_{j}\right]\right)\right]\\
 & \overset{(b_{2})}{\geq} & \inf_{U}\mathbb{E}_{j}\left[\mathcal{R}_{\mathcal{W}_{1}^{(e)}\left[U\right]}\left(\Xi_{\left\{ p(x\mid a)\right\} }\left[\mathbb{M}_{j}\right]\right)\right]\\
 & = & \mathcal{R}_{n}\left(\Xi_{\left\{ p(x\mid a)\right\} }\left[\vec{\mathbb{M}}\right]\right),
\end{eqnarray*}
where in $(b_1)$ we use the fact that generalised robustness of measurements is non-increasing under CPP whenever the free set is closed under CPP, which indeed is the case for $\mathcal{W}_{1}^{(e)}\left[\bar{U}\right]$ in the particular basis defined by $\bar{U}$ \cite{Ducuara2020} and in $(b_2)$ we simply use the definition of infimum. This shows the monotonicity under CPP:
\begin{equation}
\mathcal{R}_{n}\left(\vec{\mathbb{M}}\right)\geq\mathcal{R}_{n}\left(\Xi_{\left\{ p(x\mid a)\right\} }\left[\vec{\mathbb{M}}\right]\right).\label{eq:cpp1}
\end{equation}

We proceed now to demonstrate the monotonicity of sum-negativity based quantifiers under depolarising noise: 

\begin{eqnarray}
1+\mathcal{N}_{\left[U\right]}^{(s)}\left(\Theta_{\epsilon}\left[\rho_{j}\right]\right) & = & \sum_{\alpha\in\Omega\left[U\right]}\left|W_{\alpha}\left(\Theta_{\epsilon}\left[\rho_{j}\right]\right)\right|\nonumber \\
 & \leq & \sum_{\alpha\in\Omega\left[U\right]}\left|W_{\alpha}\left(\left(1-\epsilon\right)\rho_{j}+\epsilon I/d\right)\right|\nonumber \\
 & \overset{(c_{1})}{\leq} & \sum_{\alpha\in\Omega\left[U\right]}\left\{ \left(1-\epsilon\right)\left|W_{\alpha}\left(\rho_{j}\right)\right|+\epsilon\left|W_{\alpha}\left(I/d\right)\right|\right\} \nonumber \\
 & \overset{(c_{2})}{=} & \sum_{\alpha\in\Omega\left[U\right]}\left|W_{\alpha}\left(\rho_{j}\right)\right|-\epsilon\left[\sum_{\alpha\in\Omega\left[U\right]}\left|W_{\alpha}\left(\rho_{j}\right)\right|-1\right]\nonumber \\
 & = & 1+\mathcal{N}_{\left[U\right]}^{(s)}\left(\rho_{j}\right)-\epsilon\mathcal{N}_{\left[U\right]}^{(s)}\left(\rho_{j}\right)\nonumber \\
 & \overset{(c_{3})}{\leq} & 1+\mathcal{N}_{\left[U\right]}^{(s)}\left(\rho_{j}\right).\label{eq:mon2}
\end{eqnarray}
In $(c_1)$ we applied the triangular inequality, $(c_2)$ we used the fact that the maximally mixed state is free and in $(c_3)$ the sum negativity of any state is non-negative. Taking the average, infimum over unitaries and subtracting 1 on both sides of (\ref{eq:mon2}) we obtain: 

\begin{equation}
\inf_{U}\mathbb{E}_{j}\left[\mathcal{N}_{\left[U\right]}^{(s)}\left(\Theta_{\epsilon}\left[\rho_{j}\right]\right)\right]\leq\inf_{U}\mathbb{E}_{j}\left[\mathcal{N}_{\left[U\right]}^{(s)}\left(\rho_{j}\right)\right].
\end{equation}

Now, we proceed to show the monotonicity of the measurement sum-negativity under depolarising noise, starting with $\bar{a}$,  the output maximising the measurement sum-negativity after the operation $\Theta_{\epsilon}$:
\begin{eqnarray*}
\max_{a}\left[\frac{\mathcal{N}_{\left[U\right]}^{(e)}\left(\Theta_{\epsilon}\left[E_{a\mid j}\right]\right)}{\mathrm{Tr}\left[\Theta_{\epsilon}\left[E_{a\mid j}\right]\right]}\right] & = & \frac{\sum_{\alpha\in\Omega\left[U\right]}\left|W\left(\Theta_{\epsilon}\left[E_{\bar{a}\mid j}\right]\mid\alpha\right)\right|}{\mathrm{Tr}\left[\Theta_{\epsilon}\left[E_{\bar{a}\mid j}\right]\right]}\\
 & = & \frac{\sum_{\alpha\in\Omega\left[U\right]}\left|W\left(\left(1-\epsilon\right)E_{\bar{a}\mid j}+\epsilon I/d\mid\alpha\right)\right|}{\mathrm{Tr}\left[\left(1-\epsilon\right)E_{\bar{a}\mid j}+\epsilon I/d\right]}\\
 & \overset{(d_{1})}{\leq} & \frac{\sum_{\alpha\in\Omega\left[U\right]}\left\{ \left(1-\epsilon\right)\left|W\left(E_{\bar{a}\mid j}\mid\alpha\right)\right|+\epsilon\left|W\left(I/d\mid\alpha\right)\right|\right\} }{\left(1-\epsilon\right)\mathrm{Tr}\left[E_{\bar{a}\mid j}\right]+\epsilon\mathrm{Tr}\left[I/d\right]}\\
 & \overset{(d_{2})}{=} & \frac{\left(1-\epsilon\right)\mathrm{Tr}\left[E_{\bar{a}\mid j}\right]\frac{\sum_{\alpha\in\Omega\left[U\right]}\left|W\left(E_{\bar{a}\mid j}\mid\alpha\right)\right|}{\mathrm{Tr}\left[E_{\bar{a}\mid j}\right]}+\epsilon\mathrm{Tr}\left[I/d\right]\frac{\sum_{\alpha\in\Omega\left[U\right]}\left|W\left(I/d\mid\alpha\right)\right|}{\mathrm{Tr}\left[I/d\right]}}{\left(1-\epsilon\right)\mathrm{Tr}\left[E_{\bar{a}\mid j}\right]+\epsilon\mathrm{Tr}\left[I/d\right]},
\end{eqnarray*}
where in $(d_1)$ we apply linearity of the trace and triangular inequality, while in $(d_2)$ we rearrange terms,
 \begin{eqnarray}
 & \overset{(d_{3})}{=} & \lambda\frac{\mathcal{N}_{\left[U\right]}^{(e)}\left(E_{\bar{a}\mid j}\right)}{\mathrm{Tr}\left[E_{\bar{a}\mid j}\right]}+\left(1-\lambda\right)\frac{\mathcal{N}_{\left[U\right]}^{(e)}\left(I/d\right)}{\mathrm{Tr}\left[I/d\right]}\nonumber \\
 & \overset{(d_{4})}{\leq} & \max\left\{ \frac{\mathcal{N}_{\left[U\right]}^{(e)}\left(E_{\bar{a}\mid j}\right)}{\mathrm{Tr}\left[E_{\bar{a}\mid j}\right]},\frac{\mathcal{N}_{\left[U\right]}^{(e)}\left(I/d\right)}{\mathrm{Tr}\left[I/d\right]}\right\} \nonumber \\
 & = & \frac{\mathcal{N}_{\left[U\right]}^{(e)}\left(E_{\bar{a}\mid j}\right)}{\mathrm{Tr}\left[E_{\bar{a}\mid j}\right]}\nonumber \\
 & \overset{(d_{5})}{\leq} & \max_{a}\left[\frac{\mathcal{N}_{\left[U\right]}^{(e)}\left(E_{a\mid j}\right)}{\mathrm{Tr}\left[E_{a\mid j}\right]}\right].\label{eq:mon3}
\end{eqnarray}
Then, in $(d_3)$ we define $0\leq\lambda=\left(1-\epsilon\right)\mathrm{Tr}\left[E_{\bar{a}\mid j}\right]/\left[\left(1-\epsilon\right)\mathrm{Tr}\left[E_{\bar{a}\mid j}\right]+\epsilon\mathrm{Tr}\left[I/d\right]\right]\leq 1$. From the latter definition it follows that $(d_4)$ where we maximise over convex combinations of two non-negative numbers. In $(d_5)$ we use the fact that $\mathcal{N}_{\left[U\right]}^{(e)}\left(I/d\right)/\mathrm{Tr}\left[I/d\right]=1\leq \mathcal{N}_{\left[U\right]}^{(e)}\left(E_{\bar{a}\mid j}\right)/\mathrm{Tr}\left[E_{\bar{a}\mid j}\right]$ and finally in $(d_6$) we select the output with maximum ratio of effect sum-negativity over trace. 
 
Starting with~(\ref{eq:mon3}) we subtract 1, take average measurements and minimise over unitaries on both sides of the inequality. Then, from the definition (\ref{eq:MeasNeg}) we obtain: 
\begin{equation}
\inf_{U}\mathbb{E}_{j}\left[\mathcal{N}_{\left[U\right]}^{(\mathbf{e})}\left(\mathbb{M}_{j}\right)\right]\geq\inf_{U}\mathbb{E}_{j}\left[\mathcal{N}_{\left[U\right]}^{(\mathbf{e})}\left(\Theta_{\epsilon}\left[\mathbb{M}_{j}\right]\right)\right] ,
\end{equation}
which shows that measurement sum-negativity does not increase under depolarising noise. 

Now, we demonstrate that measurement sum-negativity also does not increase under CPP operations. We begin with the output $\bar{x}$ with effect $K_{\bar{x}\mid j}$ maximising the ration effect sum-negativity to trace after $\Xi_{\left\{ p(x\mid a)\right\}} $:

\begin{align}
    \max_x \frac{\mathcal{N}_{\left[U\right]}^{(e)}(K_{x\mid j})}{\Tr[ K_{x\mid j}]} = \frac{\sum_\alpha\abs{W(K_{\bar{x}\mid j}\mid\alpha)}}{\Tr[K_{\bar{x}\mid j}]} & = \frac{\sum_\alpha\abs{W(\sum_a p(\bar{x}\mid a) E_{a\mid j}\mid\alpha)}}{\Tr[\sum_a p(\bar{x}\mid a) E_{a\mid j}]} \nonumber\\
    & \overset{(f_1)}{=} \frac{\sum_\alpha\abs{\sum_a p(\bar{x}\mid a) W(E_{a\mid j}\mid\alpha)}}{\Tr[\sum_a p(\bar{x}\mid a) E_{a\mid j}]} \nonumber\\    
    & \overset{(f_2)}{\leq} \frac{\sum_\alpha\sum_a p(\bar{x}\mid a)\abs{W( E_{a\mid j}\mid\alpha)}}{\sum_a p(\bar{x}\mid a)\Tr[ E_{a\mid j}]} \nonumber\\
    & = \frac{\sum_a p(\bar{x}\mid a)\Tr[ E_{a\mid j}]\left(\sum_\alpha\frac{\abs{W( E_{a\mid j}\mid\alpha)}}{\Tr[ E_{a\mid j}]}\right)}{\sum_a p(\bar{x}\mid a)\Tr[ E_{a\mid j}]}\nonumber \\
    & \overset{(f_3)}{=} \frac{\sum_a p(\bar{x}\mid a)\Tr[ E_{a\mid j}]\left(\frac{\mathcal{N}_{\left[U\right]}^{(e)}(E_{a\mid j})}{\Tr[ E_{a\mid j}]}\right)}{\sum_a p(\bar{x}\mid a)\Tr[ E_{a\mid j}]} \nonumber\\
    & \overset{(f_4)}{\leq} \max_a\,\,\frac{\mathcal{N}_{\left[U\right]}^{(e)}(E_{a\mid j})}{\Tr[ E_{a\mid j}]},\label{fin}
\end{align}
where in $(f_1)$ we use the linearity of Wigner representation, $(f_2)$ the triangle inequality, $(f_3)$ the definition of negativity of effect and $(f_4)$ the basic properties of weighted mean.

Similarly, as in the previous proof, we subtract 1, take average over effects and minimum over unitaries on both sides of (\ref{fin}): 
\begin{equation}
\inf_{U}\mathbb{E}_{j}\left[\mathcal{N}_{\left[U\right]}^{(\mathbf{e})}\left(\mathbb{M}_{j}\right)\right]\geq\inf_{U}\mathbb{E}_{j}\left[\mathcal{N}_{\left[U\right]}^{(\mathbf{e})}\left(\Xi_{\left\{ p(x\mid a)\right\}}\left[\mathbb{M}_{j}\right]\right)\right].
\end{equation}
\end{widetext}
\section{Proofs for example with 4 qubit states}
\label{app:4_qbit_proofs}

\subsection{Geometrical expressions for robustness and negativity in Wigner representation}
\label{sapp:geom_exp_rob}

\begin{figure}[!htb]
    \centering
    \includegraphics{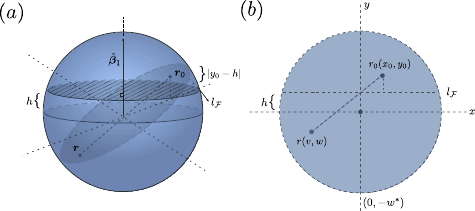}
    \caption{(a) a resource state $\v{r}_0$ over the free set's boundary  $l_\mathcal{F}$  defined by $ \v{r}\cdot \hat{\v{\beta}}_1 = 1/\sqrt{3}$. (b) A plane containing $\v{r}_0$, a vector $\v{r}$ in the region $ \v{r}\cdot \hat{\v{\beta}}_1 < 1/\sqrt{3}$ and perpendicular to  the boundary  plane $l_\mathcal{F}$.   }
    \label{robproof}
\end{figure}

In this Appendix, we derive useful expressions to apply our results to states in Wigner representation for fixed $U$ with the free set $\mathcal{W}^{s}_{1}\left[U\right]$ described in Section~\ref{sec:single_ob_neg}. The simple geometric structure of $\mathcal{W}^{s}_{1}\left[U\right]$ in the Bloch space allows obtaining analytical expressions for the quantifiers used in our theory, such as generalised robustness $\mathcal{R}_{\mathcal{W}_{1}^{s}\left[U\right]}$ or sum-negativity $\mathcal{N}_{[U]}^{s}$. 

Let us start with the generalised robustness $\mathcal{R}_{\mathcal{W}_{1}^{s}\left[I\right]}$ on the fixed initial basis: first note that the region of vectors violating one of the conditions (\ref{eq:dotproduct_rvectors}) say $ \v{r}\cdot \hat{\v{\beta}}_1 > 1/\sqrt{3}$ does not overlap with any region defined by $ \v{r}\cdot \hat{\v{\beta}}_i  > 1/\sqrt{3}$ with $i=\{2,3,4\}$ inside the Bloch sphere. This means that each Bloch vector of a resource state will violate one and only one of the conditions (\ref{eq:dotproduct_rvectors}). Second, without loss of generality, assume that a given resource vector $\v{r}_0$ satisfies $\v{r}_0\cdot \hat{\v{\beta}}_1  > 1/\sqrt{3}$. It is straightforward that a convex combination of $\v{r}_0$ with an arbitrary vector $\v{r}$ will belong to the free set if and only if $\v{r}\cdot \hat{\v{\beta}}_1  < 1/\sqrt{3}$. In Fig.\ref{robproof} (a) we show an arbitrary plane of those containing $\v{r}_0$, $\v{r}$ and perpendicular to the boundary plane $l_\mathcal{F}$ of $\mathcal{R}_{\mathcal{W}_{1}^{s}\left[I\right]}$ defined by $\v{r}\cdot \hat{\v{\beta}}_1 = 1/\sqrt{3}$. Since any of the convex combinations of $\v{r}_0$ and $\v{r}$ belong to at least one of the planes represented in Fig~\ref{robproof} (a), to find the generalised robustness, we can proceed in the following way: find the generalised robustness in an arbitrary plane of Fig.~\ref{robproof} (a) and then minimise over all possible planes. We can simplify the first step by noting that all planes have the same structure as presented in Fig.~\ref{robproof} (b). To find the generalised robustness in an arbitrary plane, we have to find the vector $\v{r}=(v,w)$ that minimizes $\lambda$ in the convex combination
\begin{equation}
\frac{1}{1+\lambda}\left(x_{0},y_{0}\right)+\frac{\lambda}{1+\lambda}\left(v,w\right)=\left(x^{\prime},h\right),
\end{equation}
with arbitrary $x^{\prime}$ and $h=1/\sqrt{3}$, since it is the condition for $\left(x^{\prime},h\right)$ to be at the closest boundary $\v{r}\cdot \hat{\v{\alpha}}_1= 1/\sqrt{3}$ of the free set ($l_{\F}$ in Fig.\ref{robproof} (b)). Solving for $\lambda$ in the second component, we obtain: 
 \begin{equation}
\lambda=\frac{y_{0}-h}{h-w}.
\end{equation}
Since $\left(x_{0},y_{0}\right)$ is a resource state, we have $y_{0}-h>0$ and $h$ as constants, which means that to minimise $\lambda$, one should choose $w$ as negative as possible. In the section of the Bloch sphere depicted in Fig.~\ref{robproof} (b) is straightforward that this happen at the most distant point from $l_{\F}$, in the region \mbox{$\v{r}\cdot \hat{\v{\beta}}_1 < 1/\sqrt{3}$}, which is of the form $(0,-w^{*})$ (See Fig.~\ref{robproof} (b)). Thus, for each plane of Fig.~\ref{robproof} (a), the generalised robustness is $(y_{0}-h)/(h+w^{*})$ which minimises for the maximum $w^{*}=1$ in the Bloch sphere. Now, since $y_0= \v{r}_0\cdot \hat{\v{\beta}}_1 $ we have 
\begin{equation}
    \mathcal{R}_{\mathcal{W}^{s}_1[U]}(\rho) = \frac{\sqrt{3}}{(1+\sqrt{3})} \:  \left(  \v{r}_0\cdot \hat{\v{\beta}}_1  - \frac{1}{\sqrt{3}} \right)  \, .
    \label{formularob}
\end{equation}

To include all possible states, we maximise the above expression over the regions defined by $\v{r}\cdot \hat{\v{\beta}}_1> 1/\sqrt{3}$, and set it to zero if there is no positive value, because in that case the state is free: 
\begin{equation}
    \mathcal{R}_{\mathcal{W}^{s}_1[U]}(\rho) = \frac{\sqrt{3}}{(1+\sqrt{3})} \: \max_k \left\{\mathcal{I}^{+} \left[  \v{r}\cdot \hat{\v{\beta}}_k[U] \! - \! \frac{1}{\sqrt{3}}\right]\right\},  
\end{equation}
with $\mathcal{I}^{+}[x]=\max\{x,0\}$ and $\hat{\v{\beta}}_k[U] $ the vectors from (\ref{eq:dotproduct_rvectors}) but in the standard basis rotated by $U$. Replacing the above expression in the definition (\ref{eq:SEmon1}), applied to a multi-object  $(\vec{\rho},\vec{\mathbb{L}})$ with $\vec{\mathbb{L}} \in  \mathcal{W}^{(e)}_n[U] $ for all $U$, we find:
\begin{equation}
    \mathcal{R}_{\mathcal{W}_n[U]}(\vec{\rho},\vec{\mathbb{L}}) \!=\! \frac{\sqrt{3}}{(1+\sqrt{3})} \mathbb{E}_{j}\!\left[\! \max_k\! \left\{\mathcal{I}^{+}\! \!\left[  \v{r}_j\!\cdot\! \hat{\v{\beta}}_k[U] \! - \!\frac{1}{\sqrt{3}}\! \right]\!\right\} \right] 
    \label{formgenrob}
\end{equation}
with $\v{r}_j$ being the Bloch representation for each $\rho_j$ in $\vec{\rho}$. Defining $\mathcal{R}_{\mathcal{W}_n[U]}(\vec{\rho},\vec{\mathbb{L}})=\mathcal{R}_{\mathcal{W}_n[U]}(\vec{\rho})$ we obtain the expression (\ref{GeomRob}) used in the main text.

Now, to see the proportionality between robustness and sum-negativity, we start assuming, as before that $\v{r}_0\cdot \hat{\v{\beta}}_1  > 1/\sqrt{3}$. Then, a direct computation gives:  

\begin{eqnarray*}
\mathcal{N}_{\left[U\right]}^{(s)}\left(\rho\right) & = & \sum_{\alpha\in\Omega\left[U\right]}\left|W_{\alpha}\left(\rho\right)\right|-1\\
 & = & \sum_{\alpha\in\Omega\left[U\right]}\!\!\left|W_{\alpha}\left(\rho\right)\right|\!+\!\left|W_{\beta_{1}}\left(\rho\right)\right|\!-\!\left|W_{\beta_{1}}\left(\rho\right)\right|\!-\!1\\
 & = & 2\left|W_{\beta_{1}}\left(\rho\right)\right|\!+\!\sum_{\alpha\neq\beta_{1}}\!\left|W_{\alpha}\left(\rho\right)\right|\!-\!\left|W_{\beta_{1}}\left(\rho\right)\right|\!-\!1\\
 & \overset{(a)}{=} & 2\left|W_{\beta_{1}}\left(\rho\right)\right|+\left(\sum_{\alpha\in\Omega\left[U\right]}W_{\alpha}\left(\rho\right)-1\right)\\
 & \overset{(b)}{=} & 2\left|W_{\beta_{1}}\left(\rho\right)\right|\\
 & = & 2\left|\frac{1}{2}\left(1-\sqrt{3}\mathbf{r}_{0}\cdot\hat{\beta_{1}}\right)\right|\\
 & = & \sqrt{3}\left|\left(\frac{1}{\sqrt{3}}-\mathbf{r}_{0}\cdot\hat{\beta_{1}}\right)\right|\\
 & \overset{(c)}{=} & \left(1+\sqrt{3}\right)\mathcal{R}_{\mathcal{W}_{1}^{(s)}\left[U\right]}\left(\rho\right),
\end{eqnarray*}
where in $(a)$ we use that assuming $\v{r}_0\cdot \hat{\v{\beta}}_1  > 1/\sqrt{3}$ implies $\left|W_{\alpha}\left(\rho\right)\right|=W_{\alpha}\left(\rho\right)$ for $\alpha\neq\beta_{1} $ and $\left|W_{\beta_{1}}\left(\rho\right)\right|=-W_{\beta_{1}}\left(\rho\right)$, in $(b)$ we cancel the parehtesis term from previous line, due to normalisation of quasiprobabilities and in $(c)$ we use formula (\ref{formularob}). Then, generalising as in equation (\ref{formgenrob}) and taking average over states $\rho_j$ we have:
\begin{equation}
\mathbb{E}_{j}\left[\mathcal{N}_{\left[U\right]}^{(s)}\left(\rho_{j}\right)\right]=\left(1+\sqrt{3}\right)\mathcal{R}_{\mathcal{W}_{n}^{(s)}\left[U\right]}\left(\vec{\rho}\right) .
\end{equation}

A relation that allows us to discuss our examples just in terms of robustness, without loss of consequences for a sum-negativity analogue study.

\subsection{Proof that three states in a regular polyhedral cone are free}
\label{app:3_vec_proof}

To properly consider the case of three pure states arranged in a regular triangular cone, we first rotate the free set in such a way that it exhibits its symmetry with respect to $2\pi/3$-rotations around selected axes, such as the axis in $(1,1,1)$ direction. We achieve this by introducing a rotation of the free set,

\begin{align}\label{eq:coordtransform}
    \mqty(x' \\ y' \\ z') = \mqty(
        \frac{1}{\sqrt{2}} & -\frac{1}{\sqrt{2}} & 0 \\
        \frac{1}{\sqrt{6}} & \frac{1}{\sqrt{6}} & -\frac{2}{\sqrt{6}} \\
        \frac{1}{\sqrt{3}} & \frac{1}{\sqrt{3}} & \frac{1}{\sqrt{3}})\mqty(x \\ y \\ z) .
\end{align}
and by using this transformation the inequalities turn into 

\begin{align}
    \qty(1 + \sqrt{3}z') \geq 0, \label{eq:conditions21}\\
    \qty(1 +\sqrt{2}x' + \frac{2}{\sqrt{6}}y' - \frac{1}{\sqrt{3}}z') \geq 0, \label{eq:conditions22}\\
    \qty(1 +\sqrt{2}x' + \frac{2}{\sqrt{6}}y' - \frac{1}{\sqrt{3}}z') \geq 0, \label{eq:conditions23}\\
    \qty(1 - \frac{4}{\sqrt{6}}y' - \frac{1}{\sqrt{3}}z') \geq 0, \label{eq:conditions24}\\
    x'^2 + y'^2 + z'^2 \leq 1. \label{eq:conditions25}
\end{align}

It is easy to see that the inequalities \eqref{eq:conditions22}-\eqref{eq:conditions24} are symmetric with respect to the $2\pi/3$-rotations around the $z$-axis, as intended. Now, we proceed to describe triplets of states arranged in a regular polyhedral cone, 
\begin{align}
    \ket{\psi_j} = \begin{pmatrix} c \\ \omega^{j-1}s \end{pmatrix},
\end{align}
where $j\in\{1, 2, 3\}$, $c \equiv \cos \frac{\theta}{4}$, $s \equiv  \sin \frac{\theta}{4}$, $\theta$ the opening angle of the polyhedral cone, and $\omega = e^{i\frac{2\pi}{3}}$. For such a set, we consider the rotation \eqref{eq:coordtransform} of the free set of Wigner negativity. Due to the $\frac{2\pi}{3}$ rotational invariance of the vectors, we can limit our considerations to just one out of three states without loss of generality. The Bloch vector of the target state can be expressed as $\qty(0,-\sin \theta, \cos \theta)$ with respect to the rotated free set. By considering the inequalities defining the set of states with positive Wigner functions, we find that two states are fixed within the free set and the target state is free for $\theta\in\qty[0,2\arccos(-\frac{1}{\sqrt{3}})]\supset\qty[0,\pi]$, which is sufficient to cover all possible angles.\qed

\subsection{The extreme points geometric derivation}
\label{app:ext_pts}

All the states saturating inequalities \eqref{eq:conditions21} and \eqref{eq:conditions25} can be described by a simple vector
\begin{equation}
    v = \mqty(\sqrt{\frac{2}{3}}\cos a \\ \sqrt{\frac{2}{3}}\sin a \\ -\frac{1}{\sqrt{3}}),
\end{equation}
and the simplest critical value of $\theta$ is derived by setting four vectors exactly in this circle,
\begin{align}
    v_1 = \mqty(\sqrt{\frac{2}{3}} \\ 0 \\ -\frac{1}{\sqrt{3}}) && ,\quad 
    v_2 = \mqty(-\sqrt{\frac{2}{3}} \\ 0 \\ -\frac{1}{\sqrt{3}} ) && \hdots
\end{align}

\begin{align}
    \theta_2 = \arccos(v_1 \cdot v_2) = \arccos(-\frac{1}{3}).
\end{align}

The remaining two solutions are found by setting the first two vectors on the primary circle and imposing the mirror symmetry. For convenience we work with a reflected free set without loss of generality,
\begin{align}
    v_1 = \mqty(\sqrt{\frac{2}{3}}\cos(a) \\ \sqrt{\frac{2}{3}}\sin(a) \\ \frac{1}{\sqrt{3}}) && , \quad 
    v_2 = \mqty(\sqrt{\frac{2}{3}}\cos(-a) \\ \sqrt{\frac{2}{3}}\sin(-a) \\ \frac{1}{\sqrt{3}}),
\end{align}
and the other two lying symmetrically on two faces of the set,

\begin{align}
    v_3 = \mqty(
        \frac{1}{3} \sqrt{2} (\sin (b)+1) \\
        \frac{1}{3} \sqrt{2} \left(-\frac{2 \sin (b)}{\sqrt{3}}+\cos (b)+\frac{1}{\sqrt{3}}\right) \\
        \frac{2 \sin (b)+2 \sqrt{3} \cos (b)-1}{3 \sqrt{3}}), \\
    v_4 = \mqty(
        -\frac{1}{3} \sqrt{2} (\sin (b)+1) \\
        \frac{1}{3} \sqrt{2} \left(-\frac{2 \sin (b)}{\sqrt{3}}+\cos (b)+\frac{1}{\sqrt{3}}\right) \\
        \frac{2 \sin (b)+2 \sqrt{3} \cos (b)-1}{3 \sqrt{3}}).
\end{align}
The next step is to require that the products be equal,

\begin{equation}
    v_1\cdot v_2 = v_3\cdot v_4 ,
\end{equation}
which yields 8 possible solutions for $b$,
\begin{equation}
    b = \pm \arccos(\pm\sqrt{ - 3\cos(a)^2 \pm 2\sqrt{3}\cos(a)}).
\end{equation}
In the last step, in order to determine the critical values, we impose condition
\begin{equation}
    v_2\cdot v_4 = v_1\cdot v_2 ,
\end{equation}
which captures the final constraint for the four vectors to form a square. By analysing all the possible values for $b$ one arrives at several cases:
\begin{itemize}
    \item The vectors may not form a square, since the solution forces $v_2$ to be closer to $v_3$ rather than $v_4$, forcing the vectors into a trapezoidal configuration and thus rendering the solution irrelevant.
    \item $a = \frac{\pi}{2}$, which renders the solution redundant, as they are known to be free from the very beginning
    \item There are no real roots of such an equation.
    \item The two solutions that properly give the actual criticals.
\end{itemize}
In particular,
\begin{align*}
    b & = -\arccos(-\sqrt{ - 3\cos(a)^2 + 2\sqrt{3}\cos(a)}) \\ \rightarrow \theta_1 & \approx 0.203171\hdots\pi \\
    b & = \arccos(-\sqrt{ - 3\cos(a)^2 + 2\sqrt{3}\cos(a)})  \\ \rightarrow \theta_3 & \approx 0.710499\hdots\pi ,
\end{align*}
give explicitly the remaining two extreme values of $\theta$, giving the full boundary.
\section{Resources from sets of three arbitrary vectors}
\label{app:3_arb_vecs}

After showing the arrangements of three vectors in regular polyhedral cones to be free, it is natural to ask if any irregular arrangements would be resourceful.  We start from a simple arrangement, which can be treated analytically:

\begin{obs}
    Take a set of three quantum states $\qty{\ket{\psi_1}, \ket{\psi_2}, \ket{\psi_3}}$ with the restriction $\braket{\psi_1}{\psi_2} = 0$. Then, for every $\ket{\psi_3}$, such a triplet is free.
\end{obs}
First, we note that an orthonormal basis formed by the first two vectors can always be made free. Such a pair forms an axis in the Bloch ball that joins midpoints of opposing edges of the free set. These two points can be joined by a continuous path that belongs to the intersection of the free set with the Bloch sphere. Along this path, one finds all possible allowed angles between the axis $\qty{\ket{\psi_1},\ket{\psi_2}}$ and the remaining state $\ket{\psi_3}$. Therefore, by rotating around the distinguished axis, one can always make $\ket{\psi_3}$ free by making it an element of such a continuous path. \qed

Let us consider now a parametrisation given as

\begin{align}
    \ket{\psi_1} &= \mqty(\cos \frac{\theta_1}{4} \\ \sin \frac{\theta_1}{4}) , &&
    \ket{\psi_2} = \mqty(\cos \frac{\theta_1}{4} \\ e^{i\phi_2}\sin \frac{\theta_1}{4}), \nonumber\\\nonumber \\
    \ket{\psi_3} &= \mqty(\cos \frac{\theta_2}{4} \\ e^{i\phi_3}\sin \frac{\theta_2}{4}),&
\end{align}
where the first two vectors lie on a cone with opening angle $\theta_1$ and the third one on a cone with opening angle $\theta_2$. For this parameterisation the triplet $\qty{\ket{\psi_1}, \ket{\psi_2}, \ket{\psi_3}}$ is free whenever $\theta_1 \leq \arccos(\frac{23}{27})$ and $\theta_2\leq \arccos(-\frac{1}{3})$ or when $\theta_1 = \arccos(-\frac{1}{3})$ and $\theta_2 \leq \arccos(\frac{23}{27})$. The conditions in $\theta_1$ come from the fact that the free set (oriented as in Appendix \ref{app:ext_pts}) and the Bloch sphere share full circles for the polar angles $\theta \in [0,\arccos(\frac{23}{27})/2]\cup\qty{\arccos(-\frac{1}{3})/2}$, therefore, for a pair laying on a cone of two times the opening angle can be rotated arbitrary around the axis of the cone and remain free. The restriction on angle $\theta_2$ comes simply from the fact that in the region $\theta >  \arccos(-\frac{1}{3})/2$ the free set does not overlap with the Bloch ball, therefore, making configurations with a greater opening angle potentially, although not necessarily, resourceful. 

Finally, for a parametrisation given by
\begin{align}
    \ket{\psi_1} &= \mqty(\cos \frac{\theta_1}{4} \\ \sin \frac{\theta_1}{4}), &
    \ket{\psi_2} = \mqty(\cos \frac{\theta_2}{4} \\ e^{i \frac{2\pi}{3}}\sin \frac{\theta_2}{4}), \nonumber\\\nonumber \\
    \ket{\psi_3} &= \mqty(\cos \frac{\theta_3}{4} \\ e^{-i \frac{2\pi}{3}}\sin \frac{\theta_3}{4}).&
\end{align}
The triplet is free whenever we have $\theta_i \leq \arccos(-\frac{1}{3})$ for all $i = 1, 2, 3$. This comes directly from the fact that the free set can be rotated as in Appendix \ref{app:ext_pts} and then such states lay on fragments of great circles which coincide with the free set because it being an intersection between the Bloch ball and a regular tetrahedron centered in the ball such that the edges are tangent to the ball at their midpoints. 
\begin{figure}[H]
    \centering
    \includegraphics{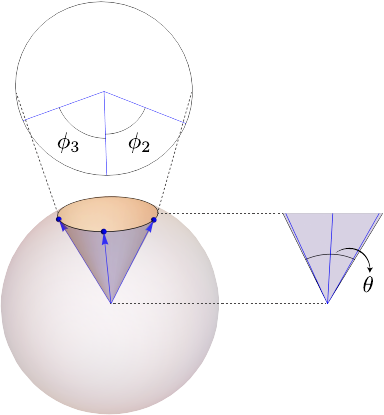}
    \caption{Arbitrary set of three pure qubit states can be parametrised by three numbers. For the sake of simulation we choose the opening angle $\theta \in [0,\pi/2]$ of the common cone on which the vectors lay together with $\phi_1$ and $\phi_2$ between first and second, and first and third vector, respectively. By choosing this parametrisation and restricting $\phi_1\in[0,2\pi/3]$ and $\phi_2\in[\phi_1, \pi - \phi_1/2]$ we avoid double counting of the configurations everywhere except for the boundaries of the considered region.}
    \label{fig:3_irr_vec}
\end{figure}
Given the families presented above and their size, one may be inclined toward thinking that all triplets of pure states are free. In order to probe this possibility we use the numerical annealing procedure from Section \ref{sec:regular_cones} and verify that there actually exist arrangements possessing non-zero absolute Wigner Negativity. We parametrise the three vectors by setting them on a common cone of opening angle $\theta$ and control the angles $\phi_2$ and $\phi_3$ between them in the projection.

\begin{align}
    \ket{\psi_1} &= \mqty(\cos\frac{\theta}{4} \\ \sin\frac{\theta}{4}),& 
    \ket{\psi_2} = \mqty(\cos\frac{\theta}{4} \\
    e^{i\phi_2}\sin\frac{\theta}{4}), \nonumber\\\nonumber \\
        \ket{\psi_3} &= \mqty(\cos\frac{\theta}{4} \\ e^{-i\phi_3}\sin\frac{\theta}{4}).& \label{eq:triplet_parametrisation}
\end{align}

\begin{figure}[H]
    \centering
    \includegraphics[width = \columnwidth]{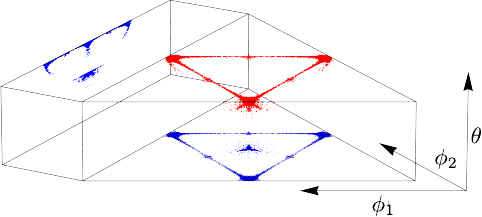}
    \caption{Plot depicting possible resourceful triplets of states as given by parametrisation \eqref{eq:triplet_parametrisation}, based on numerical annealing procedure. 
    The red points represent all triplets found to be resourceful, with blue regions showing orthogonal projections on the base and one of the sides of the prism of possibilities, respectively. It is important to note resourceful triplets which proceed from a small perturbations to triplets comprised of an orthogonal basis and an arbitrary third vector (the regular triangle in the top layer, $\theta = \pi/2$), which are qualitatively understood. 
    }
    \label{fig:3_irr_vec_sim}
\end{figure}

By taking such vectors, we need to consider only $\theta\in\qty[0,\pi/2]$, $\phi_2\in[0,2\pi/3]$ and $\phi_3\in[\phi_2, \pi - \phi_2/2]$  to avoid double counting of configurations. The results presented in Fig.~\ref{fig:3_irr_vec_sim}, suggest that almost all of the triplets of pure states are free. 

The main region of interest is located around $\theta = \pi/2$ and $\phi_1 + \phi_2 = \pi$, together with the symmetric regions. It represents perturbations of free sets with $\braket{\psi_1}{\psi_2} =  0$ and an arbitrary third vector. The existence of resourceful sets within this neighbourhood can be explained qualitatively. Given a basis consisting of $\ket{\psi_1}$ and $\ket{\psi_2}$, there exists only a single direction in which an infinitesimal change of the second vector $\ket{\psi_2} \rightarrow \ket{\psi_2} + \epsilon \ket{\phi}$ leaves it in the free set of zero standard negativity. Therefore, if we perturb it along a different direction, $\ket{\psi_2} \rightarrow \ket{\psi_2} + \epsilon \ket{\phi'}$ with $\abs{\braket{\phi}{\phi'}} = 1$, a rotation which would allow us to keep all three states free may not exist. 

A similar interpretation can be given for the states with $\theta\approx \arccos(-1/3)$ with $\phi_1\approx\phi_2\approx\pi-\phi_1-\phi_2$, located around the central axis of the prism. They represent a small perturbation of the set composed from a single vector taken from each of the three MUBs of the full set. Such a triple is easily proved to be free, but analogously, they are made free only in measure-zero subsets of all possible unitary transformations. Therefore, small perturbations of such a set may be resourceful, as proven by the simulation results.

\begin{widetext}

\section{Hierarchy for higher dimensions} 
\label{app:ndimHier}

In order to extend the hierarchy from qubits to higher-dimensional spaces, we remark that each unitary in $\mathrm{SU}(d)$ is equivalent to a certain composition of at most $d(d-1)/2$ unitaries in $\mathrm{SU}(2)$. For instance, by defining $\mathcal{V}^{(ij)}$ with $i < j$ as the unitaries acting in the subspace spanned by $\ket{i}, \ket{j}$, we can recover $\mathrm{SU}(d)$ by selecting one unitary from every $\mathcal{V}^{(ij)}$, composing them and then taking all possible compositions of such form. 

Then, we proceed to discretize every $\mathcal{V}^{(ij)}$ into a set of unitaries $O^{(ij)}_m$ by replacing $\ket{0}\rightarrow\ket{i},\ket{1}\rightarrow \ket{j}$ and following our qubit hierarchy stages 1-2 until step $m$ and defining $O^{(ij)}_m$ as in stage 3. Here, we define $O_m(d)$ as the set of all $\mathrm{SU}(d)$ unitaries composed of $d(d-1)/2$ unitaries, each  selected from a different $O^{(ij)}_m$. 

Finally, by considering the group $S_{\mathcal{W}}(d)$ which leaves the free set invariant, we further simplify the set of unitaries by taking $\mathcal{U}_{m}(d)\cong O_{m}(d)/S_{\mathcal{W}}(d)$ as the set of unitaries in the $m$ step of the hierarchy. For the Wigner representation of the main text examples, one always arrives at $d^2$ hyperplanes, which delimit the free set. Thus the reduction may be as considerable as a factor $d^2!$, taking into account the symmetries of the  $(d^2\!-\!1)$-dimensional regular simplex form by the limiting hyperplanes.

For large $d$, the total number of unitaries in  $\mathcal{U}_m(d)$ is  approximately,

\begin{align*}
    \overline{\mathcal{U}_m(d)}
    & \approx \frac{\qty[\qty(6 + 12 (3\cdot2^{m-1} -1) + 8 \binom{3\cdot 2^{m-1}-1}{2})\qty(6\cdot 2^{m})]^{d(d-1)/2}}{d^2!} \\
    & \approx \frac{\qty[\qty(6 + 12 (3\cdot2^{m-1} -1) + 8 \binom{3\cdot 2^{m-1}-1}{2})\qty(6\cdot 2^{m})]^{d(d-1)/2} e^{d^2}}{\sqrt{2\pi}d^{2d^2 + 1}} \\
    & \overset{m\gg1}{\approx} \frac{\qty[48 \binom{3\cdot 2^{m-1}}{3}]^{d(d-1)/2} e^{d^2}}{\sqrt{2\pi}d^{2d^2 + 1}} \approx \frac{72^{d(d-1)/2}2^{md(d-1)/2}e^{d^2}}{\sqrt{2\pi}d^{2d^2 + 1}} \\
    & = \frac{e^{(\log 72 + m \log 2 + 1 - 2 \log d)d^2 - (\log 72 + m \log 2)d - 2\log d}}{\sqrt{2\pi}} .
\end{align*}

From the above it turns that to keep an approximately constant number of unitaries as $d$ increases, the last level in the hierarchy must increase like $ m(d) \propto \frac{2}{\log 2} \log (\frac{d}{12})$. Since the monotone $\mathcal{M}_{n}^{(m)}$ defined by the hiearchy at level $m$  is closer to the quantifier $\mathcal{M}_{n}$ as the number of non redundant unitaries $\overline{\mathcal{U}_m(d)}$ increases, to keep similar degrees of approximation between two large dimension $d_1 < d_2$  we should have $m(d_2)/m(d_1)\approx \log (\frac{d_2}{12})/\log (\frac{d_1}{12})$.

\end{widetext}
Now, to distinguish our hierarchy from other techniques used in the state of the art of resource theory, we compare it with \cite{Oszmaniec20017}. A crucial difference with the above work is that their approximation of the polytope in question is a single approximation as opposed to the convergent hierarchy we present; moreover, in ref \cite{Oszmaniec20017}, the Bloch sphere is approximated from the outside, while our hierarchy, approximates it from the inside.  

More specifically, they consider a 100-point approximation of semicircle times a truncated icosahedron and its dual, yielding a polytope of 850 000 points. However, as far as we can observe, they are not leveraging the symmetries of the underlying space, which might reduce the number of points necessary for the precision found in their article. In our work, we leverage the symmetries of the free set with a significant reduction by a factor of $d^2!$.

Furthermore, we remark that the explicit goal of \cite{Oszmaniec20017}  is to approximate the space of measurements. In contrast, we approximate the space of possible unitaries, and the hierarchy is extendable to a large class of multi-object resource theories, like sets of measurements, states or channels.

\bibliographystyle{apsrev4-2}
\bibliography{references}

\begin{thebibliography}{56}%
\makeatletter
\providecommand \@ifxundefined [1]{%
 \@ifx{#1\undefined}
}%
\providecommand \@ifnum [1]{%
 \ifnum #1\expandafter \@firstoftwo
 \else \expandafter \@secondoftwo
 \fi
}%
\providecommand \@ifx [1]{%
 \ifx #1\expandafter \@firstoftwo
 \else \expandafter \@secondoftwo
 \fi
}%
\providecommand \natexlab [1]{#1}%
\providecommand \enquote  [1]{``#1''}%
\providecommand \bibnamefont  [1]{#1}%
\providecommand \bibfnamefont [1]{#1}%
\providecommand \citenamefont [1]{#1}%
\providecommand \href@noop [0]{\@secondoftwo}%
\providecommand \href [0]{\begingroup \@sanitize@url \@href}%
\providecommand \@href[1]{\@@startlink{#1}\@@href}%
\providecommand \@@href[1]{\endgroup#1\@@endlink}%
\providecommand \@sanitize@url [0]{\catcode `\\12\catcode `\$12\catcode
  `\&12\catcode `\#12\catcode `\^12\catcode `\_12\catcode `\%12\relax}%
\providecommand \@@startlink[1]{}%
\providecommand \@@endlink[0]{}%
\providecommand \url  [0]{\begingroup\@sanitize@url \@url }%
\providecommand \@url [1]{\endgroup\@href {#1}{\urlprefix }}%
\providecommand \urlprefix  [0]{URL }%
\providecommand \Eprint [0]{\href }%
\providecommand \doibase [0]{https://doi.org/}%
\providecommand \selectlanguage [0]{\@gobble}%
\providecommand \bibinfo  [0]{\@secondoftwo}%
\providecommand \bibfield  [0]{\@secondoftwo}%
\providecommand \translation [1]{[#1]}%
\providecommand \BibitemOpen [0]{}%
\providecommand \bibitemStop [0]{}%
\providecommand \bibitemNoStop [0]{.\EOS\space}%
\providecommand \EOS [0]{\spacefactor3000\relax}%
\providecommand \BibitemShut  [1]{\csname bibitem#1\endcsname}%
\let\auto@bib@innerbib\@empty
\bibitem [{\citenamefont {Chitambar}\ and\ \citenamefont
  {Gour}(2019)}]{Gour2019}%
  \BibitemOpen
  \bibfield  {author} {\bibinfo {author} {\bibfnamefont {E.}~\bibnamefont
  {Chitambar}}\ and\ \bibinfo {author} {\bibfnamefont {G.}~\bibnamefont
  {Gour}},\ }\href {https://doi.org/10.1103/RevModPhys.91.025001} {\bibfield
  {journal} {\bibinfo  {journal} {Rev. Mod. Phys.}\ }\textbf {\bibinfo {volume}
  {91}},\ \bibinfo {pages} {025001} (\bibinfo {year} {2019})}\BibitemShut
  {NoStop}%
\bibitem [{\citenamefont {Horodecki}\ \emph {et~al.}(2003)\citenamefont
  {Horodecki}, \citenamefont {Horodecki}, \citenamefont {Horodecki},
  \citenamefont {Horodecki}, \citenamefont {Oppenheim}, \citenamefont
  {Sen(De)},\ and\ \citenamefont {Sen}}]{Horodecki2003}%
  \BibitemOpen
  \bibfield  {author} {\bibinfo {author} {\bibfnamefont {M.}~\bibnamefont
  {Horodecki}}, \bibinfo {author} {\bibfnamefont {K.}~\bibnamefont
  {Horodecki}}, \bibinfo {author} {\bibfnamefont {P.}~\bibnamefont
  {Horodecki}}, \bibinfo {author} {\bibfnamefont {R.}~\bibnamefont
  {Horodecki}}, \bibinfo {author} {\bibfnamefont {J.}~\bibnamefont
  {Oppenheim}}, \bibinfo {author} {\bibfnamefont {A.}~\bibnamefont {Sen(De)}},\
  and\ \bibinfo {author} {\bibfnamefont {U.}~\bibnamefont {Sen}},\ }\href
  {https://doi.org/10.1103/PhysRevLett.90.100402} {\bibfield  {journal}
  {\bibinfo  {journal} {Phys. Rev. Lett.}\ }\textbf {\bibinfo {volume} {90}},\
  \bibinfo {pages} {100402} (\bibinfo {year} {2003})}\BibitemShut {NoStop}%
\bibitem [{\citenamefont {Horodecki}\ \emph {et~al.}(2009)\citenamefont
  {Horodecki}, \citenamefont {Horodecki}, \citenamefont {Horodecki},\ and\
  \citenamefont {Horodecki}}]{Horodecki2009}%
  \BibitemOpen
  \bibfield  {author} {\bibinfo {author} {\bibfnamefont {R.}~\bibnamefont
  {Horodecki}}, \bibinfo {author} {\bibfnamefont {P.}~\bibnamefont
  {Horodecki}}, \bibinfo {author} {\bibfnamefont {M.}~\bibnamefont
  {Horodecki}},\ and\ \bibinfo {author} {\bibfnamefont {K.}~\bibnamefont
  {Horodecki}},\ }\href {https://doi.org/10.1103/RevModPhys.81.865} {\bibfield
  {journal} {\bibinfo  {journal} {Rev. Mod. Phys.}\ }\textbf {\bibinfo {volume}
  {81}},\ \bibinfo {pages} {865} (\bibinfo {year} {2009})}\BibitemShut
  {NoStop}%
\bibitem [{\citenamefont {Streltsov}\ \emph {et~al.}(2017)\citenamefont
  {Streltsov}, \citenamefont {Adesso},\ and\ \citenamefont
  {Plenio}}]{Streltsov2017}%
  \BibitemOpen
  \bibfield  {author} {\bibinfo {author} {\bibfnamefont {A.}~\bibnamefont
  {Streltsov}}, \bibinfo {author} {\bibfnamefont {G.}~\bibnamefont {Adesso}},\
  and\ \bibinfo {author} {\bibfnamefont {M.~B.}\ \bibnamefont {Plenio}},\
  }\href {https://doi.org/10.1103/RevModPhys.89.041003} {\bibfield  {journal}
  {\bibinfo  {journal} {Rev. Mod. Phys.}\ }\textbf {\bibinfo {volume} {89}},\
  \bibinfo {pages} {041003} (\bibinfo {year} {2017})}\BibitemShut {NoStop}%
\bibitem [{\citenamefont {Takagi}\ and\ \citenamefont
  {Regula}(2019)}]{TakagiRegula2019}%
  \BibitemOpen
  \bibfield  {author} {\bibinfo {author} {\bibfnamefont {R.}~\bibnamefont
  {Takagi}}\ and\ \bibinfo {author} {\bibfnamefont {B.}~\bibnamefont
  {Regula}},\ }\href {https://doi.org/10.1103/PhysRevX.9.031053} {\bibfield
  {journal} {\bibinfo  {journal} {Phys. Rev. X}\ }\textbf {\bibinfo {volume}
  {9}},\ \bibinfo {pages} {031053} (\bibinfo {year} {2019})}\BibitemShut
  {NoStop}%
\bibitem [{\citenamefont {Oszmaniec}\ and\ \citenamefont
  {Biswas}(2019)}]{Oszmaniec2019}%
  \BibitemOpen
  \bibfield  {author} {\bibinfo {author} {\bibfnamefont {M.}~\bibnamefont
  {Oszmaniec}}\ and\ \bibinfo {author} {\bibfnamefont {T.}~\bibnamefont
  {Biswas}},\ }\href {https://doi.org/10.22331/q-2019-04-26-133} {\bibfield
  {journal} {\bibinfo  {journal} {{Quantum}}\ }\textbf {\bibinfo {volume}
  {3}},\ \bibinfo {pages} {133} (\bibinfo {year} {2019})}\BibitemShut {NoStop}%
\bibitem [{\citenamefont {Skrzypczyk}\ \emph {et~al.}(2019)\citenamefont
  {Skrzypczyk}, \citenamefont {\ifmmode \check{S}\else
  \v{S}\fi{}upi\ifmmode~\acute{c}\else \'{c}\fi{}},\ and\ \citenamefont
  {Cavalcanti}}]{Paul2019}%
  \BibitemOpen
  \bibfield  {author} {\bibinfo {author} {\bibfnamefont {P.}~\bibnamefont
  {Skrzypczyk}}, \bibinfo {author} {\bibfnamefont {I.}~\bibnamefont {\ifmmode
  \check{S}\else \v{S}\fi{}upi\ifmmode~\acute{c}\else \'{c}\fi{}}},\ and\
  \bibinfo {author} {\bibfnamefont {D.}~\bibnamefont {Cavalcanti}},\ }\href
  {https://doi.org/10.1103/PhysRevLett.122.130403} {\bibfield  {journal}
  {\bibinfo  {journal} {Phys. Rev. Lett.}\ }\textbf {\bibinfo {volume} {122}},\
  \bibinfo {pages} {130403} (\bibinfo {year} {2019})}\BibitemShut {NoStop}%
\bibitem [{\citenamefont {Regula}\ \emph {et~al.}(2020)\citenamefont {Regula},
  \citenamefont {Bu}, \citenamefont {Takagi},\ and\ \citenamefont
  {Liu}}]{Regula2020}%
  \BibitemOpen
  \bibfield  {author} {\bibinfo {author} {\bibfnamefont {B.}~\bibnamefont
  {Regula}}, \bibinfo {author} {\bibfnamefont {K.}~\bibnamefont {Bu}}, \bibinfo
  {author} {\bibfnamefont {R.}~\bibnamefont {Takagi}},\ and\ \bibinfo {author}
  {\bibfnamefont {Z.-W.}\ \bibnamefont {Liu}},\ }\href
  {https://doi.org/10.1103/PhysRevA.101.062315} {\bibfield  {journal} {\bibinfo
   {journal} {Phys. Rev. A}\ }\textbf {\bibinfo {volume} {101}},\ \bibinfo
  {pages} {062315} (\bibinfo {year} {2020})}\BibitemShut {NoStop}%
\bibitem [{\citenamefont {Wehner}\ \emph {et~al.}(2018)\citenamefont {Wehner},
  \citenamefont {Elkouss},\ and\ \citenamefont {Hanson}}]{Wehner2018}%
  \BibitemOpen
  \bibfield  {author} {\bibinfo {author} {\bibfnamefont {S.}~\bibnamefont
  {Wehner}}, \bibinfo {author} {\bibfnamefont {D.}~\bibnamefont {Elkouss}},\
  and\ \bibinfo {author} {\bibfnamefont {R.}~\bibnamefont {Hanson}},\ }\href
  {https://doi.org/10.1126/science.aam9288} {\bibfield  {journal} {\bibinfo
  {journal} {Science}\ }\textbf {\bibinfo {volume} {362}} (\bibinfo {year}
  {2018})}\BibitemShut {NoStop}%
\bibitem [{\citenamefont {Uola}\ \emph {et~al.}(2019)\citenamefont {Uola},
  \citenamefont {Kraft}, \citenamefont {Shang}, \citenamefont {Yu},\ and\
  \citenamefont {G\"uhne}}]{Roope2019}%
  \BibitemOpen
  \bibfield  {author} {\bibinfo {author} {\bibfnamefont {R.}~\bibnamefont
  {Uola}}, \bibinfo {author} {\bibfnamefont {T.}~\bibnamefont {Kraft}},
  \bibinfo {author} {\bibfnamefont {J.}~\bibnamefont {Shang}}, \bibinfo
  {author} {\bibfnamefont {X.-D.}\ \bibnamefont {Yu}},\ and\ \bibinfo {author}
  {\bibfnamefont {O.}~\bibnamefont {G\"uhne}},\ }\href
  {https://doi.org/10.1103/PhysRevLett.122.130404} {\bibfield  {journal}
  {\bibinfo  {journal} {Phys. Rev. Lett.}\ }\textbf {\bibinfo {volume} {122}},\
  \bibinfo {pages} {130404} (\bibinfo {year} {2019})}\BibitemShut {NoStop}%
\bibitem [{\citenamefont {Buscemi}\ \emph {et~al.}(2020)\citenamefont
  {Buscemi}, \citenamefont {Chitambar},\ and\ \citenamefont
  {Zhou}}]{Buscemi2020}%
  \BibitemOpen
  \bibfield  {author} {\bibinfo {author} {\bibfnamefont {F.}~\bibnamefont
  {Buscemi}}, \bibinfo {author} {\bibfnamefont {E.}~\bibnamefont {Chitambar}},\
  and\ \bibinfo {author} {\bibfnamefont {W.}~\bibnamefont {Zhou}},\ }\href
  {https://doi.org/10.1103/PhysRevLett.124.120401} {\bibfield  {journal}
  {\bibinfo  {journal} {Phys. Rev. Lett.}\ }\textbf {\bibinfo {volume} {124}},\
  \bibinfo {pages} {120401} (\bibinfo {year} {2020})}\BibitemShut {NoStop}%
\bibitem [{\citenamefont {Ducuara}\ \emph {et~al.}(2020)\citenamefont
  {Ducuara}, \citenamefont {Lipka-Bartosik},\ and\ \citenamefont
  {Skrzypczyk}}]{Ducuara2020}%
  \BibitemOpen
  \bibfield  {author} {\bibinfo {author} {\bibfnamefont {A.~F.}\ \bibnamefont
  {Ducuara}}, \bibinfo {author} {\bibfnamefont {P.}~\bibnamefont
  {Lipka-Bartosik}},\ and\ \bibinfo {author} {\bibfnamefont {P.}~\bibnamefont
  {Skrzypczyk}},\ }\href {https://doi.org/10.1103/PhysRevResearch.2.033374}
  {\bibfield  {journal} {\bibinfo  {journal} {Phys. Rev. Research}\ }\textbf
  {\bibinfo {volume} {2}},\ \bibinfo {pages} {033374} (\bibinfo {year}
  {2020})}\BibitemShut {NoStop}%
\bibitem [{\citenamefont {Martins}\ \emph {et~al.}(2020)\citenamefont
  {Martins}, \citenamefont {Savi},\ and\ \citenamefont {Angelo}}]{Martins2020}%
  \BibitemOpen
  \bibfield  {author} {\bibinfo {author} {\bibfnamefont {E.}~\bibnamefont
  {Martins}}, \bibinfo {author} {\bibfnamefont {M.~F.}\ \bibnamefont {Savi}},\
  and\ \bibinfo {author} {\bibfnamefont {R.~M.}\ \bibnamefont {Angelo}},\
  }\href {https://doi.org/10.1103/PhysRevA.102.050201} {\bibfield  {journal}
  {\bibinfo  {journal} {Phys. Rev. A}\ }\textbf {\bibinfo {volume} {102}},\
  \bibinfo {pages} {050201} (\bibinfo {year} {2020})}\BibitemShut {NoStop}%
\bibitem [{\citenamefont {Designolle}\ \emph {et~al.}(2021)\citenamefont
  {Designolle}, \citenamefont {Uola}, \citenamefont {Luoma},\ and\
  \citenamefont {Brunner}}]{Designole2020}%
  \BibitemOpen
  \bibfield  {author} {\bibinfo {author} {\bibfnamefont {S.}~\bibnamefont
  {Designolle}}, \bibinfo {author} {\bibfnamefont {R.}~\bibnamefont {Uola}},
  \bibinfo {author} {\bibfnamefont {K.}~\bibnamefont {Luoma}},\ and\ \bibinfo
  {author} {\bibfnamefont {N.}~\bibnamefont {Brunner}},\ }\href
  {https://doi.org/10.1103/PhysRevLett.126.220404} {\bibfield  {journal}
  {\bibinfo  {journal} {Phys. Rev. Lett.}\ }\textbf {\bibinfo {volume} {126}},\
  \bibinfo {pages} {220404} (\bibinfo {year} {2021})}\BibitemShut {NoStop}%
\bibitem [{\citenamefont {Haapasalo}\ \emph {et~al.}(2021)\citenamefont
  {Haapasalo}, \citenamefont {Kraft}, \citenamefont {Miklin},\ and\
  \citenamefont {Uola}}]{Roope2021}%
  \BibitemOpen
  \bibfield  {author} {\bibinfo {author} {\bibfnamefont {E.}~\bibnamefont
  {Haapasalo}}, \bibinfo {author} {\bibfnamefont {T.}~\bibnamefont {Kraft}},
  \bibinfo {author} {\bibfnamefont {N.}~\bibnamefont {Miklin}},\ and\ \bibinfo
  {author} {\bibfnamefont {R.}~\bibnamefont {Uola}},\ }\href
  {https://doi.org/10.22331/q-2021-06-15-476} {\bibfield  {journal} {\bibinfo
  {journal} {{Quantum}}\ }\textbf {\bibinfo {volume} {5}},\ \bibinfo {pages}
  {476} (\bibinfo {year} {2021})}\BibitemShut {NoStop}%
\bibitem [{\citenamefont {Ferrie}\ and\ \citenamefont
  {Emerson}(2009)}]{Ferrie2009}%
  \BibitemOpen
  \bibfield  {author} {\bibinfo {author} {\bibfnamefont {C.}~\bibnamefont
  {Ferrie}}\ and\ \bibinfo {author} {\bibfnamefont {J.}~\bibnamefont
  {Emerson}},\ }\href {https://doi.org/10.1088/1367-2630/11/6/063040}
  {\bibfield  {journal} {\bibinfo  {journal} {New J. Phys.}\ }\textbf {\bibinfo
  {volume} {11}},\ \bibinfo {pages} {063040} (\bibinfo {year}
  {2009})}\BibitemShut {NoStop}%
\bibitem [{\citenamefont {Ferrie}(2011)}]{Ferrie2011}%
  \BibitemOpen
  \bibfield  {author} {\bibinfo {author} {\bibfnamefont {C.}~\bibnamefont
  {Ferrie}},\ }\href {https://doi.org/10.1088/0034-4885/74/11/116001}
  {\bibfield  {journal} {\bibinfo  {journal} {Rep. Prog. Phys.}\ }\textbf
  {\bibinfo {volume} {74}},\ \bibinfo {pages} {116001} (\bibinfo {year}
  {2011})}\BibitemShut {NoStop}%
\bibitem [{\citenamefont {Titulaer}\ and\ \citenamefont
  {Glauber}(1965)}]{Glauber1965}%
  \BibitemOpen
  \bibfield  {author} {\bibinfo {author} {\bibfnamefont {U.~M.}\ \bibnamefont
  {Titulaer}}\ and\ \bibinfo {author} {\bibfnamefont {R.~J.}\ \bibnamefont
  {Glauber}},\ }\href {https://doi.org/10.1103/PhysRev.140.B676} {\bibfield
  {journal} {\bibinfo  {journal} {Phys. Rev.}\ }\textbf {\bibinfo {volume}
  {140}},\ \bibinfo {pages} {B676} (\bibinfo {year} {1965})}\BibitemShut
  {NoStop}%
\bibitem [{\citenamefont {Kenfack}\ and\ \citenamefont
  {{\.Z}yczkowski}(2004)}]{Kenfack2004}%
  \BibitemOpen
  \bibfield  {author} {\bibinfo {author} {\bibfnamefont {A.}~\bibnamefont
  {Kenfack}}\ and\ \bibinfo {author} {\bibfnamefont {K.}~\bibnamefont
  {{\.Z}yczkowski}},\ }\href {https://doi.org/10.1088/1464-4266/6/10/003}
  {\bibfield  {journal} {\bibinfo  {journal} {J. opt., B Quantum semiclass.
  opt.}\ }\textbf {\bibinfo {volume} {6}},\ \bibinfo {pages} {396} (\bibinfo
  {year} {2004})}\BibitemShut {NoStop}%
\bibitem [{\citenamefont {Tan}\ \emph {et~al.}(2020)\citenamefont {Tan},
  \citenamefont {Choi},\ and\ \citenamefont {Jeong}}]{Chuan2020}%
  \BibitemOpen
  \bibfield  {author} {\bibinfo {author} {\bibfnamefont {K.~C.}\ \bibnamefont
  {Tan}}, \bibinfo {author} {\bibfnamefont {S.}~\bibnamefont {Choi}},\ and\
  \bibinfo {author} {\bibfnamefont {H.}~\bibnamefont {Jeong}},\ }\href
  {https://doi.org/10.1103/PhysRevLett.124.110404} {\bibfield  {journal}
  {\bibinfo  {journal} {Phys. Rev. Lett.}\ }\textbf {\bibinfo {volume} {124}},\
  \bibinfo {pages} {110404} (\bibinfo {year} {2020})}\BibitemShut {NoStop}%
\bibitem [{\citenamefont {Mari}\ and\ \citenamefont
  {Eisert}(2012)}]{Eisert2012}%
  \BibitemOpen
  \bibfield  {author} {\bibinfo {author} {\bibfnamefont {A.}~\bibnamefont
  {Mari}}\ and\ \bibinfo {author} {\bibfnamefont {J.}~\bibnamefont {Eisert}},\
  }\href {https://doi.org/10.1103/PhysRevLett.109.230503} {\bibfield  {journal}
  {\bibinfo  {journal} {Phys. Rev. Lett.}\ }\textbf {\bibinfo {volume} {109}},\
  \bibinfo {pages} {230503} (\bibinfo {year} {2012})}\BibitemShut {NoStop}%
\bibitem [{\citenamefont {Pashayan}\ \emph {et~al.}(2015)\citenamefont
  {Pashayan}, \citenamefont {Wallman},\ and\ \citenamefont
  {Bartlett}}]{Pashayan2015}%
  \BibitemOpen
  \bibfield  {author} {\bibinfo {author} {\bibfnamefont {H.}~\bibnamefont
  {Pashayan}}, \bibinfo {author} {\bibfnamefont {J.~J.}\ \bibnamefont
  {Wallman}},\ and\ \bibinfo {author} {\bibfnamefont {S.~D.}\ \bibnamefont
  {Bartlett}},\ }\href {https://doi.org/10.1103/PhysRevLett.115.070501}
  {\bibfield  {journal} {\bibinfo  {journal} {Phys. Rev. Lett.}\ }\textbf
  {\bibinfo {volume} {115}},\ \bibinfo {pages} {070501} (\bibinfo {year}
  {2015})}\BibitemShut {NoStop}%
\bibitem [{\citenamefont {Wigner}(1932)}]{Wigner1932}%
  \BibitemOpen
  \bibfield  {author} {\bibinfo {author} {\bibfnamefont {E.}~\bibnamefont
  {Wigner}},\ }\href {https://doi.org/10.1103/PhysRev.40.749} {\bibfield
  {journal} {\bibinfo  {journal} {Phys. Rev.}\ }\textbf {\bibinfo {volume}
  {40}},\ \bibinfo {pages} {749} (\bibinfo {year} {1932})}\BibitemShut
  {NoStop}%
\bibitem [{\citenamefont {Peres}(1995)}]{peres1995quantum}%
  \BibitemOpen
  \bibfield  {author} {\bibinfo {author} {\bibfnamefont {A.}~\bibnamefont
  {Peres}},\ }\href {https://books.google.pl/books?id=rMGqMyFBcL8C} {\emph
  {\bibinfo {title} {Quantum Theory: Concepts and Methods}}},\ Fundamental
  Theories of Physics\ (\bibinfo  {publisher} {Springer Netherlands},\ \bibinfo
  {year} {1995})\BibitemShut {NoStop}%
\bibitem [{\citenamefont {Keyl}(2002)}]{Keyl2002}%
  \BibitemOpen
  \bibfield  {author} {\bibinfo {author} {\bibfnamefont {M.}~\bibnamefont
  {Keyl}},\ }\href
  {https://doi.org/https://doi.org/10.1016/S0370-1573(02)00266-1} {\bibfield
  {journal} {\bibinfo  {journal} {Phys. Rep.}\ }\textbf {\bibinfo {volume}
  {369}},\ \bibinfo {pages} {431} (\bibinfo {year} {2002})}\BibitemShut
  {NoStop}%
\bibitem [{\citenamefont {Hillery}\ \emph {et~al.}(1984)\citenamefont
  {Hillery}, \citenamefont {O'Connell}, \citenamefont {Scully},\ and\
  \citenamefont {Wigner}}]{hillery1984}%
  \BibitemOpen
  \bibfield  {author} {\bibinfo {author} {\bibfnamefont {M.}~\bibnamefont
  {Hillery}}, \bibinfo {author} {\bibfnamefont {R.}~\bibnamefont {O'Connell}},
  \bibinfo {author} {\bibfnamefont {M.}~\bibnamefont {Scully}},\ and\ \bibinfo
  {author} {\bibfnamefont {E.}~\bibnamefont {Wigner}},\ }\href
  {https://doi.org/https://doi.org/10.1016/0370-1573(84)90160-1} {\bibfield
  {journal} {\bibinfo  {journal} {Phys. Rep.}\ }\textbf {\bibinfo {volume}
  {106}},\ \bibinfo {pages} {121} (\bibinfo {year} {1984})}\BibitemShut
  {NoStop}%
\bibitem [{\citenamefont {Lee}(1995)}]{Lee1995}%
  \BibitemOpen
  \bibfield  {author} {\bibinfo {author} {\bibfnamefont {H.-W.}\ \bibnamefont
  {Lee}},\ }\href
  {https://doi.org/https://doi.org/10.1016/0370-1573(95)00007-4} {\bibfield
  {journal} {\bibinfo  {journal} {Phys. Rep.}\ }\textbf {\bibinfo {volume}
  {259}},\ \bibinfo {pages} {147} (\bibinfo {year} {1995})}\BibitemShut
  {NoStop}%
\bibitem [{\citenamefont {Leonhardt}(1996)}]{Leonhardt1996}%
  \BibitemOpen
  \bibfield  {author} {\bibinfo {author} {\bibfnamefont {U.}~\bibnamefont
  {Leonhardt}},\ }\href {https://doi.org/10.1103/PhysRevA.53.2998} {\bibfield
  {journal} {\bibinfo  {journal} {Phys. Rev. A}\ }\textbf {\bibinfo {volume}
  {53}},\ \bibinfo {pages} {2998} (\bibinfo {year} {1996})}\BibitemShut
  {NoStop}%
\bibitem [{\citenamefont {Gibbons}\ \emph {et~al.}(2004)\citenamefont
  {Gibbons}, \citenamefont {Hoffman},\ and\ \citenamefont
  {Wootters}}]{Gibbons2004}%
  \BibitemOpen
  \bibfield  {author} {\bibinfo {author} {\bibfnamefont {K.~S.}\ \bibnamefont
  {Gibbons}}, \bibinfo {author} {\bibfnamefont {M.~J.}\ \bibnamefont
  {Hoffman}},\ and\ \bibinfo {author} {\bibfnamefont {W.~K.}\ \bibnamefont
  {Wootters}},\ }\href {https://doi.org/10.1103/PhysRevA.70.062101} {\bibfield
  {journal} {\bibinfo  {journal} {Phys. Rev. A}\ }\textbf {\bibinfo {volume}
  {70}},\ \bibinfo {pages} {062101} (\bibinfo {year} {2004})}\BibitemShut
  {NoStop}%
\bibitem [{\citenamefont {Klimov}\ \emph {et~al.}(2006)\citenamefont {Klimov},
  \citenamefont {Mu{\~{n}}oz},\ and\ \citenamefont {Romero}}]{Klimov2006}%
  \BibitemOpen
  \bibfield  {author} {\bibinfo {author} {\bibfnamefont {A.~B.}\ \bibnamefont
  {Klimov}}, \bibinfo {author} {\bibfnamefont {C.}~\bibnamefont
  {Mu{\~{n}}oz}},\ and\ \bibinfo {author} {\bibfnamefont {J.~L.}\ \bibnamefont
  {Romero}},\ }\href {https://doi.org/10.1088/0305-4470/39/46/016} {\bibfield
  {journal} {\bibinfo  {journal} {J. Phys. A Math. Theor.}\ }\textbf {\bibinfo
  {volume} {39}},\ \bibinfo {pages} {14471} (\bibinfo {year}
  {2006})}\BibitemShut {NoStop}%
\bibitem [{\citenamefont {Gross}(2007)}]{Gross2007}%
  \BibitemOpen
  \bibfield  {author} {\bibinfo {author} {\bibfnamefont {D.}~\bibnamefont
  {Gross}},\ }\href {https://doi.org/10.1007/s00340-006-2510-9} {\bibfield
  {journal} {\bibinfo  {journal} {Appl. Phys. B}\ }\textbf {\bibinfo {volume}
  {86}},\ \bibinfo {pages} {367} (\bibinfo {year} {2007})}\BibitemShut
  {NoStop}%
\bibitem [{\citenamefont {Takagi}\ \emph {et~al.}(2019)\citenamefont {Takagi},
  \citenamefont {Regula}, \citenamefont {Bu}, \citenamefont {Liu},\ and\
  \citenamefont {Adesso}}]{Takagi2019}%
  \BibitemOpen
  \bibfield  {author} {\bibinfo {author} {\bibfnamefont {R.}~\bibnamefont
  {Takagi}}, \bibinfo {author} {\bibfnamefont {B.}~\bibnamefont {Regula}},
  \bibinfo {author} {\bibfnamefont {K.}~\bibnamefont {Bu}}, \bibinfo {author}
  {\bibfnamefont {Z.-W.}\ \bibnamefont {Liu}},\ and\ \bibinfo {author}
  {\bibfnamefont {G.}~\bibnamefont {Adesso}},\ }\href
  {https://doi.org/10.1103/PhysRevLett.122.140402} {\bibfield  {journal}
  {\bibinfo  {journal} {Phys. Rev. Lett.}\ }\textbf {\bibinfo {volume} {122}},\
  \bibinfo {pages} {140402} (\bibinfo {year} {2019})}\BibitemShut {NoStop}%
\bibitem [{\citenamefont {Ekert}(1991)}]{Ekert1991}%
  \BibitemOpen
  \bibfield  {author} {\bibinfo {author} {\bibfnamefont {A.~K.}\ \bibnamefont
  {Ekert}},\ }\href {https://doi.org/10.1103/PhysRevLett.67.661} {\bibfield
  {journal} {\bibinfo  {journal} {Phys. Rev. Lett.}\ }\textbf {\bibinfo
  {volume} {67}},\ \bibinfo {pages} {661} (\bibinfo {year} {1991})}\BibitemShut
  {NoStop}%
\bibitem [{\citenamefont {Bravyi}\ \emph {et~al.}(2018)\citenamefont {Bravyi},
  \citenamefont {Gosset},\ and\ \citenamefont {K{\"o}nig}}]{bravy2018}%
  \BibitemOpen
  \bibfield  {author} {\bibinfo {author} {\bibfnamefont {S.}~\bibnamefont
  {Bravyi}}, \bibinfo {author} {\bibfnamefont {D.}~\bibnamefont {Gosset}},\
  and\ \bibinfo {author} {\bibfnamefont {R.}~\bibnamefont {K{\"o}nig}},\ }\href
  {https://doi.org/10.1126/science.aar3106} {\bibfield  {journal} {\bibinfo
  {journal} {Science}\ }\textbf {\bibinfo {volume} {362}},\ \bibinfo {pages}
  {308} (\bibinfo {year} {2018})}\BibitemShut {NoStop}%
\bibitem [{\citenamefont {Brunner}\ \emph {et~al.}(2013)\citenamefont
  {Brunner}, \citenamefont {Navascu\'es},\ and\ \citenamefont
  {V\'ertesi}}]{Brunner2013}%
  \BibitemOpen
  \bibfield  {author} {\bibinfo {author} {\bibfnamefont {N.}~\bibnamefont
  {Brunner}}, \bibinfo {author} {\bibfnamefont {M.}~\bibnamefont
  {Navascu\'es}},\ and\ \bibinfo {author} {\bibfnamefont {T.}~\bibnamefont
  {V\'ertesi}},\ }\href {https://doi.org/10.1103/PhysRevLett.110.150501}
  {\bibfield  {journal} {\bibinfo  {journal} {Phys. Rev. Lett.}\ }\textbf
  {\bibinfo {volume} {110}},\ \bibinfo {pages} {150501} (\bibinfo {year}
  {2013})}\BibitemShut {NoStop}%
\bibitem [{\citenamefont {Ku\ifmmode~\acute{s}\else \'{s}\fi{}}\ and\
  \citenamefont {\ifmmode~\dot{Z}\else \.{Z}\fi{}yczkowski}(2001)}]{Karol2001}%
  \BibitemOpen
  \bibfield  {author} {\bibinfo {author} {\bibfnamefont {M.}~\bibnamefont
  {Ku\ifmmode~\acute{s}\else \'{s}\fi{}}}\ and\ \bibinfo {author}
  {\bibfnamefont {K.}~\bibnamefont {\ifmmode~\dot{Z}\else
  \.{Z}\fi{}yczkowski}},\ }\href {https://doi.org/10.1103/PhysRevA.63.032307}
  {\bibfield  {journal} {\bibinfo  {journal} {Phys. Rev. A}\ }\textbf {\bibinfo
  {volume} {63}},\ \bibinfo {pages} {032307} (\bibinfo {year}
  {2001})}\BibitemShut {NoStop}%
\bibitem [{\citenamefont {Popko}(2012)}]{Popko2012}%
  \BibitemOpen
  \bibfield  {author} {\bibinfo {author} {\bibfnamefont {E.}~\bibnamefont
  {Popko}},\ }\href {https://doi.org/10.1201/b12253} {\emph {\bibinfo {title}
  {Divided Spheres: Geodesics and the Orderly Subdivision of the Sphere}}}\
  (\bibinfo {year} {2012})\ pp.\ \bibinfo {pages} {1--404}\BibitemShut
  {NoStop}%
\bibitem [{\citenamefont {Oszmaniec}\ \emph {et~al.}(2017)\citenamefont
  {Oszmaniec}, \citenamefont {Guerini}, \citenamefont {Wittek},\ and\
  \citenamefont {Ac\'{\i}n}}]{Oszmaniec20017}%
  \BibitemOpen
  \bibfield  {author} {\bibinfo {author} {\bibfnamefont {M.}~\bibnamefont
  {Oszmaniec}}, \bibinfo {author} {\bibfnamefont {L.}~\bibnamefont {Guerini}},
  \bibinfo {author} {\bibfnamefont {P.}~\bibnamefont {Wittek}},\ and\ \bibinfo
  {author} {\bibfnamefont {A.}~\bibnamefont {Ac\'{\i}n}},\ }\href
  {https://doi.org/10.1103/PhysRevLett.119.190501} {\bibfield  {journal}
  {\bibinfo  {journal} {Phys. Rev. Lett.}\ }\textbf {\bibinfo {volume} {119}},\
  \bibinfo {pages} {190501} (\bibinfo {year} {2017})}\BibitemShut {NoStop}%
\bibitem [{\citenamefont {Wilde}(2017)}]{wilde2017}%
  \BibitemOpen
  \bibfield  {author} {\bibinfo {author} {\bibfnamefont {M.}~\bibnamefont
  {Wilde}},\ }\href {https://books.google.pl/books?id=gYcHDgAAQBAJ} {\emph
  {\bibinfo {title} {Quantum Information Theory}}}\ (\bibinfo  {publisher}
  {Cambridge University Press},\ \bibinfo {year} {2017})\BibitemShut {NoStop}%
\bibitem [{\citenamefont {Hoeffding}(1963)}]{Hoeff1963}%
  \BibitemOpen
  \bibfield  {author} {\bibinfo {author} {\bibfnamefont {W.}~\bibnamefont
  {Hoeffding}},\ }\href {https://doi.org/10.1080/01621459.1963.10500830}
  {\bibfield  {journal} {\bibinfo  {journal} {Journal of the American
  Statistical Association}\ }\textbf {\bibinfo {volume} {58}},\ \bibinfo
  {pages} {13} (\bibinfo {year} {1963})}\BibitemShut {NoStop}%
\bibitem [{\citenamefont {Bae}\ and\ \citenamefont {Kwek}(2015)}]{Bae2015}%
  \BibitemOpen
  \bibfield  {author} {\bibinfo {author} {\bibfnamefont {J.}~\bibnamefont
  {Bae}}\ and\ \bibinfo {author} {\bibfnamefont {L.-C.}\ \bibnamefont {Kwek}},\
  }\href {https://doi.org/10.1088/1751-8113/48/8/083001} {\bibfield  {journal}
  {\bibinfo  {journal} {J. Phys. A Math. Theor.}\ }\textbf {\bibinfo {volume}
  {48}},\ \bibinfo {pages} {083001} (\bibinfo {year} {2015})}\BibitemShut
  {NoStop}%
\bibitem [{\citenamefont {Bravyi}\ and\ \citenamefont
  {Gosset}(2016)}]{Bravyi2016}%
  \BibitemOpen
  \bibfield  {author} {\bibinfo {author} {\bibfnamefont {S.}~\bibnamefont
  {Bravyi}}\ and\ \bibinfo {author} {\bibfnamefont {D.}~\bibnamefont
  {Gosset}},\ }\href {https://doi.org/10.1103/PhysRevLett.116.250501}
  {\bibfield  {journal} {\bibinfo  {journal} {Phys. Rev. Lett.}\ }\textbf
  {\bibinfo {volume} {116}},\ \bibinfo {pages} {250501} (\bibinfo {year}
  {2016})}\BibitemShut {NoStop}%
\bibitem [{\citenamefont {Bravyi}\ \emph {et~al.}(2019)\citenamefont {Bravyi},
  \citenamefont {Browne}, \citenamefont {Calpin}, \citenamefont {Campbell},
  \citenamefont {Gosset},\ and\ \citenamefont {Howard}}]{Bravyi2019}%
  \BibitemOpen
  \bibfield  {author} {\bibinfo {author} {\bibfnamefont {S.}~\bibnamefont
  {Bravyi}}, \bibinfo {author} {\bibfnamefont {D.}~\bibnamefont {Browne}},
  \bibinfo {author} {\bibfnamefont {P.}~\bibnamefont {Calpin}}, \bibinfo
  {author} {\bibfnamefont {E.}~\bibnamefont {Campbell}}, \bibinfo {author}
  {\bibfnamefont {D.}~\bibnamefont {Gosset}},\ and\ \bibinfo {author}
  {\bibfnamefont {M.}~\bibnamefont {Howard}},\ }\href
  {https://doi.org/10.22331/q-2019-09-02-181} {\bibfield  {journal} {\bibinfo
  {journal} {{Quantum}}\ }\textbf {\bibinfo {volume} {3}},\ \bibinfo {pages}
  {181} (\bibinfo {year} {2019})}\BibitemShut {NoStop}%
\bibitem [{\citenamefont {Pashayan}\ \emph {et~al.}(2022)\citenamefont
  {Pashayan}, \citenamefont {Reardon-Smith}, \citenamefont {Korzekwa},\ and\
  \citenamefont {Bartlett}}]{pashayan2021}%
  \BibitemOpen
  \bibfield  {author} {\bibinfo {author} {\bibfnamefont {H.}~\bibnamefont
  {Pashayan}}, \bibinfo {author} {\bibfnamefont {O.}~\bibnamefont
  {Reardon-Smith}}, \bibinfo {author} {\bibfnamefont {K.}~\bibnamefont
  {Korzekwa}},\ and\ \bibinfo {author} {\bibfnamefont {S.~D.}\ \bibnamefont
  {Bartlett}},\ }\href {https://doi.org/10.1103/PRXQuantum.3.020361} {\bibfield
   {journal} {\bibinfo  {journal} {PRX Quantum}\ }\textbf {\bibinfo {volume}
  {3}},\ \bibinfo {pages} {020361} (\bibinfo {year} {2022})}\BibitemShut
  {NoStop}%
\bibitem [{\citenamefont {Shapira}\ \emph {et~al.}(2003)\citenamefont
  {Shapira}, \citenamefont {Biham}, \citenamefont {Bracken},\ and\
  \citenamefont {Hackett}}]{Shap2003}%
  \BibitemOpen
  \bibfield  {author} {\bibinfo {author} {\bibfnamefont {D.}~\bibnamefont
  {Shapira}}, \bibinfo {author} {\bibfnamefont {O.}~\bibnamefont {Biham}},
  \bibinfo {author} {\bibfnamefont {A.~J.}\ \bibnamefont {Bracken}},\ and\
  \bibinfo {author} {\bibfnamefont {M.}~\bibnamefont {Hackett}},\ }\href
  {https://doi.org/10.1103/PhysRevA.68.062315} {\bibfield  {journal} {\bibinfo
  {journal} {Phys. Rev. A}\ }\textbf {\bibinfo {volume} {68}},\ \bibinfo
  {pages} {062315} (\bibinfo {year} {2003})}\BibitemShut {NoStop}%
\bibitem [{\citenamefont {Govia}\ \emph {et~al.}(2020)\citenamefont {Govia},
  \citenamefont {Bunandar}, \citenamefont {Lin}, \citenamefont {Englund},
  \citenamefont {L\"utkenhaus},\ and\ \citenamefont {Krovi}}]{Govia2020}%
  \BibitemOpen
  \bibfield  {author} {\bibinfo {author} {\bibfnamefont {L.~C.~G.}\
  \bibnamefont {Govia}}, \bibinfo {author} {\bibfnamefont {D.}~\bibnamefont
  {Bunandar}}, \bibinfo {author} {\bibfnamefont {J.}~\bibnamefont {Lin}},
  \bibinfo {author} {\bibfnamefont {D.}~\bibnamefont {Englund}}, \bibinfo
  {author} {\bibfnamefont {N.}~\bibnamefont {L\"utkenhaus}},\ and\ \bibinfo
  {author} {\bibfnamefont {H.}~\bibnamefont {Krovi}},\ }\href
  {https://doi.org/10.1103/PhysRevA.101.062318} {\bibfield  {journal} {\bibinfo
   {journal} {Phys. Rev. A}\ }\textbf {\bibinfo {volume} {101}},\ \bibinfo
  {pages} {062318} (\bibinfo {year} {2020})}\BibitemShut {NoStop}%
\bibitem [{\citenamefont {Spekkens}(2008)}]{Spekkens2008}%
  \BibitemOpen
  \bibfield  {author} {\bibinfo {author} {\bibfnamefont {R.~W.}\ \bibnamefont
  {Spekkens}},\ }\href {https://doi.org/10.1103/PhysRevLett.101.020401}
  {\bibfield  {journal} {\bibinfo  {journal} {Phys. Rev. Lett.}\ }\textbf
  {\bibinfo {volume} {101}},\ \bibinfo {pages} {020401} (\bibinfo {year}
  {2008})}\BibitemShut {NoStop}%
\bibitem [{\citenamefont {Li}\ \emph {et~al.}(2020)\citenamefont {Li},
  \citenamefont {Bu},\ and\ \citenamefont {Wu}}]{LiLu2020}%
  \BibitemOpen
  \bibfield  {author} {\bibinfo {author} {\bibfnamefont {L.}~\bibnamefont
  {Li}}, \bibinfo {author} {\bibfnamefont {K.}~\bibnamefont {Bu}},\ and\
  \bibinfo {author} {\bibfnamefont {J.}~\bibnamefont {Wu}},\ }\href
  {https://doi.org/10.1103/PhysRevA.101.012120} {\bibfield  {journal} {\bibinfo
   {journal} {Phys. Rev. A}\ }\textbf {\bibinfo {volume} {101}},\ \bibinfo
  {pages} {012120} (\bibinfo {year} {2020})}\BibitemShut {NoStop}%
\bibitem [{\citenamefont {Ducuara}\ and\ \citenamefont
  {Skrzypczyk}(2022)}]{liuwinter2019}%
  \BibitemOpen
  \bibfield  {author} {\bibinfo {author} {\bibfnamefont {A.~F.}\ \bibnamefont
  {Ducuara}}\ and\ \bibinfo {author} {\bibfnamefont {P.}~\bibnamefont
  {Skrzypczyk}},\ }\href {https://doi.org/10.1103/PRXQuantum.3.020366}
  {\bibfield  {journal} {\bibinfo  {journal} {PRX Quantum}\ }\textbf {\bibinfo
  {volume} {3}},\ \bibinfo {pages} {020366} (\bibinfo {year}
  {2022})}\BibitemShut {NoStop}%
\bibitem [{\citenamefont {Prakash}\ and\ \citenamefont
  {Gupta}(2020)}]{Praka2020}%
  \BibitemOpen
  \bibfield  {author} {\bibinfo {author} {\bibfnamefont {S.}~\bibnamefont
  {Prakash}}\ and\ \bibinfo {author} {\bibfnamefont {A.}~\bibnamefont
  {Gupta}},\ }\href {https://doi.org/10.1103/PhysRevA.101.010303} {\bibfield
  {journal} {\bibinfo  {journal} {Phys. Rev. A}\ }\textbf {\bibinfo {volume}
  {101}},\ \bibinfo {pages} {010303} (\bibinfo {year} {2020})}\BibitemShut
  {NoStop}%
\bibitem [{\citenamefont {Skrzypczyk}\ and\ \citenamefont
  {Linden}(2019)}]{PaulLind2019}%
  \BibitemOpen
  \bibfield  {author} {\bibinfo {author} {\bibfnamefont {P.}~\bibnamefont
  {Skrzypczyk}}\ and\ \bibinfo {author} {\bibfnamefont {N.}~\bibnamefont
  {Linden}},\ }\href {https://doi.org/10.1103/PhysRevLett.122.140403}
  {\bibfield  {journal} {\bibinfo  {journal} {Phys. Rev. Lett.}\ }\textbf
  {\bibinfo {volume} {122}},\ \bibinfo {pages} {140403} (\bibinfo {year}
  {2019})}\BibitemShut {NoStop}%
\bibitem [{\citenamefont {Ducuara}\ and\ \citenamefont
  {Skrzypczyk}(2020)}]{DucaraPaul2020}%
  \BibitemOpen
  \bibfield  {author} {\bibinfo {author} {\bibfnamefont {A.~F.}\ \bibnamefont
  {Ducuara}}\ and\ \bibinfo {author} {\bibfnamefont {P.}~\bibnamefont
  {Skrzypczyk}},\ }\href {https://doi.org/10.1103/PhysRevLett.125.110401}
  {\bibfield  {journal} {\bibinfo  {journal} {Phys. Rev. Lett.}\ }\textbf
  {\bibinfo {volume} {125}},\ \bibinfo {pages} {110401} (\bibinfo {year}
  {2020})}\BibitemShut {NoStop}%
\bibitem [{\citenamefont {Tan}\ \emph {et~al.}(2021)\citenamefont {Tan},
  \citenamefont {Narasimhachar},\ and\ \citenamefont {Regula}}]{Regula2021}%
  \BibitemOpen
  \bibfield  {author} {\bibinfo {author} {\bibfnamefont {K.~C.}\ \bibnamefont
  {Tan}}, \bibinfo {author} {\bibfnamefont {V.}~\bibnamefont {Narasimhachar}},\
  and\ \bibinfo {author} {\bibfnamefont {B.}~\bibnamefont {Regula}},\ }\href
  {https://doi.org/10.1103/PhysRevLett.127.200402} {\bibfield  {journal}
  {\bibinfo  {journal} {Phys. Rev. Lett.}\ }\textbf {\bibinfo {volume} {127}},\
  \bibinfo {pages} {200402} (\bibinfo {year} {2021})}\BibitemShut {NoStop}%
\bibitem [{\citenamefont {Ducuara}\ and\ \citenamefont
  {Skrzypczyk}(2021)}]{Ducuara2021}%
  \BibitemOpen
  \bibfield  {author} {\bibinfo {author} {\bibfnamefont {A.~F.}\ \bibnamefont
  {Ducuara}}\ and\ \bibinfo {author} {\bibfnamefont {P.}~\bibnamefont
  {Skrzypczyk}},\ }\href@noop {} {\bibinfo {title} {Characterisation of quantum
  betting tasks in terms of arimoto mutual information}} (\bibinfo {year}
  {2021}),\ \Eprint {https://arxiv.org/abs/2106.12711} {arXiv:2106.12711
  [quant-ph]} \BibitemShut {NoStop}%
\bibitem [{\citenamefont {Yu}\ \emph {et~al.}(2021)\citenamefont {Yu},
  \citenamefont {Jayachandran}, \citenamefont {Burchardt}, \citenamefont {Cai},
  \citenamefont {Brunner},\ and\ \citenamefont {Scarani}}]{Yu2021}%
  \BibitemOpen
  \bibfield  {author} {\bibinfo {author} {\bibfnamefont {B.}~\bibnamefont
  {Yu}}, \bibinfo {author} {\bibfnamefont {P.}~\bibnamefont {Jayachandran}},
  \bibinfo {author} {\bibfnamefont {A.}~\bibnamefont {Burchardt}}, \bibinfo
  {author} {\bibfnamefont {Y.}~\bibnamefont {Cai}}, \bibinfo {author}
  {\bibfnamefont {N.}~\bibnamefont {Brunner}},\ and\ \bibinfo {author}
  {\bibfnamefont {V.}~\bibnamefont {Scarani}},\ }\href
  {https://doi.org/10.1103/PhysRevA.104.032414} {\bibfield  {journal} {\bibinfo
   {journal} {Phys. Rev. A}\ }\textbf {\bibinfo {volume} {104}},\ \bibinfo
  {pages} {032414} (\bibinfo {year} {2021})}\BibitemShut {NoStop}%
\bibitem [{\citenamefont {Wootters}(1987)}]{Wootters1987}%
  \BibitemOpen
  \bibfield  {author} {\bibinfo {author} {\bibfnamefont {W.~K.}\ \bibnamefont
  {Wootters}},\ }\href
  {https://doi.org/https://doi.org/10.1016/0003-4916(87)90176-X} {\bibfield
  {journal} {\bibinfo  {journal} {Ann. Phys.}\ }\textbf {\bibinfo {volume}
  {176}},\ \bibinfo {pages} {1} (\bibinfo {year} {1987})}\BibitemShut {NoStop}%
\end{thebibliography}%

\end{document}